\pdfoutput=1
\documentclass[11pt,a4paper]{article}
\usepackage[T1]{fontenc}
\usepackage{xcolor}
\usepackage{jheppub}
\usepackage{cleveref}
\usepackage[inline]{enumitem}
\usepackage{array}
\usepackage{xspace}
\usepackage{arydshln}
\usepackage{mathtools}
\usepackage{mathabx}
\usepackage{mathrsfs}
\usepackage{multirow}
\usepackage{leftidx}
\usepackage{slashed}
\usepackage{cleveref}
\usepackage[final]{showlabels}
\usepackage{subfig}
\usepackage{tikz}
\usepackage{cancel}

\usetikzlibrary{calc,arrows.meta,decorations.markings,topaths}

\definecolor{myblue}{rgb}{0,0,0.7}
\definecolor{alexgreen}{rgb}{0.2,0.7,0.2}

\def\yz#1\yz {{\color{blue} [[YZ: #1]] }}

\def\dd{\textrm{d}}

\DeclareMathOperator*{\Tr}{Tr}

\DeclareMathOperator*{\cut}{Cut}

\def\trV{\tr}

\newcommand{\fpol}{\chi}
\newcommand{\tpol}{\xi}

\def\nn{\nonumber}

\def\Cdot{\!\cdot\!}

\newcommand{\tr}{\ensuremath{\text{tr}}}
\newcommand{\trVS}{\ensuremath{\tr_{(\textnormal{max})}}}
\newcommand{\trvec}{\ensuremath{\tr_{(\textnormal{vec})}}}
\newcommand{\trhyp}{\ensuremath{\tr_{(\textnormal{hyp})}}}

\newcommand{\pol}{\ensuremath{\epsilon}}
\newcommand{\graph}{G}
\newcommand{\sym}{SYM\xspace}
\newcommand{\msym}{\mbox{MSYM}\xspace}
\newcommand{\hsym}{\mbox{1/2-MSYM}\xspace}
\newcommand{\idv}{\ensuremath{\mathbf{1}_{\text{Vec}}}}
\newcommand{\ids}{\ensuremath{\mathbf{1}_{\text{Dirac}}}}
\newcommand{\fsl}{\ensuremath{\cancel{\hspace{1pt}f\hspace{1pt}}}}

\newcommand{\ls}[1]{\ensuremath{\ell_{#1}^2}}
\newcommand{\gmten}[1]{\ensuremath{\Gamma_{#1}^{[10]}}}
\newcommand{\gmsix}[1]{\ensuremath{\Gamma_{#1}^{[6]}}}

\newcommand{\nmeven}[1]{\ensuremath{N^{\text{m-even}}_{#1}}}
\newcommand{\nmodd}[1]{\ensuremath{N^{\text{m-odd}}_{#1}}}
\newcommand{\nmax}[1]{\ensuremath{N^{\text{max}}_{#1}}}

\newcommand{\ntmeven}[1]{\ensuremath{\widetilde{N}^{\text{m-even}}_{#1}}}
\newcommand{\ntmodd}[1]{\ensuremath{\widetilde{N}^{\text{m-odd}}_{#1}}}

\newcommand{\nheven}[1]{\ensuremath{N^{\text{h-even}}_{#1}}}
\newcommand{\nhodd}[1]{\ensuremath{N^{\text{h-odd}}_{#1}}}
\newcommand{\nh}[1]{\ensuremath{N^{\text{hyp}}_{#1}}}
\newcommand{\nhb}[1]{\ensuremath{N^{\overline{\text{hyp}}}_{#1}}}

\newcommand{\ntheven}[1]{\ensuremath{\widetilde{N}^{\text{h-even}}_{#1}}}
\newcommand{\nthodd}[1]{\ensuremath{\widetilde{N}^{\text{h-odd}}_{#1}}}

\newcommand{\nveven}[1]{\ensuremath{N^{\text{v-even}}_{#1}}}
\newcommand{\nvodd}[1]{\ensuremath{N^{\text{v-odd}}_{#1}}}
\newcommand{\nv}[1]{\ensuremath{N^{\text{vec}}_{#1}}}
\newcommand{\nvb}[1]{\ensuremath{N^{\text{vec}'}_{#1}}}

\tikzset{
    cuts/.style={dash pattern=on 2pt off 1pt,draw=blue},
    sexchange/.pic={\tikzset{every node/.style={font=\scriptsize}}
    \pgfmathsetmacro{\h}{0.5}
    \pgfmathsetmacro{\w}{0.75}
    \pgfmathsetmacro{\l}{0.5}
    \coordinate (A) at (0,\h);
    \coordinate (B) at (0,-\h);
    \coordinate (C) at (\w,\h);
    \coordinate (D) at (\w,-\h);
    \draw [thick] (A) -- (B);
    \draw [thick] (A) -- (C);
    \draw [thick] (B) -- (D);
    \draw [thick] (A) -- ++ (135:\l) node[left=-1pt]{$1$};
    \draw [thick] (B) -- ++ (-135:\l) node[left=-1pt]{$2$};
    \draw [thick] (C) -- ++ (45:\l);
    \draw [thick] (D) -- ++ (-45:\l);
    \draw [thick] (\w,0) -- ++(\l,0);
    \filldraw [fill=gray!50!white] (\w,0) ellipse (0.25cm and 0.7cm);
    },
    uexchange/.pic={\tikzset{every node/.style={font=\scriptsize}}
    \pgfmathsetmacro{\h}{0.5}
    \pgfmathsetmacro{\w}{0.75}
    \pgfmathsetmacro{\l}{0.5}
    \coordinate (A) at (0,\h);
    \coordinate (B) at (0,-\h);
    \coordinate (C) at (\w,\h);
    \coordinate (D) at (\w,-\h);
    \draw [thick] (A) -- (B);
    \draw [thick] (A) -- (C);
    \draw [thick] (B) -- (D);
    \draw [thick] (A) -- ++ (135:\l) node[left=-1pt]{$2$};
    \draw [thick] (B) -- ++ (-135:\l) node[left=-1pt]{$1$};
    \draw [thick] (C) -- ++ (45:\l);
    \draw [thick] (D) -- ++ (-45:\l);
    \draw [thick] (\w,0) -- ++(\l,0);
    \filldraw [fill=gray!50!white] (\w,0) ellipse (0.25cm and 0.7cm);
    },
    texchange/.pic={\tikzset{every node/.style={font=\scriptsize}}
    \pgfmathsetmacro{\h}{0.5}
    \pgfmathsetmacro{\w}{0.75}
    \pgfmathsetmacro{\l}{0.5}
    \coordinate (A) at (0,0);
    \coordinate (C) at (\w,\h);
    \coordinate (D) at (\w,-\h);
    \draw [thick] (A) -- (C);
    \draw [thick] (A) -- (D);
    \draw [thick] (A) -- ++ (-\l,0) -- ++(135:\l) node[left=-1pt]{$1$};
    \draw [thick] (A) ++ (-\l,0) -- ++ (-135:\l) node[left=-1pt]{$2$};
    \draw [thick] (C) -- ++ (45:\l);
    \draw [thick] (D) -- ++ (-45:\l);
    \draw [thick] (\w,0) -- ++(\l,0);
    \filldraw [fill=gray!50!white] (\w,0) ellipse (0.25cm and 0.7cm);
    \path (0,\h) ++ (135:\l) node [left=-1pt]{$\phantom{2}$};
    \path (0,-\h) ++ (-135:\l) node [left=-1pt]{$\phantom{1}$};
    }
}

\crefname{figure}{figure}{figures}
\Crefname{figure}{Figure}{Figures}

\title{Perfecting one-loop BCJ numerators in SYM and supergravity}

\author[a,b,c]{Alex Edison,}
\author[d,e]{Song He,}
\author[b,f,c]{Henrik Johansson,}
\author[b,c]{Oliver Schlotterer,}
\author[g,c]{Fei Teng}
\author[h]{and Yong Zhang}

\affiliation[a]{Department of Physics and Astronomy, Northwestern University, Evanston, IL 60208, USA}
\affiliation[b]{Department of Physics and Astronomy, Uppsala University, Box 516, 75120 Uppsala, Sweden}
\affiliation[c]{Kavli Institute for Theoretical Physics,
University of California, Santa Barbara, CA 93106, USA}
\affiliation[d]{CAS Key Laboratory of Theoretical Physics, Institute of Theoretical Physics, Chinese Academy of Sciences, Beijing 100190, China}
\affiliation[e]{School of Fundamental Physics and Mathematical Sciences, Hangzhou Institute for Advanced Study \&  ICTP-AP, UCAS, Hangzhou 310024, China}
\affiliation[f]{Nordita, Stockholm University and KTH Royal Institute of Technology, Hannes Alfv\'{e}ns v\"{a}g 12, 10691 Stockholm, Sweden}
\affiliation[g]{Institute for Gravitation and the Cosmos, \\ Pennsylvania State University, University Park, PA 16802, USA}
\affiliation[h]{Perimeter Institute for Theoretical Physics, Waterloo, ON N2L 2Y5, Canada.}

\emailAdd{alexander.edison@northwestern.edu}
\emailAdd{songhe@itp.ac.cn}
\emailAdd{henrik.johansson@physics.uu.se}
\emailAdd{oliver.schlotterer@physics.uu.se}
\emailAdd{fei.teng@psu.edu}
\emailAdd{yzhang@perimeterinstitute.ca}

\abstract{
We take a major step towards computing $D$-dimensional one-loop amplitudes in general gauge theories, compatible with the principles of unitarity and the color-kinematics duality. For $n$-point amplitudes with either supersymmetry multiplets or generic non-supersymmetric matter in the loop, simple all-multiplicity expressions are obtained for the maximal cuts of kinematic numerators of $n$-gon diagrams. At $n=6,7$ points with maximal supersymmetry, we extend the cubic-diagram numerators to encode all contact terms, and thus solve the long-standing problem of \emph{simultaneously} realizing the following properties: color-kinematics duality, manifest locality, optimal power counting of loop momenta, quadratic rather than linearized Feynman propagators, compatibility with double copy as well as all graph symmetries. Color-kinematics dual representations with similar properties are presented in the half-maximally supersymmetric case at $n=4,5$ points. The resulting gauge-theory integrands and their supergravity counterparts obtained from the double copy are checked to reproduce the expected ultraviolet divergences.
\vskip 0.35in
\begin{center}
{\it In Memory and Spirit of Professor Lars Brink}
\end{center}
}

\preprint{UUITP--45/22 \\
\phantom{~} \hfill NORDITA 2022-074} 

\begin{document}

\addtocontents{toc}{\protect\setcounter{tocdepth}{2}}

\maketitle

\section{Introduction}
\label{sec:1}

In a variety of situations, perturbative computations in gravitational theories have been
reduced to \emph{double copies} of
gauge-theory building blocks. 
The double-copy notion has expanded into
a rich research program of steadily
growing scope which unravels surprising connections between different types of quantum field theories and string theories. The reader is referred to the
reviews~\cite{Bern:2019prr, Bern:2022wqg} and the white paper~\cite{Adamo:2022dcm} for an overview.

The gravitational double copy is crucially driven by the color-kinematics duality in gauge theories~\cite{Bern:2008qj} and
dramatically improved our structural understanding of and
computational reach for gravitational scattering amplitudes.  A well-established manifestation of the color-kinematics duality and
gravitational double copy is based on an organization of loop
integrands in terms of diagrams with trivalent vertices or \emph{cubic graphs} for short.

In the loop integrand of gauge theories, each cubic graph is decorated by Feynman propagators, color factors and \emph{kinematic numerators}.  These numerators depend on polarization data and momenta and furnish the loop integrals with structure that cannot be straightforwardly read off from the graph. The color-kinematics duality is manifest if the kinematic numerators share both
the symmetry properties and, crucially, the Jacobi identities of the corresponding color factors.
Gravitational loop integrands then follow from their gauge-theory counterparts by trading the color factors for another copy of such kinematic numerators, possibly from a different gauge theory~\cite{Bern:2010ue}.
%such kinematic numerators, possibly from a second gauge-theory copy~\cite{Bern:2010ue}.

The general existence of gauge-theory numerators at loop level that satisfy all kinematic Jacobi relations is of conjectural status, but it has been broadly observed for different theories of varying spacetime dimension and with or without supersymmetry. For super-Yang-Mills (SYM) theory with maximal supersymmetry, Jacobi-satisfying numerators
are for instance known at the four-point level
through four loops~\cite{Green:1982sw,Bern:1997nh,Bern:2010ue,Bern:2012uf} and for five-point amplitudes at least through two loops~\cite{Carrasco:2011mn, Mafra:2015mja}.
Five- and six-loop non-planar four-point integrands are known in the literature~\cite{Bern:2017ucb,Carrasco:2021otn}, but finding the corresponding Jacobi-satisfying numerators remains a technically difficult open problem. However, for maximally supersymmetric form factors there are substantial results for such representations through five loops~\cite{Boels:2012ew,Yang:2016ear,Lin:2021kht,Lin:2021lqo}.

The state of the art for higher-multiplicity one-loop amplitudes, which admit color-kinematics duality and the double copy, depends on propagator choices and spacetime dimension.  In ten-dimensional SYM~\cite{Gliozzi:1976qd,Brink:1976bc}, five-point numerators are compatible with standard Feynman propagators quadratic in loop momentum~\cite{Mafra:2014gja}. However, supergravity loop integrands at $n \geq 6$ points in  $D\geq 4$ dimensions are until now only available in a double-copy form where the propagators are linearized with respect to loop momenta~\cite{He:2017spx} as prescribed by ambitwistor string theories~\cite{Geyer:2015bja,Geyer:2015jch}.
For the one-loop six-point SYM numerators found in the literature~\cite{He:2017spx}, the conversion to the traditional quadratic propagators breaks some of the kinematic Jacobi identities. An interesting variant of kinematic Jacobi relations was solved  by the six- and seven-point SYM numerators on quadratic propagators in ref.~\cite{Bridges:2021ebs}, but it remains an open problem to derive one-loop supergravity double-copy integrands from these representations beyond five points.

The apparent challenges of finding Jacobi-satisfying numerators, both for four-point amplitudes at five loops and beyond~\cite{Bern:2017ucb,Carrasco:2021otn}, and for one-loop amplitudes at six points and beyond~\cite{He:2017spx,Bridges:2021ebs}, begs the question whether the standard color-kinematics duality and double copy~\cite{Bern:2010ue} are limited to low multiplicities and loop orders,
%low orders and multiplicities in perturbation theory, 
or if these challenges are only of technical nature to be resolved by further progress.\footnote{Note that in certain cases color-kinematics duality is simpler to manifest in four dimensions, where numerators either can be built using Gram determinants \cite{Carrasco:2011mn, Bjerrum-Bohr:2013iza}, or can be found to admit compact one-loop all-multiplicity formulae for maximally-helicity-violating amplitudes~\cite{He:2015wgf} and for self-dual Yang-Mills theory~\cite{Monteiro:2011pc,Boels:2013bi}.
Furthermore, in any dimension, the loop-level double copy admits a generalized form~\cite{Bern:2017yxu}, where the need for kinematic Jacobi identities and cubic graphs is relaxed, and this was used in the construction of the five-loop supergravity integrand~\cite{Bern:2017ucb, Bern:2018jmv}.} We here provide evidence in support of the latter.

In this work, we overcome the mentioned one-loop obstacles and present six- and seven-point numerators for external gluons in ten-dimensional SYM. These numerators
satisfy all kinematic Jacobi identities and graph symmetries, belong to diagrams with quadratic propagators, and double copy to type II supergravity integrands with any combination of external gravitons, $B^{\mu \nu}$-fields and dilatons. Furthermore, the numerators are local and written in dimension-agnostic notation using polarization vectors and linearized field strengths. As a crucial new ingredient in comparison to earlier $D$-dimensional numerators at $n \geq 6$ points, we encounter inverse propagators in the (one-particle irreducible) $n$-gon numerators. There is no known smoking gun for these contact terms in string-theory approaches\footnote{We nevertheless borrow the characteristic $t_8$-tensor (defined in \cref{teight}) known from string amplitudes to organize the Lorentz-contractions of the linearized field strengths in the $n$-gon numerators. Together with the complete dihedreal symmetry we impose, this led to compact ansaetze with considerably fewer free parameters for the $D$-dimensional polarization dependence in comparison to earlier field-theory constructions.} which might be part of the reasons that the numerators in this work have not been constructed before.

Corresponding numerators of maximally supersymmetric gauge theories in $D<10$ dimensions follow from straightforward dimensional reduction. We have crosschecked the result by computing the ultraviolet (UV) divergences in $D=8$ and $D=10$ for the six-point supergravity amplitude obtained from double copy. The correct $D=8$ result is obtained through a delicate interplay of the parity even and parity odd parts of the gauge-theory numerators, and the calculation relies on
%was computationally straightforward due to 
the transparent UV properties of Feynman integrals with quadratic (rather than linearized) propagators.

At one loop, half-maximally supersymmetric $n$-point integrands are known to inherit structural properties of the corresponding maximally supersymmetric $(n{+}2)$-point integrands~\cite{Berg:2016wux, Berg:2016fui}. Accordingly, the new results of ten-dimensional SYM provide information that allows us to explicitly construct the six-dimensional four- and five-point gauge-theory numerators that preserve eight supercharges and manifest color-kinematics duality on quadratic propagators. In the literature, Jacobi-satisfying four-point numerators for half-maximal supersymmetry have previously been computed at one~\cite{Chiodaroli:2013upa,Johansson:2014zca} and two loops~\cite{Johansson:2017bfl,Duhr:2019ywc}, both for four-dimensional external gluons and manifest four-dimensional supersymmetry. For gluons in $D \geq 4$ dimensions, kinematic Jacobi relations of the earlier four-point numerators in ref.~\cite{Berg:2016fui} were tied to linearized propagators \cite{He:2017spx}.

Finally, as a first stepping stone towards obtaining one-loop numerators beyond seven points, we provide compact all-multiplicity expressions for $n$-gon numerator contributions to the maximal cuts with general internal matter. The $n$-gon is the {\it basis diagram} of the one-loop integrand at $n$ points in the sense that it determines all other diagrams via kinematic Jacobi identities~\cite{DelDuca:1999rs, Bern:2011rj}. The
maximal cuts are derived by sewing cubic trees for gauge theories with any number of supercharges between 0 and 16. Tentative supersymmetry cancellations are manifested via simple relations between traces over the vector and spinor representations of the Lorentz group, akin to the forward-limit construction of gauge-theory numerators in ref.~\cite{Edison:2020uzf}.

As a guide to the reader: kinematic numerators that satisfy all the same Jacobi identities as  the color factors will be referred to as \emph{BCJ numerators}. Furthermore, we will use the following acronyms for the theories of chief interest to this work: 

{\bf \sym} denotes a generic supersymmetric Yang-Mills theory;

{\bf \msym} denotes the maximal \sym theories that preserve 16 supercharges;

{\bf \hsym} denotes the half-maximal \sym theories that preserve 8 supercharges. \vskip0.2cm

\noindent
This work is organized as follows: in \cref{sec:review}, we review the color-kinematics duality and double copy for one-loop amplitudes, as well as the method of generalized unitarity and state sewing.  
In \cref{sec:kin-build}, we review the needed kinematic building blocks for maximal supersymmetry, including Lorentz traces of linearized field strengths and multiparticle polarizations, which are key ingredients for compact BCJ numerators. In \cref{sec:maxSYM}, we spell out the new results for BCJ numerators for \msym in $D\leq 10$ dimensions. Based on compact general formulae for maximal-cut contributions with any number of external gluons, we present local six- and seven-point one-loop BCJ numerators that respect all graph symmetries, obey expected loop-momentum power counting, and belong to cubic graphs with standard (quadratic) Feynman propagators. In \cref{sec:UV}, we carry out important consistency checks by explicitly computing the UV divergences of \msym and type II supergravity amplitudes up to six points. In \cref{sec:half-max}, we present similar perfected numerators for four- and five-point one-loop amplitudes in $(D\leq 6)$-dimensional \hsym and crosscheck the UV divergences of the supergravity integrands obtained from the double copy.
Conclusions and outlook are given in \cref{sec:6}, 
%after which follows 
followed by four appendices with further details that complement the main text.

\section{Review: color-kinematics duality and unitarity at one loop}\label{sec:review}

Consider a one-loop $n$-point amplitude for a generic purely-adjoint SYM theory in $D$ spacetime dimensions. The color-dressed amplitude can be written as a sum over all cubic $n$-point one-loop graphs,\footnote{For later convenience, we use a somewhat non-standard overall normalization and remove the coupling.
} 
\begin{align}\label{eq:integrand}
    \mathcal{A}^{\text{1-loop}}=\sum_{\graph\in\text{cubic}}\int\frac{\dd^{D}\ell}{i\pi^{D/2}}\frac{1}{S_{\graph}}\frac{C_{\graph}N_{\graph}}{D_{\graph}}\,,
\end{align}
where for each cubic graph $\graph$ the color factor $C_{\graph}$ is given by the contraction of structure constants $f^{abc}$ associated to the vertices, and $D_{\graph}^{-1}$ is the product of Feynman propagators associated to each internal edge. The kinematic numerator $N_{\graph}$ encodes the remaining nontrivial kinematic data -- polarizations, external momenta and loop momenta. In principle, the numerators can be obtained from conventional Feynman rules, after resolving contact vertices into cubic ones, for instance using auxiliary fields in the adjoint representation of the gauge group. However, by relaxing the use of Feynman rules and auxiliary fields, we are free to find more beneficial properties and explicit formulae for the numerators.  Finally, the symmetry factor $S_{\graph}$ removes the overcounting coming either from the  sum over redundant graph permutations or from the integral phase space of each graph. 

\subsection{Color-kinematics duality and double copy}
\label{sec:ck-dual}

The color factors associated with different graphs 
%are linearly dependent. 
obey linear relations. In particular, if we have a triplet of graphs that differ by only one internal edge, then their color factors are related by the Jacobi identity,
\begin{align}\label{eq:jac}
    C\left[\vcenter{\hbox{\begin{tikzpicture}
        \pic{sexchange};
    \end{tikzpicture}}}\right]-C\left[\vcenter{\hbox{\begin{tikzpicture}
        \pic{uexchange};
    \end{tikzpicture}}}\right] &=C\left[\vcenter{\hbox{\begin{tikzpicture}
        \pic{texchange};
    \end{tikzpicture}}}\right]\,,
\end{align}
where the notation assumes insertions of structure constants where cubic vertices are exposed, and the connected lines represent contractions over adjoint indices. 
An amplitude is said to satisfy the \emph{color-kinematics duality} if its kinematic numerators satisfy the same identities as the color factors~\cite{Bern:2008qj, Bern:2010ue},
\begin{align}\label{eq:Njac}
    N\left[\vcenter{\hbox{\begin{tikzpicture}
        \pic{sexchange};
    \end{tikzpicture}}}\right]-N\left[\vcenter{\hbox{\begin{tikzpicture}
        \pic{uexchange};
    \end{tikzpicture}}}\right] &=N\left[\vcenter{\hbox{\begin{tikzpicture}
        \pic{texchange};
    \end{tikzpicture}}}\right]\,.
\end{align}
We call the duality satisfying kinematic numerators \emph{BCJ numerators}, and a loop integrand constructed out of BCJ numerators is called a \emph{BCJ representation} (of the loop integrand).

At loop level, a generic constructive derivation of BCJ numerators is currently unknown (apart from the one-loop ambitwistor construction \cite{He:2017spx} leading to linearized propagators), instead one typically needs to proceed using an ansatz-based approach. A major advantage of the BCJ representation is that gavitational amplitudes can be obtained with minimal work. If a set of BCJ numerators is known, say $\widetilde{N}_{\graph}$, then one can simply replace the color factors $C_{\graph}\rightarrow \widetilde{N}_{\graph}$ in \cref{eq:integrand} to obtain a gravitational amplitude,
\begin{align}
\label{sgamplitude}
    M^{\text{1-loop}}=\sum_{\graph\in\text{cubic}}\int\frac{\dd^D\ell}{i\pi^{D/2}}\frac{1}{S_{\graph}}\frac{\widetilde{N}_{\graph}N_{\graph}}{D_{\graph}}\,.
\end{align}
The two copies of numerators, $\widetilde{N}_{\graph}$ and $N_{\graph}$, can be identical to each other, or encode different external and internal particles, or belong to entirely different gauge theories. Thus the double copy has the potential to give a large variety of different gravitational amplitudes.
See the review~\cite{Bern:2019prr} for a comprehensive discussion on known examples of color-kinematics duality and double-copy constructions.

The color-factor Jacobi relations can be systematically reduced as in \cref{eq:jac}, and a minimal basis can be obtained. For $n$-point amplitudes at tree level, a well-known minimal basis corresponds to the set of half-ladder graphs with legs $1$ and $n$ fixed, which is often called the Del Duca-Dixon-Maltoni (DDM) basis~\cite{DelDuca:1999rs}. At one loop, a similar minimal DDM basis consists of the set of
$n$-gon graphs, or equivalently all one-particle irreducible (1PI) cubic graphs. Because of color-kinematics duality, the BCJ numerators can be expressed in same DDM-type bases. Thus for one-loop $n$-point amplitudes, the $n$-gon numerators provide the independent BCJ \emph{basis numerators}.

In the following, $n$-gon basis numerators will be denoted by $N_{12\ldots n}$, where the subscript specifies the external-leg ordering $(12\ldots n)$. More generally, we will use
$N_{\mathsf{A}_1\mathsf{A}_2\ldots\mathsf{A}_m}$ with $m<n$ to denote the numerator of a one-particle reducible (1PR) graph with an $m$-gon
structure. Each word $\mathsf{A}_i$ of length greater than one represents a dangling tree connected to the
$m$-gon whose structure will be specified by nested brackets. The external-leg ordering is simply given by the
concatenation $\mathsf{A}_1\mathsf{A}_2\ldots\mathsf{A}_m$ with all
the brackets removed. For example, $N_{[12]345}$ stands for the
five-point box numerator in which leg $1$ and $2$ appear in a dangling
three-point tree, and we will also encounter six-point box numerators $N_{[[12]3]456}$ and $N_{[1[23]]456}$ associated with four-point trees.

\subsubsection{Color-ordered gauge-theory amplitudes}

The kinematic numerators are not unique and usually contain a large amount of redundancy, which is called generalized gauge freedom. It is useful to assemble gauge-invariant color-ordered amplitudes to check if free parameters are indeed part of the gauge redundancy. A one-loop single-trace color-ordered gauge-theory amplitude is given by
\begin{align}
A^\text{1-loop}_{\rm gauge}(1,2,\ldots,n)=
\int \frac{ \dd^{D} \ell}{i\pi^{D/2}} \bigg[
\frac{N_{12\ldots n}(\ell)}{\ell^2 (\ell{+}k_1)^2 (\ell{+}k_{12})^2 \ldots
(\ell{+}k_{12\ldots n-1})^2}+\ldots
\bigg] \, ,
\label{strconv}
\end{align}
where $k_{12\ldots i}= k_1+k_2+\ldots+k_i$, and the ellipsis includes all cubic 1PR graphs that are compatible with the prescribed color
ordering.  We make a global choice for the position of the loop momentum $\ell$ for all graphs, to be located between legs $1$ and $n$ such that all loop propagators are of the form $1/(\ell +k_{12\ldots i})^{2}$. With this choice the integrand in~\cref{strconv} can be made to algebraically satisfy linearized gauge invariance $\pol_i\rightarrow k_i$ (in $D\le 10$ and $D\le 6$ for \msym and \hsym, respectively). 
In this paper, we will not discuss one-loop double-trace amplitudes since they are obtained from linear combinations of their single-trace
counterparts~\cite{Bern:1994zx}.

\subsubsection{Crossing symmetry}

We demand the numerators to satisfy the same symmetry properties as the color factors under relabeling of external legs. Suppose the $n$-gon basis numerator $N_{12\ldots n}$ is known, then any other basis numerator is obtained by simply relabeling the external legs, 
\begin{align}\label{eq:relabel}
    N_{\rho(1)\rho(2)\ldots\rho(n)}=N_{12\ldots n} \, \big|_{i\rightarrow\rho(i)}\,,
   \quad \rho \in S_n \,,
\end{align}
where the position of loop momentum is unchanged under the permutation $\rho$. The $n$-gons have a dihedral symmetry $D_n$, and we expect the numerators to satisfy the same symmetry. Namely, the action of $D_n$ includes cyclic shifts and reflections, which act non-trivially on the loop momentum,
\begin{align}
\label{crossing_symmetry}
    & N_{i,i+1,\ldots,n,1,2,\ldots,i-1}(\ell)=N_{12\ldots n}(\ell-k_{12\ldots i-1})\,,\nonumber\\
    & N_{1,n,n-1,\ldots,2}(\ell) = (-1)^n N_{12\ldots n}(-\ell-k_1)\,.
\end{align}
The properties~\eqref{eq:relabel}
and~\eqref{crossing_symmetry} are equivalent to imposing manifest \emph{crossing symmetry} for
the integrand. Imposing the above constraints significantly reduces the number of free parameters in the numerator functions.\footnote{Strictly speaking, the crossing symmetry of the integrand is not necessary, since the crossing symmetry of the full amplitude can be recovered after integration. However, relaxing it would lead to a proliferating number of free parameters related to the position of the loop momentum, since the Jacobi relations will explore numerators with arbitrarily shifted $\ell$. Moreover, imposing the crossing symmetry of the amplitudes would involve integration, which is more difficult to accomplish in practice.}
%
%Note that the cyclic symmetry is necessary to impose in order to have a finite number of numerator functions, since the Jacobi relations will otherwise explore numerators with arbitrarily shifted arguments. 
%
After imposing color-kinematics duality, all other numerators of 1PR graphs are related to the $n$-gons through Jacobi identities. Note that for one-loop graphs, the 1PR numerators inherit a complete set of crossing-symmetry requirements from the $n$-gons via the  kinematic Jacobi relations.

For non-planar gauge-theory and gravity integrands, as given in~\cref{eq:integrand,sgamplitude}, a good global notion of loop momentum is difficult to obtain. This is sometimes referred to as the ``labeling problem'' as it is unclear how to assign a loop momentum $\ell$ such that required cancellations (from gauge redundancy) between different graphs happens algebraically, without shifting $\ell$ locally for each graph. Since the kinematic Jacobi relation~\eqref{eq:Njac} relates planar and non-planar graphs it is not possible to work under the assumption that $\ell$ is fixed, instead one must allow for local shifts in $\ell$ in order to make sure that all possible Jacobi relations are satisfied.
The labeling problem is well-understood in the field-theory limit of closed-string amplitudes, where one can extract certain global definitions of loop momenta in gravitational loop integrands from a canonical dissection of the string worldsheet~\cite{Tourkine:2019ukp}, also see~\cite{Casali:2020knc} for a discussion in the context of the color-kinematics duality. 

We may sidestep the labeling problem by imposing the full crossing symmetry. The three graphs that form a Jacobi triplet, as in \cref{eq:Njac}, may have different loop-momentum labelings, but due to the dihedral part of the crossing symmetry we can always perform local shifts of $\ell$ for individual graphs to align them. With the crossing symmetry imposed all the kinematic Jacobi identities can be considered to hold by construction, as they are simply used to define the reducible numerators. The nontrivial part of construction is instead delegated to the matching of the ansatz with the physical information coming from unitarity cuts. Cuts both in gravity and for color-dressed non-planar gauge theory can be conveniently performed using the crossing symmetric numerators, as it is now easy to shift $\ell$ to align it with the loop momentum in the cut expressions. See \cref{sec:grcut} for a more detailed discussion.

\subsubsection{Supersymmetry}

Incorporating supersymmetry brings further simplification to kinematic numerators. 
In the presence of maximal supersymmetry, i.e.\ 16 supercharges in gauge theory~\cite{Gliozzi:1976qd,Brink:1976bc} or 32 supercharges in supergravity~\cite{Cremmer:1978km}, it is expected that one can find numerator representations where the highest power of loop momentum for an $m$-gon \msym subgraph is no worse than $(m{-}4)$~\cite{Bern:1992ad,Bern:1993tz}. For type II supergravity one expects twice that number, $2(m{-}4)$, consistent with the no-triangle property~\cite{Bjerrum-Bohr:2006xbk} and the double copy. Hence, the one-loop $n$-point \msym amplitudes can be written as linear combinations of scalar boxes and higher $(m{\le} n)$-gon tensor integrals of rank $(m{-}4)$.
We require that our \msym BCJ  numerators manifest this good power counting, implying that triangles, bubbles and tadpoles are absent. As a consequence, one can find box numerators $\nmax{\mathsf{A}_1\mathsf{A}_2\mathsf{A}_3\mathsf{A}_4}$ that are independent of loop momentum, and must be totally symmetric under permutations of the arguments $\mathsf{A}_i$,
\begin{align} 
\label{eq:no-tri-jac}    
\nmax{\mathsf{A}_1\mathsf{A}_2\mathsf{A}_3\mathsf{A}_4}=\nmax{\mathsf{A}_{\rho(1)}\mathsf{A}_{\rho(2)}\mathsf{A}_{\rho(3)}\mathsf{A}_{\rho(4)}}\,,\quad ~~~ \rho \in S_4 \,,
\end{align}
such that no triangle numerators are generated by anti-symmetrization of the arguments via the Jacobi identity. 

For half-maximal supersymmety with 8 supercharges~\cite{Brink:1976bc}, one can find integrand representations where $m$-gon numerators have at most $(m{-}2)$ powers of loop momentum. This implies that triangle and bubble graphs now contribute~\cite{Bern:1994zx,Bern:1994cg}. At the one-loop level, we expect that one can impose a no-tadpole\footnote{Two-loop amplitudes with internal hypermutiplets, however, give rise to certain tadpole diagrams~\cite{Johansson:2017bfl}.} condition, 
$\nh{\mathsf{A}_1\mathsf{A}_2}=\nh{\mathsf{A}_2\mathsf{A}_1}$, on bubble
numerators. As indicated by the superscript, we here consider numerators with only hypermutiplets running in the loop, as these are sufficient for obtaining general multiplet contributions, see section~\ref{sec:half-max}.

Crossing symmetry and color-kinematics duality reduce the ansatz to a single unknown basis numerator, $N_{12\ldots n}$, and all the remaining numerators are related by relabeling and Jacobi identities. The assumptions of good power counting and no-triangle (no-tadpole) property further significantly simplify the ansatz. Although these assumptions are not strictly necessary, whenever it is possible to impose strict power counting it will greatly simplify the computation.\footnote{See ref.~\cite{Mogull:2015adi}, where color-kinematics duality is realized using
  worse-than-naive power counting.}
In sections~\ref{sec:maxSYM} and~\ref{sec:half-max}, we give the maximal-cut contributions to $n$-gon numerators of \msym and \hsym as closed all-multiplicity formulae, with the aforementioned properties. Hence, an ansatz will only be required for the remaining leftover numerator terms, which vanish on the maximal cut.
%Hence, only for the remaining leftover numerator terms, which vanish on the maximal cut, will an ansatz be required. 

\subsection{Unitarity cuts}
\label{sec:cuts}
While the color-kinematics duality does an excellent job of relating numerators to one another, the numerators still need to be constrained further using physical input from the desired theory.
This is most conveniently done by matching the numerators against gauge-invariant generalized unitarity cuts~\cite{Bern:1994zx,Bern:1994cg,Britto:2004nc}.

At one loop, generalized unitarity cuts are relatively straightforward to construct directly by sewing together tree amplitudes of appropriate states. For a gauge-theory amplitude we may without loss of generality assume that external legs follow a certain color ordering. Then to calculate the cut associated with a (generally non-cubic) graph $\graph$
with edges $E(\graph)$ and vertices $V(\graph)$, we take
\begin{equation}
  \text{Cut}_{\graph}
  = \sum_{\substack{\text{states}\\\text{of } E(\graph)}} \prod_{v \in V(\graph)} A(v) \,,
  \label{cutgg}
\end{equation}
where $A(v)=A^{\text{tree}}(v)$ are
color-ordered tree amplitudes, and the ``tree'' superscript is suppressed throughout this section to avoid cluttering.
Gauge invariance requires that the momenta associated to all edges of $\graph$ must satisfy on-shell conditions.

\subsubsection{Polarization variables}

We will mainly consider one-loop amplitudes with external gluons, but allow arbitrary (massless) matter to run in the loop. To obtain maximal cuts associated with $n$-gon diagrams at $n$ points, we need to use color-stripped three-point amplitudes for the gluon self-coupling, as well as the interactions with fermions and scalars,
\begin{subequations}
\label{eq:3ptree}
\begin{align}
  A(1^g,2^g,3^g)
  &= -\; \pol_1 \cdot f_2 \cdot \pol_3 - (\pol_2 \cdot k_3) \pol_1 \cdot \pol_3\,, \label{eq:a3g}\\
  A(1^\psi,2^g,3^\psi)
  &
  =-\frac{1}{2}\bar{\fpol}_1 \slashed{\epsilon}_2\fpol_3 = -\;\bar{\fpol}_1 \fsl_2 \tpol_3 - (\pol_2 \cdot k_3) \bar{\fpol}_1 \tpol_3 \label{eq:a3f} \,, \\
   A(1^s,2^g,3^s) &= -\;\pol_2\cdot k_3\,,
\end{align}
\end{subequations}
where the polarizations $\pol_i$ and the spinor wavefunctions $\fpol_i$ satisfy the transversality conditions $\pol_i \cdot k_i=0$ and massless Dirac equations $\slashed{k}_i \fpol_i=0$, respectively. In \cref{eq:a3f}, we have massaged the fermion amplitude into a slightly unconventional form by introducing another spinor $\tpol_i = \frac{ \slashed{q}\fpol_i}{2q \cdot k_i}$, through the use of a null reference momentum $q$, such that $\slashed{k}_i\tpol_i=\fpol_i$. In both the fermion and pure-gluon amplitude we have exposed a
linearized field strength $f_i^{\mu\nu}= k_i^\mu \epsilon_i^\nu- k_i^\nu \epsilon_i^\mu$. The natural contractions between a vector $v^\mu$, or tensor $f_i^{\mu\nu}$, and Dirac gamma matrices are defined as
\begin{align}\label{eq:fsl}
\slashed{v} = v^\mu \Gamma_\mu
\, , \qquad
    \fsl_i=\frac{1}{4}f_i^{\mu\nu}\Gamma_{\mu\nu}=\frac{1}{2}\slashed{k}{}_i\slashed{\pol}{}_{i}\,,
\end{align}
where $\Gamma_{\mu\nu}=\frac{1}{2}(\Gamma_{\mu}\Gamma_{\nu}-\Gamma_{\nu}\Gamma_{\mu})$. We will mainly work in even dimensions $d$, where $\Gamma^\mu$ are $2^{d/2}\times 2^{d/2}$ matrices subject to the Clifford algebra $\Gamma_{\mu}\Gamma_{\nu}+ \Gamma_{\nu}\Gamma_{\mu}=2 \eta_{\mu \nu}$, and where we can introduce the chiral gamma matrix
\begin{align}\label{eq:g5}
    \Gamma=i^{-d/2+1}\Gamma^0\Gamma^1\ldots\Gamma^{d-1}\,,
\end{align}
to obtain the chiral projectors $P_{\pm}=\frac{1}{2}(\ids\pm\Gamma)$. 
Left- and right-handed Weyl spinors
$\chi_{\rm l}$ and $\chi_{\rm r}$ satisfy
$P_{-}(\chi_{\rm l},\chi_{\rm r})=(\chi_{\rm l},0)$
and $P_{+}(\chi_{\rm l},\chi_{\rm r})=(0,\chi_{\rm r})$.

The state sum on each internal edge includes a sum over
particle types crossing the edge as well as a sum over polarization
states for those particles.  Focusing on \sym,
the sum over states will generally be split into vector, fermion, and
scalar components,
\begin{subequations}
\begin{align}
  & \text{vector:}& & \sum_{\text{states}} \pol_{-\ell}^{\mu} \pol{}_{\ell}^{\vphantom{\mu}\nu}
  = \eta^{\mu \nu} - \frac{\ell^{\mu}q^{\nu}+\ell^{\nu}q^{\mu}}{\ell \cdot q} \,, \label{eq:pol-phys} \\
  & \text{fermion:}& & \sum_{\text{states}} \fpol_{-\ell} \bar{\fpol}_{\ell} = -\slashed{\ell}\,, \label{eq:ferm-phys}\\
  & \text{scalar:}& & \sum_{\text{states}} 1 = 1 \,.
\end{align}
\end{subequations}
In the presence of chiral fermions, the state sum~\eqref{eq:ferm-phys} is supplemented by a chiral projector $P_{\pm}$.
The null reference momentum $q$ introduced through \cref{eq:pol-phys} for the gluon state sum will cancel out in the final expression for the cut, except for a contribution equivalent to the Faddeev-Popov ghost running in the loop. We adopt the regularization scheme that all the external momenta and polarizations are strictly in $d \in \mathbb N$ dimensions, and the loop momentum $\ell$ is in non-integer dimensions $D=d{-}2\epsilon$.  All internal bosonic and fermionic polarizations are defined in $D_s>D$ dimensions. 
We will take the smooth limit $D_s=d$ after all the Lorentz and gamma-matrix algebra has been performed. This prescription is similar to FDH~\cite{Bern:2002zk}, but carried out in general dimensions.

\subsubsection{State sewing for internal gluons}
\label{sec:state_sewing_gluons}

The state sewing for $n$-gon maximal cuts for 1PI graphs is straightforward to perform using the expressions \cref{eq:3ptree} for three-point tree amplitudes. We align all tree amplitudes such that the state sewings are between the last leg of one amplitude and the first of the next. First, the sewing for scalars corresponds to trivial multiplication. For gluonic states, the sewing of two three-point amplitudes is effectively implemented by repeated matrix products of linearized field strengths $f_j$,
%the concatenation of field strength matrix products, 
for example,
\begin{align}
 \sum_{\substack{\text{states}\\\text{across } \ell_2}}\!\!
    A(\ell_1^{g}, 2^{g}, -\ell_2^{g})A(\ell_2^{g}, 3^{g} ,-\ell_3^{g}) = \pol_{\ell_1}\Cdot\Big[(\pol_2\Cdot\ell_2)\idv\!-\!f_2\Big]\Cdot\Big[(\pol_3\Cdot\ell_3)\idv\!-\!f_3\Big]\Cdot\pol_{-\ell_3}\,,
\end{align}
where $\idv = (\delta_{\mu}^{\nu})$ is the identity matrix. The gluons labelled by $\pm\ell_j^g$ will later on be associated with the edges of an $n$-gon diagram carrying momenta $\ell_j = \ell + k_{12\ldots j}$. Notably, the $q$-dependence in the state sum~\eqref{eq:pol-phys} drops out since both three-point amplitudes are gauge invariant. Therefore, we can effectively use $\sum_{\text{states}}\pol_{-\ell}^{\mu}\pol_{\ell}^{\nu}\rightarrow\eta^{\mu\nu}$ to sew all but the last edge. For the final state sewing, on the last edge $\ell_n$, one encounters two cases of tensor structures sandwiched between two polarizations:
\begin{subequations}
\begin{align}
    \label{eq:lastsewb}
    \sum_{\text{states}}\pol_{\ell_n} \Cdot \pol_{-\ell_n} = d - 2 &= \tr(\idv)-2\,,\\
    \sum_{\text{states}}\pol_{\ell_n} \Cdot f_i \Cdot f_j \Cdot \ldots \Cdot \pol_{-\ell_n} &= \tr \left( f_i f_j \ldots\right)+\mathcal{O}(q\Cdot k_i)+\mathcal{O}(q\Cdot\pol_i)
    \,.
    \label{eq:lastsewcde}
\end{align}
\end{subequations}
The first case gives the trace of the state projector, $d{-}2$, which is $q$-independent. As indicated, in the second case one potentially has terms linear in the independent variables $q\cdot k_i$ and $q\cdot \pol_i$. However, by gauge invariance, the unitarity cut must be independent of the reference vector $q$. These terms are bound to drop out in the full cut expression,
%%%
due to kinematic identities applied to the coefficients.
Therefore, it is legitimate to simply use $\eta_{\mu\nu}$ in all state sums that involve the linearized field strengths, and to effectively drop the contributions of $q\Cdot k_i$ and $q\Cdot\pol_i$ in \cref{eq:lastsewcde}.

The maximal cut of a box (the quadruple cut) with a gluon loop is a good example for the simple
structure of cuts constructed in this manner,
\begin{align}\label{eq:ncut_gluon}
    \cut_{\text{vec}}\!\left[\!\!\vcenter{\hbox{
    \begin{tikzpicture}[scale=0.7,every node/.style={font=\scriptsize}]
    \pgfmathsetmacro{\r}{1.1}
        \pgfmathsetmacro{\l}{0.6}
        \pgfmathsetmacro{\y}{-2.2}
            \foreach \x in {1,2,3,4} {
                \pgfmathsetmacro{\a}{135-90*(\x-1)}
                \pgfmathsetmacro{\b}{90-90*(\x-1)}
                \coordinate (\x) at ( \a : \r);
                \fill (\x) circle (1pt);
                \draw[thick] (\x) -- ++ ( \a : \l) node[label={[label distance=-8pt]\a:\x}] {};
                \draw[cuts] (\b:{\r-0.6}) -- (\b:{\r});
                \node at ( \b: {\r + 0.35} ) {$\ell_{\x}$};
            }
            \draw[thick] (2) -- (1);
            \draw[thick] (3) -- (2);
            \draw[thick] (4) -- (3);
            \draw[thick] (1) -- (4);
            \node at (0,0) {\scalebox{1.75}{$\circlearrowleft$}};
    \end{tikzpicture}
    }}\!\!\right]&=\sum_{\text{states}}A(-\ell_4^g,1^g,\ell_1^g)A(-\ell_1^g,2^g,\ell_2^g)A(-\ell_2^g,3^g,\ell_3^g)A(-\ell_3^g,4^g,\ell_4^g)\nonumber\\
    &=\tr\prod_{i=1}^{4}\Big[(\pol_i\Cdot\ell_i)\idv - f_i\Big]-2\prod_{i=1}^4\pol_{i}\Cdot\ell_i \,,
\end{align}
where the last term comes from the $-2$ in \cref{eq:lastsewb}, which is equivalent to the Faddeev-Popov ghost contribution. 
Here and after, we assume that all external momenta are outgoing, and $\ell$ runs anti-clockwise in the loop as indicated by the arrow. The 
edges that are intersected by a dashed line indicate cut propagators.
%edge that is intersected by a dashed line indicates a cut propagator.

\subsubsection{State sewing for arbitrary internal states}

The state sewing of fermions requires manipulations of gamma-matrix identities. Since our external particles are all gluons, the final result should be given by Lorentz dot products and traces. The chiral matrix $\Gamma$ is defined using only the first $d$ gamma matrices following \cref{eq:g5}. The chiral trace
%projects out all the extra-dimensional components and 
results in a $d$-dimensional totally antisymmetric Levi-Civita tensor (following 't Hooft and Veltman~\cite{tHooft:1972tcz}),
\begin{align}\label{eq:ctrace}
\tr(\Gamma\Gamma^{\mu_1}\Gamma^{\mu_2}\ldots\Gamma^{\mu_{d}})=i(-2i)^{d/2}\varepsilon^{\mu_1\mu_2\ldots\mu_d}\,,\qquad 0\leqslant\mu_i\leqslant d{-}1\,,
\end{align}
which projects out all the extra-dimensional components of the contracting vectors and
satisfies the Schouten (over-antisymmetrization) identity
\begin{align}\label{eq:schouten}
\varepsilon_{[\mu_1\mu_2\ldots\mu_d}v_{\mu_{d+1}]}=0\,.
\end{align}
We also define for convenience the following notations when the Levi-Civita tensor is contracted with vectors and linearized field strengths,
\begin{align}
    & \varepsilon_{d}(v_1,v_2,\ldots,v_{d})=\varepsilon_{\mu_1\mu_2\ldots\mu_{d}}v_1^{\mu_1}v_2^{\mu_2}\ldots v_{d}^{\mu_{d}}\,, \nonumber\\
    & \varepsilon_{d}^{\mu}(v_2,\ldots,v_{d})=\varepsilon^{\mu}{}_{\mu_2\ldots\mu_{d}}v_2^{\mu_2}\ldots v_{d}^{\mu_{d}}\,,\\
    &\varepsilon_{d}(\ldots,f_i,\ldots)=2\,\varepsilon_{d}(\ldots,k_i,\pol_{i},\ldots)\,.\nonumber
\end{align}
Further details on computing the fermionic cuts are relegated to \cref{sec:fcut}.

Using the above sewing method, we can immediately write down the $n$-gon maximal cut contributed by a spectrum of $n_{\text{v}}$ gluons, $n_{\text{s}}$ scalars, $n_{\text{l}}$ left-handed and $n_{\text{r}}$ right-handed chiral fermions running in the loop, 
\begin{align}
  \cut_{n_i}\!\left[\!\!\vcenter{\hbox{
    \begin{tikzpicture}[scale=0.75,every node/.style={font=\scriptsize}]
    \pgfmathsetmacro{\r}{1.1}
    \pgfmathsetmacro{\l}{0.6}
    \pgfmathsetmacro{\y}{-2.2}
    \foreach \x in {1,2,...,8} {
        \pgfmathsetmacro{\a}{200.5-45*(\x-1)}
        \pgfmathsetmacro{\b}{180-45*(\x-1)}
        \coordinate (\x) at ( \a : \r);
        \fill (\x) circle (1pt);
    }
    \foreach \x in {1,2,3,4} {
        \pgfmathsetmacro{\a}{200.5-45*(\x-1)}
        \draw[thick] (\x) -- ++ ( \a : \l) node[label={[label distance=-8pt]\a:\x}] {};
    }
    \foreach \x in {1,2,3} {
        \pgfmathsetmacro{\a}{200.5-45*(\x-1)}
        \pgfmathsetmacro{\b}{180-45*(\x-1)}
        \draw[cuts] (\b:{\r-0.4}) -- (\b:{\r+0.2});
        \node at ( \b: {\r + 0.5} ) {$\ell_{\x}$};
    }
    \draw[thick] (8) -- ++ ( 245.5 : \l) node[label={[label distance=-8pt]245.5:$n$}] {};
    \draw[cuts] (225:{\r-0.4}) -- (225:{\r+0.2});
    \node at ( 225: {\r + 0.5} ) {$\ell_{n}$};
    \draw[thick] (8) -- (1);
    \draw[thick] (2) -- (1);
    \draw[thick] (3) -- (2);
    \draw[thick] (4) -- (3);
    \draw[thick,dashed] (8) -- (7) -- (6) -- (5) -- (4);
    \node at (0,0) {\scalebox{1.75}{$\circlearrowleft$}};
    \end{tikzpicture}
    }}\!\!\right]
  &= n_{\text{v}} \tr \prod_{i=1}^n \Big((\pol_i \cdot \ell_i)\idv - f_i \Big) + (n_{\text{s}}-2n_{\text{v}})\prod_{i=1}^n (\pol_i \cdot \ell_i)
   \nonumber \\[-1cm]
  &\quad - \frac{n_{\text{l}}+n_{\text{r}}}{4}\tr \prod_{i=1}^n \Big(( \pol_i \cdot \ell_i)\ids - \fsl_i \Big) \nonumber\\
  &\quad + \frac{n_{\text{l}}-n_{\text{r}}}{4}\tr\Big[\Gamma\slashed{\pol}{}_1\slashed{\ell}{}_1\prod_{i=2}^{n}\Big((\pol_i\cdot\ell_i)\ids - \fsl_i\Big)\Big]\,,
    \label{eq:ngon-mc}
\end{align}
with the dimensional dependence encoded in the traces of the two identity matrices $\idv$ and $\ids$, namely, $\tr(\idv)=d$ and $\tr(\ids)=2^{d/2}$. A convenient closed formula to evaluate both parity even and odd gamma traces is given in ref.~\cite{Edison:2020uzf}.
Significant simplifications happen
for various supersymmetric configurations, which we will elaborate in
\cref{sec:kin-build,sec:maxSYM,sec:half-max} below. The same expression can also be obtained by further cutting the forward-limit integrand constructed in ref.~\cite{Edison:2020uzf}.

\subsubsection{Beyond maximal cuts}

The construction of non-maximal cuts associated with 1PR diagrams is also straightforward: we just insert higher-point tree amplitudes following the
prescribed color ordering. For example, the following one-mass box cut at five points can be computed by
\begin{align}
    \text{Cut}\!\left[\!\!\!\vcenter{\hbox{
    \begin{tikzpicture}[scale=0.7,every node/.style={font=\scriptsize}]
    \pgfmathsetmacro{\r}{1.1}
        \pgfmathsetmacro{\l}{0.6}
        \pgfmathsetmacro{\y}{-2.2}
        \coordinate (1) at (135:\r);
        \draw[cuts] (90:{\r-0.6}) -- (90:{\r});
        \node at ( 90: {\r + 0.35} ) {$\ell_{2}$};
        \draw[thick] (1) -- ++ (-\l,0) node[label={[label distance=-6pt]180:1}] {};
        \draw[thick] (1) -- ++ (0,\l) node[label={[label distance=-6pt]90:2}] {};
        \fill (1) circle (1pt);
        \foreach \x in {3,4,5} {
            \pgfmathsetmacro{\a}{135-90*(\x-2)}
            \pgfmathsetmacro{\b}{90-90*(\x-2)}
            \coordinate (\x) at ( \a : \r);
            \fill (\x) circle (1pt);
            \draw[thick] (\x) -- ++ ( \a : \l) node[label={[label distance=-8pt]\a:\x}] {};
            \draw[cuts] (\b:{\r-0.6}) -- (\b:{\r});
            \node at ( \b: {\r + 0.35} ) {$\ell_{\x}$};
        }
        \draw[thick] (3) -- (1);
        \draw[thick] (4) -- (3);
        \draw[thick] (5) -- (4);
        \draw[thick] (1) -- (5);
        \node at (0,0) {\scalebox{1.75}{$\circlearrowleft$}};
    \end{tikzpicture}
    }}\!\!\!\right]=\sum_{\text{states}}A(\ell_5,1^g,2^g,-\ell_2)A(\ell_2,3^g,-\ell_3)A(\ell_3,4^g,-\ell_4)A(\ell_4,5^g,-\ell_5)\,, \nonumber
    \\[-.8cm]
\end{align}
where the state sum is computed in the same way as
discussed above.  Even with the inclusion of higher-point amplitudes,
the process of resolving the product of trees into the generalized
unitarty cut can be greatly simplified by picking appropriate
kinematic bases for the tree amplitudes. In this work, we use a
representation for tree-level amplitudes that is particularly well
suited to one-loop unitarity constructions~\cite{Edison:2020ehu}: the
polarizations of the first and last particle in a particular color
ordering, which can be either fermions or gluons, always appear as the
beginning and end of a chain of field-strength contractions. The
sewing process is thus identical to the maximal-cut case, and we will
get various Lorentz traces of $f_i$ in the result. The major
difference is that the prefactors of the traces become more
complicated. They are no longer simply $\pol_i \cdot \ell_i$, but
instead include terms like $\pol_i \cdot f_j \cdots p_m$ where $i$ and
$j$ are external labels, and $p_m$ can be either a loop or external
momentum.

Once the target expressions \eqref{cutgg} for cuts are calculated from tree amplitudes, we can assemble our
numerators associated with the relevant one-loop diagrams into the same cuts. Imposing these two assemblies of a given cut to match results in constraints on the numerator ansatz.  Continuing with
the one-mass-box example, the cut can be constructed from
numerators as 
\begin{align}\label{eq:omb-num}
    \text{Cut}\!\left[\!\!\!\vcenter{\hbox{
    \begin{tikzpicture}[scale=0.7,every node/.style={font=\scriptsize}]
    \pgfmathsetmacro{\r}{1.1}
        \pgfmathsetmacro{\l}{0.6}
        \pgfmathsetmacro{\y}{-2.2}
        \coordinate (1) at (135:\r);
        \draw[cuts] (90:{\r-0.6}) -- (90:{\r});
        \node at ( 90: {\r + 0.35} ) {$\ell_{2}$};
        \draw[thick] (1) -- ++ (-\l,0) node[label={[label distance=-6pt]180:1}] {};
        \draw[thick] (1) -- ++ (0,\l) node[label={[label distance=-6pt]90:2}] {};
        \fill (1) circle (1pt);
        \foreach \x in {3,4,5} {
            \pgfmathsetmacro{\a}{135-90*(\x-2)}
            \pgfmathsetmacro{\b}{90-90*(\x-2)}
            \coordinate (\x) at ( \a : \r);
            \fill (\x) circle (1pt);
            \draw[thick] (\x) -- ++ ( \a : \l) node[label={[label distance=-8pt]\a:\x}] {};
            \draw[cuts] (\b:{\r-0.6}) -- (\b:{\r});
            \node at ( \b: {\r + 0.35} ) {$\ell_{\x}$};
        }
        \draw[thick] (3) -- (1);
        \draw[thick] (4) -- (3);
        \draw[thick] (5) -- (4);
        \draw[thick] (1) -- (5);
        \node at (0,0) {\scalebox{1.75}{$\circlearrowleft$}};
    \end{tikzpicture}
    }}\!\!\!\right]=\left[\frac{N_{12345}}{(\ell_5+k_1)^2}+\frac{N_{[12]345}}{(k_1+k_2)^2}\right]_{\ls{2}=\ls{3}=\ls{4}=\ls{5}=0}\,.
\end{align}  
The two terms on the right correspond to the two ways of blowing up the four-point contact vertex while preserving the color ordering.
Matching the cuts in all physical channels is a necessary and sufficient condition for obtaining valid diagram numerators of an amplitude~\cite{Bern:1994zx,Bern:1994cg}. We will consider cuts of gravity integrands in \cref{sec:grcut}. 

In practice, if a theory has manifest $\ell^{m-k}$ power counting for $m$-gon numerators, one only needs to check cuts down to $k$-gons, namely, box cuts for \msym and bubble cuts for \hsym. We demonstrate this point by showing that the integrands that satisfy the above condition are gauge invariant. To start with, the gauge variation $\pol_i\rightarrow k_i$ does not increase the power of $\ell$. It can only interact with propagators through turning $\pol_i\cdot\ell$ into $k_i\cdot\ell$, which can then be expressed in terms of inverse propagators $\frac{1}{2}(\ell_i^2-\ell_{i-1}^2)-k_i\Cdot k_{1\ldots i-1}$ and feeds down to a lower-gon topology. Let us illustrate an important implication for \msym as an example theory. The box numerators of \msym are independent of $\ell$, such that their gauge variation cannot generate triangles, namely, the gauge variations can only land and get canceled on boxes at most. Therefore, matching box cuts guarantees the gauge invariance for \msym. 

For generic theories, if we match all possible cuts down to tadpoles, then the \emph{only} terms we might miss are those with no propagators.
However, these terms integrate to zero in dimensional regularization, so we can freely add a compensating term to the numerator without changing the amplitudes. This is only a concern for theories with higher-derivative operators, which are beyond the scope of the current work.

\section{Kinematic building blocks 
for maximal supersymmetry}
\label{sec:kin-build}

In this section, we review the kinematic building blocks that yield the compact BCJ numerators in later sections. The kinematic numerators of \msym may contain both parity even and odd parts,
\begin{align}
    \nmax{} = \nmeven{} + \nmodd{}\,,
\end{align}
and a relative minus sign for the opposite chirality.
At $n$ points, the
parity even parts of \msym $n$-gon numerators will involve traces over
linearized gluon field strengths in special combinations of vector-
and spinor-representations of the Lorentz group to be reviewed in \cref{sec:traces}. The
numerators of lower-gon topologies in turn are most
conveniently represented via local multiparticle polarizations, see \cref{sec:mp-pol}. Gauge theories with maximal supersymmetry are chiral only in ten dimensions. Thus the parity odd part $\nmodd{}$ is proportional to the Levi-Civita tensor $\varepsilon_{10}$ and absent in lower dimensions.

\subsection{Lorentz traces for maximal supersymmetry}
\label{sec:traces}

In the construction of $d$-dimensional gauge-theory integrands from ambitwistor-string
methods \cite{Geyer:2015bja, Geyer:2015jch, He:2017spx, Geyer:2017ela,
  Edison:2020uzf}, the forward limits in gluons and gluinos introduce
traces over linearized field strengths in the vector and spinor
representations of $SO(1,d{-}1)$, respectively. 
By \cref{eq:ngon-mc} with $n_{\textrm{v}}= n_{\textrm{l}}=1$
and $n_{\textrm{s}}= n_{\textrm{r}}=0$, the loop integrand of ten-dimensional \msym can be conveniently expressed in terms of the following fine-tuned linear
combination of vector and Dirac-spinor traces \cite{Edison:2020uzf}:
\begin{align}\label{deftrvs}
\trVS(f_1f_2\ldots f_n)=2\Big[&\trV(f_1f_2\ldots f_n)
\\
&  - \frac{1}{4^{n+1}}f_1^{\mu_1 \nu_1} f_2^{\mu_2 \nu_2}\ldots f_n^{\mu_n \nu_n}\tr\big(
\gmten{\mu_1 \nu_1}
\gmten{\mu_2 \nu_2}
\ldots
\gmten{\mu_n \nu_n}
\big)\Big]\,.\nonumber
\end{align}
We use $\gmten{\mu}$ to denote Dirac gamma matrices in ten dimensions and $\gmten{\mu\nu}$ is defined in \cref{eq:fsl}.\footnote{
We note that the $\trVS$ structure arises naturally when a ten-dimensional  gauge supermultiplet is running in the loop, see~\cite{Edison:2020uzf} for a demonstration from the forward-limit perspective. In particular, $\trVS$ in this work has the following relation with
the 
%objects
vector and spinor traces denoted by $\textrm{tr}_{\rm V}$ and $\textrm{tr}_{\rm S}$ in the reference:
\begin{align*}
    \trVS(f_1f_2\ldots f_n)=2  \, \textrm{tr}_{\rm V}(1,2,\ldots,n)- \textrm{tr}_{\rm S}(1,2,\ldots,n) \,.
\end{align*}} 
Using the Clifford algebra and the spinor trace $\tr\big(\gmten{\mu}\gmten{\nu}\big)=32\eta_{\mu\nu}$, one can show that the combination of traces in \cref{deftrvs} vanishes up to length three,
\begin{align}
\trVS(f_1 )  = \trVS(f_1 f_2 )  = \trVS(f_1 f_2 f_3) =0 \,,
\label{vanish}
\end{align}
and reduces to the famous $t_8$-tensor at four points,
\begin{align}
\trVS(f_1 f_2 f_3 f_4) &=  t_8(f_1,f_2,f_3,f_4)\,,
\label{teight} 
\\
t_8(f_1,f_2,f_3,f_4) &= 
\trV(f_1 f_2 f_3 f_4)
- \frac{1}{4} \trV(f_1 f_2 ) \trV(f_3 f_4) + {\rm cyclic}(2,3,4)\,,\nonumber
\end{align}
which is symmetric in $\{f_1,f_2,f_3,f_4\}$. We will see in \cref{sec:maxSYM} that \cref{vanish} leads to the improved $\ell$ power counting for \msym.

At higher multiplicity, we adapt the group-theory
decomposition of traces \cite{vanRitbergen:1998pn, Bandiera:2020aqn} to the Lorentz group and decompose 
$\trVS$
into matrix commutators of $f_j$ and totally symmetric tensors $t_{2n}$,
\begin{align}\label{t2nten}
    t_{2n}(f_1,f_2,\ldots,f_n)= \frac{1}{(n{-}1)!}\sum_{\rho\in S_{n-1}}\trVS(f_1f_{\rho(2)}\ldots f_{\rho(n)})\,.
\end{align}
For example, one can show that \cite{Edison:2020uzf}
\begin{align}
    t_{12}(f_1,f_2,f_3,f_4,f_5,f_6)&=\frac{1}{24} \trV(f_1f_2)t_8(f_3,f_4,f_5,f_6)+(1,2|1,2,3,4,5,6)\,. \label{t12ten}
\end{align}
Here and below, the notation $+\;(1,2,\ldots,k|1,2,\ldots,m)$ denotes a sum over all the permutations of the form
%\begin{align}
%\left(\begin{array}{*{4}{wc{1.2cm}}:*{4}{wc{1.2cm}}}
 %   1 & 2 & \cdots & k & k{+}1 & k{+}2 & \cdots & m \\
 %   \downarrow & \downarrow & \cdots & \downarrow & \downarrow & \downarrow & \cdots & \downarrow \\
 %   i_1 & i_2 & \cdots & i_k & j_{k+1} & j_{k+2} & \cdots & j_m
%\end{array}\right) \, ,    
%\end{align}
\begin{align}
\left(\begin{array}{cccc|cccc }
    1 \ &\ 2 \ &\ \cdots \ &\ k \ \ &\ \ k{+}1 \ & \ k{+}2 \ &\ \cdots \ & \ m \\
    \downarrow \ & \ \downarrow\ &\ \cdots \ &\ \downarrow \ \ &\ \ \downarrow \ & \ \downarrow \ & \  \cdots \ &\ \downarrow \\
    i_1 \ &\ i_2 \ & \ \cdots \ &\ i_k \ \ &\ \ j_{k+1} \ & \ j_{k+2} \ &\ \cdots \ &\ j_m
\end{array}\right) \, ,    
\end{align}
where $(i_1,\ldots,i_k,j_{k+1},\ldots,j_m)$ is a permutation of $(1,2,\ldots,m)$ that satisfies $i_a<i_{a+1}$ and $j_a<j_{a+1}$. In all, there are $\genfrac{(}{)}{0pt}{1}{m}{k}$ such permutations, for instance $\genfrac{(}{)}{0pt}{1}{6}{2}=15$ terms in \cref{t12ten}. 

At five and six points, the group-theoretical decomposition yields
\begin{align}
\trVS(f_1 f_2 f_3 f_4 f_5)&= \frac{1}{2} t_8(f_1,[f_2,f_3],f_4,f_5)+(2,3|2,3,4,5)\,, \label{5ptex} \\
\trVS(f_1 f_2 f_3 f_4 f_5 f_6)&= t_{12}(f_1,f_2,f_3,f_4,f_5,f_6)
\label{6ptex}  \\
&\hspace{-3cm} +
\frac{1}{6}\Big[t_8(f_1,
[[f_2,f_3],f_4],f_5,f_6)+t_8(f_1,
[[f_4,f_3],f_2],f_5,f_6)+(2,3,4|2,3,4,5,6)\Big]
\nonumber \\
&\hspace{-3cm}+ \frac{1}{4}\Big[t_8(f_1,
[f_2,f_3],[f_4,f_5],f_6)
    +t_8(f_1,
    [f_2,f_4],[f_3,f_5],f_6)\nonumber \\
&\hspace{-1cm}
    +t_8(f_1,
    [f_2,f_5],[f_3,f_4],f_6)+(2,3,4,5|2,3,4,5,6)\Big]\,,
\nonumber
\end{align}
and a similar expression for $\trVS(f_1 f_2 \ldots f_7)$ in
terms of $t_8$ and $t_{12}$ tensors
can be found in \cref{app:7trace}. The brackets
$[f_i,f_j]^{\mu \nu} = f_i^{\mu \lambda} (f_j)_{\lambda}{}^{\nu}- f_i^{\nu \lambda} (f_j)_{\lambda}{}^{\mu}$ refer to matrix commutators of linearized field strengths.
%The group-theory decomposition of \cite{vanRitbergen:1998pn, Bandiera:2020aqn} drastically simplifies 
The relative factors of vector and spinor traces in $\trVS$ drastically simplify
the group-theory decomposition of \cite{vanRitbergen:1998pn, Bandiera:2020aqn}
at any multiplicity $n$ since terms with $n{-}2$ nested brackets vanish by \cref{vanish}. Moreover, the reflection property $\trVS(f_1 f_2 \ldots f_n) = (-1)^n \trVS(f_n \ldots f_2 f_1)$ only allows for terms with an even (odd) number of brackets if $n$ is even (odd).

Although the definition \eqref{deftrvs} of $\trVS$ involves gamma matrices in $d=10$, after converting spinorial traces to vectorial ones, $\trVS$  no longer refers to the number of spacetime dimensions.
Namely, as combinations of Lorentz traces, $\trVS$ takes the same form under dimensional reductions.
Thus, $\trVS$ written in terms of $\trV(f_i \ldots)$ provides a compact and dimension-agnostic way to package external gluon polarizations of \msym in any $D\leqslant 10$. On the other hand, the \msym is chiral specifically in ten dimensions. There exist parity odd terms contributed by chiral traces of gamma matrices, which lead to a ten-dimensional Levi-Civita tensor in pure gluon numerators. Such terms are absent for \msym in lower dimensions, which are always non-chiral.

\subsection{Multiparticle polarizations}
\label{sec:mp-pol}

As we will see, the numerators of 1PR diagrams admit compact representations in terms of multiparticle polarizations
$\pol^\mu_{\mathsf{P}},f^{\mu \nu}_{\mathsf{P}}$
labelled by ordered sets $\mathsf{P}=12\ldots p$. They descend from multiparticle superfields in the pure-spinor formalism \cite{Mafra:2014oia} by extracting the components as in
\cite{Mafra:2015vca}, and we will discard the contributions from external fermions in this work. Multiparticle polarizations and superfields were already used in earlier one-loop numerators for 1PR diagrams \cite{Mafra:2014gja, Mafra:2015mja, He:2017spx, Edison:2020ehu, Bridges:2021ebs}, and they form the backbone of the $n$-point BCJ numerators at tree level in \cite{Mafra:2011kj, Mafra:2015vca}, see \cite{Mafra:2022wml} for a review. 

To begin with a brief review of the key equations, the two-particle polarization and field strength are given by
\begin{align}
    \pol_{12}^{\mu}&=(\pol_1\Cdot k_2)\pol_2^{\mu}-(\pol_2\Cdot k_1)\pol_1^{\mu}+\frac{1}{2}(\pol_1\Cdot\pol_2)(k_1-k_2)^{\mu}\,,\nonumber\\
    f_{12}^{\mu\nu}&=(\pol_1\Cdot k_2)f_2^{\mu\nu}-(\pol_2\Cdot k_1)f_1^{\mu\nu}+[f_1,f_2]^{\mu\nu}
    \label{2part} \\
    &= k_{12}^{\mu}\pol_{12}^{\nu}- k_{12}^{\nu}\pol_{12}^{\mu}-(k_{1}\Cdot k_2)(\pol_{1}^{\mu}\pol_2^{\nu}-\pol_{1}^{\nu}\pol_2^{\mu})\,, \notag
\end{align}
where we use the notation $k_{12\ldots p}=k_1 {+}k_2{+}\ldots{+}k_p$ for multiparticle momenta here and below.
The three-particle polarization and field strength are defined as
\begin{align}
\pol_{123}^{\mu}&=\frac{1}{2}\Big[(\pol_{12}\Cdot k_3)\pol_3^{\mu}-(\pol_3\Cdot k_{12})\pol_{12}^{\mu}+\pol_{12,\nu}f_3^{\nu\mu}-\pol_{3,\nu}f_{12}^{\nu\mu}\Big] -k_{123}^{\mu}h_{123}\,,
\notag \\
f_{123}^{\mu\nu}&=k_{123}^{\mu}\pol_{123}^{\nu}-(k_{12}\Cdot k_3)\pol_{12}^{\mu}\pol_3^{\nu}-(k_1\Cdot k_2)(\pol_{1}^{\mu}\pol_{23}^{\nu}-\pol_2^{\mu}\pol_{13}^{\nu})-(\mu\leftrightarrow \nu)\,,
     \label{3part}
\end{align}
where $h_{123}=\frac{1}{12}\pol_{1,\mu}f_2^{\mu\nu}\pol_{3,\nu}+\text{cyclic}(1,2,3)$. Their four-particle counterparts are for instance reviewed in section 6.1 of \cite{Edison:2020uzf}.

The multiparticle field strength can appear on the same footing as the linearized field strength in $t_8$-tensors. Kinematic factors like $t_8(f_{12},f_3,f_4,f_5)$ will be used to express numerators of 1PR diagrams at five points and beyond. In addition, $t_8$-tensors can also be dressed with multiparticle polarizations. For example, we will later use
\begin{align}
    t_8^{\mu}(1,2,3,4,5)&=\pol_{1}^{\mu}t_8(f_2,f_3,f_4,f_5)+\textrm{cyclic}(1,2,3,4,5)\,, \nonumber\\
    t_8^{\mu}(12,3,4,5,6)&=\pol_{12}^{\mu}t_8(f_3,f_4,f_5,f_6)+\pol_{3}^{\mu}t_8(f_{12},f_4,f_5,f_6)+\pol_{4}^{\mu}t_8(f_{12},f_3,f_5,f_6)\nonumber\\
    &\quad+\pol_{5}^{\mu}t_8(f_{12},f_3,f_4,f_6)+\pol_{6}^{\mu}t_8(f_{12},f_3,f_4,f_5)\,.
    \label{deft8vec}
\end{align}
In general, we can define the vectorial generalization of the $t_8$-tensor as
\begin{align}
    t_8^{\mu}(\mathsf{A},\mathsf{B},\mathsf{C},\mathsf{D},\mathsf{E})=\pol_{\mathsf{A}}^{\mu}t_8(f_\mathsf{B},f_\mathsf{C},f_\mathsf{D},f_\mathsf{E})+\textrm{cyclic}(\mathsf{A},\mathsf{B},\mathsf{C},\mathsf{D},\mathsf{E})\,,
\end{align}
where $\{\mathsf{A,B,C,D,E}\}$ are generic multiparticle labels.

\section{BCJ numerators with maximal supersymmetry}
\label{sec:maxSYM}

With the kinematic building blocks established, we turn our attention
to the construction of multiparticle BCJ numerators in \msym at one loop.
From the discussion of \cref{sec:ck-dual}, we know that
a BCJ-numerator representation can be specified in terms of
only the $n$-gon numerator $\nmax{12\ldots n}$.  This numerator must
match the $n$-gon maximal cut, but will also contain contact degrees
of freedom that are set to zero by the maximal-cut conditions.

For the parity even sector, the generic maximal cut given in \cref{eq:ngon-mc} is already crossing symmetric. We can directly lift it off the cut and realize it as part of the $n$-gon numerators. We further specialize to ten dimensions and \msym, which has a particle spectrum given by $n_{\text{v}}=1$, $(n_{\text{l}},n_{\text{r}})=(1,0)$ or $(0,1)$ depending on the chirality, and $n_{\text{s}}=0$.
Doing so in \cref{eq:ngon-mc}, we find that the $n$-gon maximal cut is seeded by terms involving
$\trVS(f_i \ldots)$ with at least four field strengths in the trace.
The $n$-gon maximal cut sets all of the
inverse propagators $\ell_i^2 = (\ell {+} k_{12 \ldots i})^2$ to zero.  Thus all contact terms will
carry at least one factor of $\ell_i^2$.  This leads us to the general
form of parity even \msym $n$-gon numerators,\footnote{Note that there is a factor of $2$ difference between the numerators given here and the maximal cut~\eqref{eq:ngon-mc} due to our normalization of one-loop amplitudes.}
\begin{align}\label{nptmaxcut}
    \nmeven{12\ldots n}&=\sum_{k=0}^{n-4}(-1)^{n-k}\bigg[\trVS(f_{k+1}\ldots f_n)\prod_{j=1}^{k}\pol_{j}\cdot\ell_j+(1,2,\ldots,k|1,2,\ldots,n)\bigg]\nonumber\\
    &\quad+\mathcal{O}(\ell_i^2)\text{ terms that vanish on the maximal cut} \,.
\end{align}
The truncation to $k \leq n{-}4$, which manifests the $\ell^{n-4}$ power counting of $n$-gon numerators, is due to the vanishing of $\trVS$ below length four according to \cref{vanish}.

While the numerator \cref{nptmaxcut} manifestly matches the $n$-gon maximal cut, the
lower-gon numerators that are generated via Jacobi relations require specific choices of the contact terms to satisfy their
own maximal cuts.  To see how this works, consider the $k=n{-}4$ terms
from \cref{nptmaxcut} with $\pol_1$ and $\pol_2$ not in the Lorentz traces, and the terms
they generate on the $(n{-}1)$-gon by applying the Jacobi relation on
the edge between $1$ and $2$, $\nmeven{12\ldots n} -\nmeven{21\ldots n}=\nmeven{[12]\ldots n}$. In the resulting numerator $\nmeven{[12]\ldots n}$ for the $(n{-}1)$-gon with the $[12]$-dangling tree, we have the following contribution with $\pol_1$ and $\pol_2$ not in the traces,
\begin{align}\label{eq:nm1-num}
    &\trVS(f_3f_4f_5f_6)\bigg[ (\pol_1 \Cdot \ell_1 )(\pol_2 \Cdot \ell_2 ) - (\pol_1 \Cdot \ell'_1 )(\pol_2 \Cdot \ell'_2 )\bigg] \prod_{j=7}^n \pol_j \Cdot \ell_j + (3,4,5,6|3,4,\ldots, n)\nonumber\\
    &= \bigg[\trVS(f_3 f_4 f_5 f_6) \prod_{j=7}^n \pol_j \Cdot \ell_j + (3,4,5,6|3,\ldots n)\bigg]\Big[(\pol_1 \Cdot \ell_2)(\pol_2 \Cdot k_1) {-} (\pol_1 \Cdot k_2) (\pol_2 \Cdot \ell_2) \Big] \,,
\end{align}
where $\ell'_1=\ell_2$ and $\ell'_2=\ell_2-k_1$ are the edge momenta in the cyclic ordering $213\ldots n$.
Since $\ell_1$ and $\ell_2'$ are not edge momenta in the $(n{-}1)$-gon,
we have rewritten the $\pol \cdot \ell$ terms using
the actual edge and external momenta in the second line.

This expression \emph{does not} obey the gauge invariance of legs $1$
or $2$, even on the support of the maximal cut of the dangling tree
($\ls{2} =\dots = \ell_n^2 = s_{12}= 0$)
since $(k_1{-}k_2) \cdot \ell_2$ can be chosen as the irreducible
product that is not removed by the cut conditions.  The failure of the
gauge invariance provides strong guidance for which contact terms are
needed.  For instance, it is straightforward to see that including a contact term in the $n$-gon numerator,
\begin{equation}
  -\frac{1}{4} \ls{1} (\pol_1 \cdot \pol_2) \trVS(f_3f_4f_5f_6) \prod_{j=7}^n (\pol_j \cdot \ell_j) + (3,4,5,6|3,4,\ldots ,n) \;\;\subset\;\; \nmeven{123\ldots}\,,
  \label{eq:nm1-ward-cont}
\end{equation}
compensates for the gauge variation of \cref{eq:nm1-num} on the dangling tree maximal cut. We will indeed encounter 
\cref{eq:nm1-ward-cont} as a contribution to the six-point numerator in \cref{originhex} below. 

Continuing the analysis to deeper topologies sheds light on all of the required contact terms. By iterating the above process, one can show that the gauge variations contain at least a length-four $\trVS(f_if_jf_kf_l)$. It is thus natural to conjecture that the ansatz for $\mathcal{O}(\ell_i^2)$ contact terms will at least be proportional to a length-four $\trVS$. We will show up to seven points that, under the assumption of color-kinematics duality and manifest $\ell$ power counting, gauge invariance on cuts will uniquely fix the physical amplitudes and color-ordered integrands. Although individual numerators might still feature undetermined free parameters after we have used all the constraints, they are part of the generalized gauge freedom and will cancel at the level of color-ordered loop integrand. 
We expect that this procedure will yield BCJ representations at any multiplicity, though it will be important to check its compatibility with the no-triangle property under repeated Jacobi identities in each case.

It is more difficult to perform an in-depth multiplicity-agnostic analysis on the parity odd sector.
The Schouten identity~\eqref{eq:schouten}
makes disentangling $(\pol \cdot \ell)(\pol \cdot \ell)$ and
$\ell^2 (\pol\cdot\pol)$ terms into a minimal basis difficult for all but the simplest
cases.  As such, we limit our current investigation only to the six-point
parity odd terms, and leave seven points and beyond to a future study.

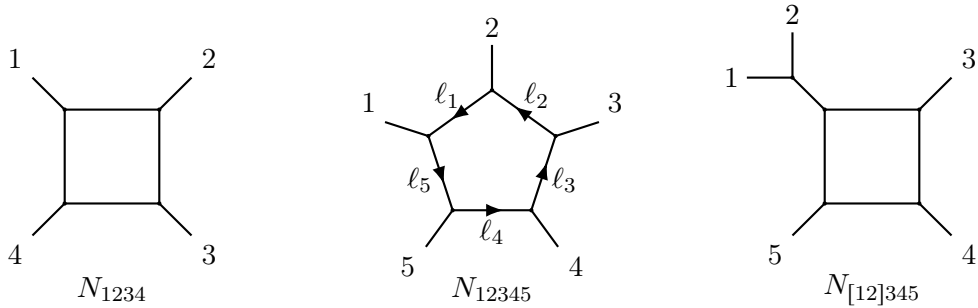
\begin{figure}
    \centering
    \begin{tikzpicture}[decoration={markings,mark=at position 0.65 with {\arrow{Latex}}}]
        \pgfmathsetmacro{\r}{1.1}
        \pgfmathsetmacro{\l}{0.75}
        \pgfmathsetmacro{\y}{-2.2}
        \pgfmathsetmacro{\s}{0.8}
        \begin{scope}[xshift=0,scale=\s]
            \foreach \x in {1,2,3,4} {
                \pgfmathsetmacro{\a}{135-90*(\x-1)}
                \coordinate (\x) at ( \a : \r);
                \fill (\x) circle (1pt);
                \draw[thick] (\x) -- ++ ( \a : \l) node[label={[label distance=-6pt]\a:\x}] {};
            }
            \draw[thick] (2) -- (1);
            \draw[thick] (3) -- (2);
            \draw[thick] (4) -- (3);
            \draw[thick] (1) -- (4);
            \node at (0,\y) {$N_{1234}$};
        \end{scope}
        \begin{scope}[xshift=5cm,scale=\s]
            \foreach \x in {1,2,...,5} {
                \pgfmathsetmacro{\a}{162-72*(\x-1)}
                \pgfmathsetmacro{\b}{126-72*(\x-1)}
                \coordinate (\x) at ( \a : \r);
                \fill (\x) circle (1pt);
                \draw[thick] (\x) -- ++ ( \a : \l) node[label={[label distance=-5pt]\a:\x}] {};
                \node at ( \b: {\r + 0.15} ) {$\ell_{\x}$};
            }
            \draw[thick,postaction={decorate}] (2) -- (1);
            \draw[thick,postaction={decorate}] (3) -- (2);
            \draw[thick,postaction={decorate}] (4) -- (3);
            \draw[thick,postaction={decorate}] (5) -- (4);
            \draw[thick,postaction={decorate}] (1) -- (5);
            \node at (0,\y) {$N_{12345}$};
        \end{scope}
        \begin{scope}[xshift=10cm,scale=\s]
            \foreach \x in {1,2,...,4} {
                \pgfmathsetmacro{\a}{135-90*(\x-1)}
                \coordinate (\x) at ( \a : \r);
                \fill (\x) circle (1pt);
            }
            \draw[thick] (2) -- (1);
            \draw[thick] (2) -- ++( 45 : \l) node[label={[label distance=-6pt]45:3}] {};
            \draw[thick] (3) -- (2);
            \draw[thick] (3) -- ++( -45 : \l) node[label={[label distance=-6pt]-45:4}] {};
            \draw[thick] (4) -- (3);
            \draw[thick] (4) -- ++( -135 : \l) node[label={[label distance=-6pt]-135:5}] {};
            \draw[thick] (1) -- (4);
            \draw[thick] (1) -- ++( 135 : \l) -- ++( 180 : \l) node[left=-1pt]{$1$};
            \draw[thick] (1) ++( 135 : \l ) -- ++( 90 : \l) node[above=-1pt]{$2$};
            \fill (1) ++( 135 : \l ) circle (1pt);
            \node at (0,\y) {$N_{[12]345}$};
        \end{scope}
    \end{tikzpicture}
    \caption{Box and pentagon numerators at four and five points. Here and below, we label the loop momenta in a diagram only when the corresponding numerator depends on them. 
}
    \label{fig:boxpenta}
\end{figure}

\subsection{Recap at four and five points}
\label{sec:msym-45}

Since the highest power of loop momentum in local kinematic numerators
of \msym is $\ell^{n-4}$, the $\mathcal{O}(\ell_i^2)$ terms in \cref{nptmaxcut} are absent
for four- and five-point numerators, and the maximal cuts give the full results. The relevant diagrams are shown
in \cref{fig:boxpenta}. Reading off the $n=4,5$ parity even part from
\cref{nptmaxcut} and the $n=5$ parity odd part from \cref{eq:ngon-mc}, we
get\footnote{We note that the parity odd terms that appear at five
  points and beyond \emph{cannot} be obtained by turning the gamma
  matrix trace in $\trVS$ into a chiral trace. In the current work, we
  solve the parity odd contribution by imposing color-kinematics and
  cut conditions following section~\ref{sec:cuts}. See
  ref.~\cite{Edison:2020uzf} for an alternative treatment from the
  forward-limit perspective.}
\begin{align}
   \nmax{1234}&=
  \trVS(f_1f_2f_3f_4)\,,
  \label{4ptalt} \\
\nmax{12345}&= \Big[ \pol_1\Cdot\ell_{1} \trVS(f_2f_3f_4f_5)+\textrm{cyclic}(1,2,3,4,5)\Big]-\trVS( f_1f_2f_3f_4f_5) 
   \notag \\
 &\quad + \frac{1}{16} \varepsilon_{10}(\ell_1,\pol_{1},f_2,f_3,f_4,f_5)\,.
  \label{5ptalt}
\end{align}
One can use \cref{teight,5ptex} to replace $\trVS$ with $t_8$-tensors such that the numerators reduce to a more familiar form,
\begin{align}\label{4ptreview}
   \nmax{1234}&=
   t_8(f_1,f_2,f_3,f_4)\,, \\
    \nmax{12345}&=\ell_{\mu}t_8^{\mu}(1,2,3,4,5)-\frac{1}{2}\Big[t_8(f_{12},f_3,f_4,f_5)+(1,2|1,2,3,4,5)\Big]\nonumber\\
    &\quad  + \frac{1}{16} \varepsilon_{10}(\ell_1,\pol_{1},f_2,f_3,f_4,f_5)\,,
    \label{eq:pen5p}
\end{align}
where we have picked $\ell=\ell_5$ for the pentagon numerator, further expanded out the ${\pol_i\Cdot\ell_i}$ factors via $\ell_i = \ell{+} k_{12 \ldots i}$ and repackaged into multiparticle field strengths defined in \cref{sec:mp-pol}. Here, we of course have full agreement with the known results~\cite{Mafra:2014gja,He:2017spx, Edison:2020uzf}. Finally, the box numerator with legs $1$ and $2$ in a dangling tree is obtained by a Jacobi identity,
\begin{align}
\nmax{[12]345} &= \nmax{12345}- \nmax{21345}=- t_8(f_{12},f_3,f_4,f_5)\,.
\label{box5pt}
\end{align}
The parity odd term can be inferred from the last line of the cut in \cref{eq:ngon-mc}, enjoys permutation invariance $ \varepsilon_{10}(\ell_1,\pol_{1},f_2,f_3,f_4,f_5) = \varepsilon_{10}(\ell_2,\pol_{2},f_1,f_3,f_4,f_5)$ by momentum conservation and therefore does not contribute to the box numerator as expected.

\subsection{The six-point parity even numerators}
\label{sec:3.1}

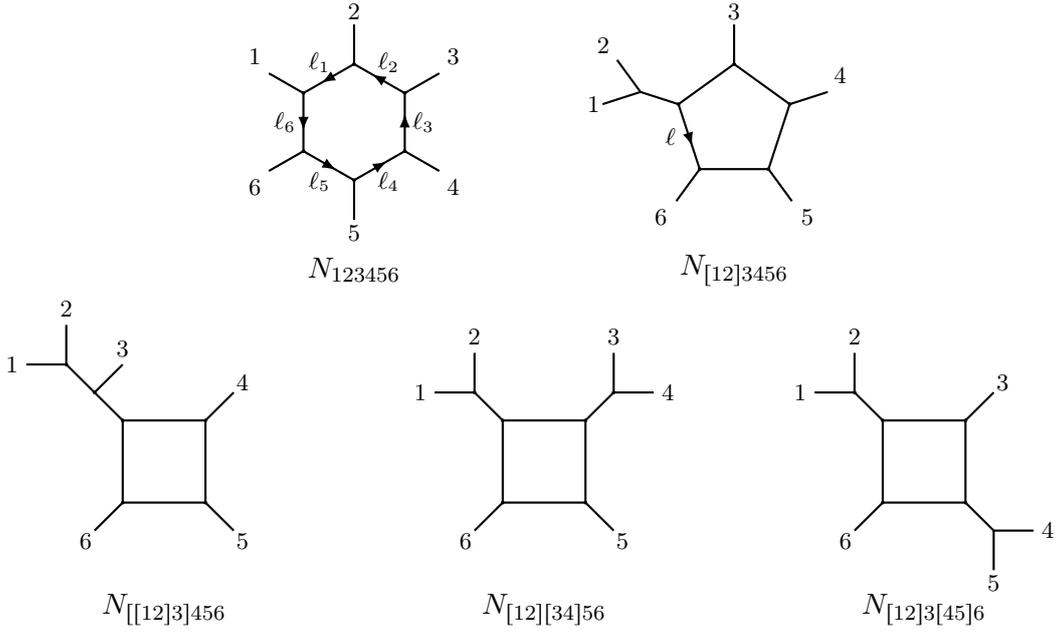
\begin{figure}
    \centering
    \begin{tikzpicture}[decoration={markings,mark=at position 0.65 with {\arrow[scale=0.8]{Latex}}},every node/.style={font=\footnotesize}]
        \pgfmathsetmacro{\r}{1.1}
        \pgfmathsetmacro{\l}{0.75}
        \pgfmathsetmacro{\s}{0.7}
        \pgfmathsetmacro{\cp}{-2.8}
        \pgfmathsetmacro{\ys}{-4.5}
        \begin{scope}[xshift=0cm,scale=\s]
            \foreach \x in {1,2,...,6} {
                \pgfmathsetmacro{\a}{150-60*(\x-1)}
                \pgfmathsetmacro{\b}{120-60*(\x-1)}
                \coordinate (\x) at ( \a : \r);
                \fill (\x) circle (1pt);
                \draw[thick] (\x) -- ++ ( \a : \l) node[label={[label distance=-6pt]\a:\x}] {};
                \node at ( \b: {\r + 0.2} ) {$\ell_{\x}$};
            }
            \draw[thick,postaction={decorate}] (2) -- (1);
            \draw[thick,postaction={decorate}] (3) -- (2);
            \draw[thick,postaction={decorate}] (4) -- (3);
            \draw[thick,postaction={decorate}] (5) -- (4);
            \draw[thick,postaction={decorate}] (6) -- (5);
            \draw[thick,postaction={decorate}] (1) -- (6);
            \node[font=\normalsize] at (0,\cp) {$N_{123456}$};
        \end{scope}
        \begin{scope}[xshift=5cm,scale=\s]
            \coordinate (1) at ( 162 : \r);
            \fill (1) circle (1pt);
            \draw[thick] (1) -- ++( 162 : \l) -- ++( 126 : \l) node[label={[label distance=-7pt]126:2}]{};
            \draw[thick] (1) ++( 162 : \l ) -- ++( 198 : \l) node[label={[label distance=-7pt]180:1}]{};
            \foreach \x in {2,...,5} {
                \pgfmathsetmacro{\a}{162-72*(\x-1)}
                \pgfmathsetmacro{\b}{126-72*(\x-1)}
                \pgfmathsetmacro{\lb}{\x+1}
                \coordinate (\x) at ( \a : \r);
                \fill (\x) circle (1pt);
                \draw[thick] (\x) -- ++ ( \a : \l) node[label={[label distance=-6pt]\a:\pgfmathprintnumber[precision=1]\lb}] {};
            }
            \draw[thick] (2) -- (1);
            \draw[thick] (3) -- (2);
            \draw[thick] (4) -- (3);
            \draw[thick] (5) -- (4);
            \draw[thick,postaction={decorate}] (1) -- (5) node[pos=0.5,left=1pt]{$\ell$};
            \node[font=\normalsize] at (0,\cp) {$N_{[12]3456}$};
        \end{scope}
        \begin{scope}[xshift=-2.5cm,yshift=\ys cm,scale=\s]
            \foreach \x in {1,2,...,4} {
                \pgfmathsetmacro{\a}{135-90*(\x-1)}
                \coordinate (\x) at ( \a : \r);
                \fill (\x) circle (1pt);
            }
            \draw[thick] (2) -- (1);
            \draw[thick] (2) -- ++( 45 : \l) node[label={[label distance=-10pt]45:4}] {};
            \draw[thick] (3) -- (2);
            \draw[thick] (3) -- ++( -45 : \l) node[label={[label distance=-10pt]-45:5}] {};
            \draw[thick] (4) -- (3);
            \draw[thick] (4) -- ++( -135 : \l) node[label={[label distance=-10pt]-135:6}] {};
            \draw[thick] (1) -- (4);
            \draw[thick] (1) -- ++( 135 : 2*\l) -- ++( 180 : \l) node[left=-1pt]{$1$};
            \draw[thick] (1) ++( 135 : 2*\l ) -- ++( 90 : \l) node[above=-1pt]{$2$};
            \draw[thick] (1) ++( 135 : \l) -- ++( 45 : \l) node[above=-1pt]{$3$};
            \fill (1) ++( 135 : \l ) circle (1pt) ++( 135 : \l ) circle (1pt);
            \node[font=\normalsize] at (0,\cp) {$N_{[[12]3]456}$};
        \end{scope}
        \begin{scope}[xshift=2.5cm,yshift=\ys cm,scale=\s]
            \foreach \x in {1,2,...,4} {
                \pgfmathsetmacro{\a}{135-90*(\x-1)}
                \coordinate (\x) at ( \a : \r);
                \fill (\x) circle (1pt);
            }
            \draw[thick] (2) -- (1);
            \draw[thick] (2) -- ++( 45 : \l) -- ++( 90 : \l) node[above=-1pt] {$3$};
            \draw[thick] (2) ++ ( 45: \l ) -- ++( \l,0 ) node[right=-1pt]{$4$};
            \fill (2) ++ ( 45: \l ) circle (1pt);
            \draw[thick] (3) -- (2);
            \draw[thick] (3) -- ++( -45 : \l) node[label={[label distance=-10pt]-45:5}] {};
            \draw[thick] (4) -- (3);
            \draw[thick] (4) -- ++( -135 : \l) node[label={[label distance=-10pt]-135:6}] {};
            \draw[thick] (1) -- (4);
            \draw[thick] (1) -- ++( 135 : \l) -- ++( 180 : \l) node[left=-1pt]{$1$};
            \draw[thick] (1) ++( 135 : \l ) -- ++( 90 : \l) node[above=-1pt]{$2$};
            \fill (1) ++( 135 : \l ) circle (1pt);
            \node[font=\normalsize] at (0,\cp) {$N_{[12][34]56}$};
        \end{scope}
        \begin{scope}[xshift=7.5cm,yshift=\ys cm,scale=\s]
            \foreach \x in {1,2,...,4} {
                \pgfmathsetmacro{\a}{135-90*(\x-1)}
                \coordinate (\x) at ( \a : \r);
                \fill (\x) circle (1pt);
            }
            \draw[thick] (2) -- (1);
            \draw[thick] (2) -- ++( 45 : \l) node[label={[label distance=-10pt]45:3}] {};
            \draw[thick] (3) -- (2);
            \draw[thick] (3) -- ++( -45 : \l) -- ++ ( -90 : \l) node[below=-2pt] {$5$};
            \draw[thick] (3) ++ ( -45 : \l ) -- ++ ( \l,0 ) node[right=-1pt]{$4$};
            \draw[thick] (4) -- (3);
            \draw[thick] (4) -- ++( -135 : \l) node[label={[label distance=-10pt]-135:6}] {};
            \draw[thick] (1) -- (4);
            \draw[thick] (1) -- ++( 135 : \l) -- ++( 180 : \l) node[left=-1pt]{$1$};
            \draw[thick] (1) ++( 135 : \l ) -- ++( 90 : \l) node[above=-1pt]{$2$};
            \fill (1) ++( 135 : \l ) circle (1pt);
            \node[font=\normalsize] at (0,\cp) {$N_{[12]3[45]6}$};
        \end{scope}
    \end{tikzpicture}
    \caption{Non-vanishing topologies for six-point numerators.}
    \label{fig:hexagon}
\end{figure}

Six-point one-loop integrands of \msym can be written in terms of numerators of five
topologies, shown in \cref{fig:hexagon}: a hexagon, a pentagon, and
three different types of boxes.  Applying the discussion at the beginning of
this section to the six-point case, we find that we only need contact
corrections on the hexagon of the form
\begin{equation}
  \pol_i \cdot \pol_j \trVS(f_af_bf_cf_d) \otimes \{\ell_{i-1}^2,\ell_i^2,\ell_{j-1}^2,\ell_j^2\} \,.
  \label{cont6pt}
\end{equation}
By imposing dihedral symmetry $D_6$ on the contributions of \cref{cont6pt} to $\nmeven{123456}$, we arrive at the five-parameter ansatz
\begin{align}
  \nmeven{123456}\,\Big|_{\ell_i^2}&=
  \pol_1\Cdot\pol_2(a_1 \ls{2}+a_2\ls{1}+a_1\ls{6})\trVS(f_3f_4f_5f_6)\nonumber\\
    &\quad+\pol_1\Cdot\pol_3(a_3 \ls{3}+a_4\ls{2}+a_4\ls{1}+a_3 \ls{6})\trVS(f_2f_4f_5f_6)\nonumber\\
    &\quad+\pol_{1}\Cdot\pol_4 a_5(\ls{1}+\ls{6})\trVS(f_2f_3f_5f_6)
      +\text{cyclic}(1,2,3,4,5,6) \,,
      \label{eq:six-ans}
\end{align}
where the expected $\pol_1 \cdot \pol_4\otimes \{\ls{3},\ls{4}\}$
terms are generated as images of the $\ls{6}$ and $\ls{1}$ terms under
the cyclic permutations. 
First, imposing gauge invariance on the maximal cuts of the pentagon in \cref{fig:hexagon} 
obtained from $\nmeven{123456}-\nmeven{213456}$ uniquely fixes
$a_2= -\frac{1}{4}$, as in \cref{eq:nm1-ward-cont}. Then, the no-triangle Jacobi
relations (\ref{eq:no-tri-jac}) between box numerators relate the
remaining coefficients of \cref{eq:six-ans} to $a_2$. The resulting parity even hexagon
numerator is
\begin{align} \label{originhex}
    \nmeven{123456}&= \Big[ \pol_1\Cdot\ell_{1}\pol_2\Cdot\ell_{2}
    \trVS(f_3 f_4 f_5 f_6)+(1,2|1,2,3,4,5,6)\Big] \\
    &\quad - \Big[\pol_1\Cdot\ell_1 \trVS(f_2 f_3 f_4 f_5 f_6)+\text{cyclic}(1,2,3,4,5,6)\Big]+\trVS(f_1 f_2 f_3 f_4 f_5f_6)\nonumber\\
    &\quad  +\frac{1}{40}\Big[\pol_1\Cdot\pol_2 \, (3\ls{2}-10\ls{1}+3\ls{6})\trVS(f_3f_4f_5f_6)\nonumber\\
    &\qquad\quad\quad+\pol_1\Cdot\pol_3 \, (\ls{3}-3\ls{2}-3\ls{1}+\ls{6})\trVS(f_2f_4f_5f_6)\nonumber\\
    &\qquad\quad\quad-\pol_{1}\Cdot\pol_4 \, (\ls{1}+\ls{6})\trVS(f_2f_3f_5f_6)
     +\text{cyclic}(1,2,3,4,5,6)  \Big]\,.
    \nonumber
\end{align}
After unpacking the traces via \cref{teight,5ptex,6ptex}
with $t_{12}$ given by \cref{t12ten}, 
each term in the numerator
features a factor of $t_{8}$ contracting linearized field strengths
and their commutators.

The lower-topology numerators can be extracted from the hexagon via
the kinematic Jacobi relations as in \cref{sec:ck-dual}, and repackaged
into the multiparticle field strengths from \cref{sec:mp-pol}.  The
pentagon numerator is (see \cref{deft8vec} for the definition of $t_8^{\mu}(12,3,4,5,6)$)
\begin{align}
    \nmeven{[12]3456}&=-\ell_{\mu}t_8^{\mu}(12,3,4,5,6)+\frac{1}{2}\Big[t_8(f_{123},f_4,f_5,f_6)+\textrm{cyclic}(3,4,5,6)\Big]\nonumber\\
    &\quad +\frac{1}{2}\Big[t_8(f_{12},f_{34},f_5,f_6)+(3,4|3,4,5,6)\Big]\nonumber\\
    &\quad +\frac{1}{20}\Big(2\ell\Cdot(k_1+k_2-k_3)-s_{13}-s_{23}\Big)\Big[\epsilon_{1}\Cdot\epsilon_{3}t_8(f_2,f_4,f_5,f_6)-(1\leftrightarrow 2)\Big]\nonumber\\
    &\quad +\frac{1}{20}\Big(2\ell\Cdot(k_1+k_2-k_4)-s_{12}+s_{45}+s_{46}\Big)\Big[\epsilon_{1}\Cdot\epsilon_{4}t_8(f_2,f_3,f_5,f_6)-(1\leftrightarrow 2)\Big]\nonumber\\
    &\quad +\frac{1}{20}\Big(2\ell\Cdot(k_1+k_2-k_5)-2s_{12}+s_{56}\Big)\Big[\epsilon_{1}\Cdot\epsilon_{5}t_8(f_2,f_3,f_4,f_6)-(1\leftrightarrow 2)\Big]\nonumber\\
    &\quad +\frac{1}{20}\Big(2\ell\Cdot(k_1+k_2-k_6)-3s_{12}\Big)\Big[\epsilon_{1}\Cdot\epsilon_{6}t_8(f_2,f_3,f_4,f_5)-(1\leftrightarrow 2)\Big]\,,
\end{align}
where $s_{ij\ldots}{=}(k_i{+}k_j{+}\ldots)^2$ and $\ell$ points from leg $1$ to $6$.  The numerators of the three $D_6$-inequivalent box diagrams are
\begin{subequations}
\label{6ptboxnums}
\begingroup
\allowdisplaybreaks
\begin{align}
    \nmeven{[[12]3]456}&=t_8(f_{123},f_4,f_5,f_6)-\frac{1}{10}s_{123}\Big[\pol_1\Cdot\pol_3 \,t_8(f_2,f_4,f_5,f_6)-(1\leftrightarrow 2)\Big]\nonumber\\*
    &\quad -\frac{1}{20}s_{123}\Big[\Big(\pol_1\Cdot\pol_4 \,t_8(f_2,f_3,f_5,f_6)-(1\leftrightarrow 2)\Big)+\text{cyclic}(4,5,6)\Big]\,,
    \label{6ptbox1}\\
    \nmeven{[12][34]56}&=t_8({f}_{12},{f}_{34},f_5,f_6)\nonumber\\*
    &\quad +\frac{1}{20}(s_{12}+s_{34})\Big[\Big(\pol_1\Cdot\pol_3 \,t_8(f_2,f_4,f_5,f_6)-(1\leftrightarrow2)\Big)-(3\leftrightarrow 4)\Big]\,,\label{6ptbox2}\\
    \nmeven{[12]3[45]6}&=t_8(f_{12},f_3,f_{45},f_6)\nonumber\\*
    &\quad +\frac{1}{20}(s_{12}+s_{45})\Big[\Big(\pol_1\Cdot\pol_4 \,t_8(f_2,f_3,f_5,f_6)-(1\leftrightarrow2)\Big)-(4\leftrightarrow 5)\Big]\,. \label{6ptbox3}
\end{align}
\endgroup
\end{subequations}
By construction, this representation satisfies the hexagon maximal cut and each numerator is gauge invariant on its maximal cuts, including those with
dangling trees. We have furthermore explicitly checked that the representation matches the pentagon and box cuts (with dangling trees pulled into the contact amplitude), as calculated via \cref{sec:cuts}.

\subsection{The six-point parity odd numerators and gauge anomaly}
\label{sec:3.2}

At six points, the parity odd contribution to the hexagon numerator $N_{123456}$ consists of one Lorentz dot product and one Levi-Civita tensor. We can put parity odd contributions of this type into a minimal basis by recursively using momentum conservation and Schouten identities (\ref{eq:schouten}) such that 
\begin{enumerate}[label=(\arabic*)]
    \item $k_6$ does not appear;
    \item $k_1\Cdot k_5$ and $\pol_6\Cdot k_1$ do not appear;
    \item $\pol_{1}$ always contracts with the Levi-Civita tensor;
    \item $\pol_{2}\Cdot k_1$ does not appear.
\end{enumerate}
Note that the last two properties are due to Schouten identities~\eqref{eq:schouten}. The most general ansatz for $\nmodd{123456}$ that contains one dot product, one Levi-Civita tensor and at most two powers of $\ell$ is composed of the following types of terms,
\begin{align}
\label{parityodd4an}
    & \pol_{i}\Cdot\ell\,\varepsilon_{10}(\ell,k^4,\pol^5)\, , && \pol_{i}\Cdot\ell\,\varepsilon_{10}(k^5,\pol^5)\, , & \pol_{i}\Cdot k_j\,\varepsilon_{10}(\ell,k^4,\pol^5)\, , && \pol_{i}\Cdot k_j\,\varepsilon_{10}(k^5,\pol^5)\, , \nonumber\\
    & \pol_i\Cdot\pol_{j}\,\varepsilon_{10}(\ell,k^5,\pol^4)\, , && \ell_i^2\varepsilon_{10}(k^4,\pol^6)\, , & k_i\Cdot k_j\,\varepsilon_{10}(\ell,k^3,\pol^6)\, , && k_i\Cdot k_j\,\varepsilon_{10}(k^4,\pol^6)\, ,
\end{align}
where, for example, $\varepsilon_{10}(\ell,k^4,\pol^5)$ refers to all possible contractions with one loop momentum, four external momenta out of the five independent ones, and five polarizations.\footnote{One may question if terms like $k_i\Cdot\ell\,\varepsilon_{10}(\ell,k^3,\pol^6)$ should be considered. We do not include them since the relation $2k_i\Cdot\ell=\ell_i^2-\ell_{i-1}^2-2k_i\Cdot k_{1\ldots i-1}$ makes them formally violate the $\ell$ power counting. Even if we include them, they will eventually be set to zero by imposing cut conditions and crossing symmetry.} In the minimal basis specified by (1) to (4) above, the ansatz contains in total $339$ free parameters. 

After imposing the dihedral crossing symmetry $D_6$,
we solve for the parameters in the ansatz by matching the hexagon, pentagon and box cuts following the generic prescription given in \cref{sec:cuts}. Similar to the parity even sector, we also get a unique solution. The resulting hexagon numerator is
\begin{align}
\label{eq:N6ptoddsym}
    \nmodd{123456}&=\frac{1}{96}\Big[\pol_{2}\Cdot\ell_2 \, \varepsilon_{10}(\ell_1,\pol_{1},f_3,f_4,f_5,f_6)\nonumber\\
    &\qquad\quad+\pol_{1}\Cdot\ell_1 \, \varepsilon_{10}(\ell_2,\pol_{2},f_3,f_4,f_5,f_6)+(1,2|1,2,3,4,5,6)\Big]\nonumber\\
    &\quad+\frac{1}{384}\Big[\pol_{1}\Cdot\ell_1 \, \varepsilon_{10}(f_2,f_3,f_4,f_5,f_6)+\textrm{cyclic}(1,2,3,4,5,6)\Big]\nonumber\\
    &\quad -\frac{1}{192}\Big[\Big(\varepsilon_{10}(\ell_3,\pol_{3},[f_1,f_2],f_4,f_5,f_6)+\textrm{cyclic}(1,2,3)\Big)+(1,2,3|1,2,3,4,5,6)\Big]\nonumber\\
    &\quad +\frac{1}{64}\Big[\ls{1} \, \varepsilon_{10}(\pol_{1},\pol_{2},f_3,f_4,f_5,f_6)+\textrm{cyclic}(1,2,3,4,5,6)\Big]\,.
\end{align}
We note that the $\ell_i^2$ appearing in the last line live in $D=10{-}2\epsilon$
dimensions. Since the cut matching is carried out in the minimal
basis, various Schouten identities have been used. When the loop
momentum is involved in such identities, it will be put into strictly
ten dimensions (see \cref{sec:mu_term}). Thus there is an ambiguity in $\mu^2$-terms, the $-2\epsilon$ dimensional contribution of $\ell^2$, which would affect the behavior of the hexagon gauge anomaly. 

Using Jacobi identities, we can obtain numerators of 1PR diagram
topologies. The pentagon numerator is most conveniently written in terms of two-particle polarizations,
\begin{align}
    \nmodd{[12]3456}&=-\frac{1}{48}\varepsilon_{10}(\ell,\pol_{12},f_3,f_4,f_5,f_6)-\frac{1}{96}\Big[\varepsilon_{10}(\ell,\pol_{3},f_{12},f_4,f_5,f_6)+\textrm{cyclic}(3,4,5,6)\Big]\nonumber\\
    &\quad +\frac{1}{192}\Big(2\ell\Cdot(k_1{+}k_2)-s_{12}\Big)\varepsilon_{10}(\pol_{1},\pol_{2},f_3,f_4,f_5,f_6)\,,
\end{align}
and there is only one nonzero box topology,
\begin{align}\label{nboxodd}
    \nmodd{[[12]3]456}&=-\frac{1}{192}s_{123}\varepsilon_{10}(\pol_{1},\pol_{2},f_3,f_4,f_5,f_6)\,,\nonumber \\
    \nmodd{[12][34]56}&=\nmodd{[12]3[45]6}= 0\,.
\end{align}

Before moving on, we note that the strategy used here to obtain a
minimal basis in the presence of Levi-Civita tensors only works for a
single power of dot products. In other words, there is no straightforward uplift of (1) to (4) above to decompose seven-point parity odd numerators to a kinematic basis, which would be a more involved task.

It is well known that the gauge symmetry of ten-dimensional \msym is anomalous at six points~\cite{Frampton:1983ah, Frampton:1983nr}. Thus, we consider the gauge variation
$\pol_{1}\rightarrow k_1$ of the six-point color ordered amplitude,
and find that the hexagon diagram
in $D=10{-}2\epsilon$ dimensions supports the gauge anomaly
\begin{align}
    A_{\text{\msym}}^{\text{1-loop}}(1,2,3,4,5,6)\Big|_{\pol_{1}\rightarrow k_1} &=\frac{1}{192}\varepsilon_{10}(f_2,f_3,f_4,f_5,f_6) I_6^{10-2\epsilon} \, , \label{tendimanom}\\
   I_6^{10-2\epsilon} &=\int\frac{\dd^{10 - 2\epsilon }\ell}{i\pi^{(10 - 2\epsilon)/2}}\frac{(-\mu^2)}{\ls{1}\ls{2}\ls{3}\ls{4}\ls{5}\ls{6}} = 
    \frac{1}{5!} \, , \nonumber
\end{align}
where $\mu^2=\ell^2{-}\ell^2_{(10)}$ is the
$-2\epsilon$ dimensional contribution of $\ell^2$. 
We can then identify $\varepsilon_{10}(f_2,f_3,f_4,f_5,f_6)$ as the linearized anomaly appearing in the path integral, and the full anomaly  $\Tr(F^5)$ affecting higher-point gauge variations follows
as the non-abelian completion.\footnote{Here $\Tr$ acts on Lie algebra
  generators, and $F^5$ is defined using wedge products.} In other
words, we do not need any higher-point calculations to determine the full
anomaly. Because our numerators manifest crossing symmetry, we observe the same gauge
anomaly for all external legs.

In contrast, string-based methods single out one external leg that
carries an anomalous ghost
picture~\cite{Gross:1987pd,Berkovits:2006bk, Mafra:2014gja,He:2017spx}. For such calculations, the gauge anomaly 
%is also the obstruction to 
also reflects an obstruction to
the cyclic invariance of color-ordered one-loop amplitudes.  
We can reproduce this behavior by adding $\mu^2$ terms to our hexagon numerator in order to push all the gauge anomalies to a single leg. On the other hand, we will show in \cref{sec:mu_term} that imposing crossing symmetry fixes the $\mu^2$-term ambiguity, and results in a unique crossing symmetric gauge anomaly as shown above.

\subsection{The seven-point numerators}
\label{sec:7pt}

\begin{figure}
  \centering
  \begin{tikzpicture}[decoration={markings,mark=at position 0.65 with
      {\arrow{Latex}}},scale=0.8]
    \pgfmathsetmacro{\r}{1.1}
    \pgfmathsetmacro{\l}{0.75}
    \begin{scope}[xshift=0cm]
      \foreach \x in {1,2,...,7} {
        \pgfmathsetmacro{\a}{90+360/7-360/7*(\x-1)}
        \pgfmathsetmacro{\b}{90+180/7-360/7*(\x-1)}
        \coordinate(\x) at ( \a : \r);
        \fill (\x) circle (1pt);
        \draw[thick](\x) -- ++ (\a : \l) node[label={[label distance=-5pt]\a:\x}] {};
        \node at ( \b: {\r + 0.25} ) {$\ell_{\x}$};
      }
      \draw[thick,postaction={decorate}] (2) --(1);
      \draw[thick,postaction={decorate}] (3) -- (2);
      \draw[thick,postaction={decorate}] (4) -- (3);
      \draw[thick,postaction={decorate}] (5) -- (4);
      \draw[thick,postaction={decorate}] (6) -- (5);
      \draw[thick,postaction={decorate}] (7) -- (6);
      \draw[thick,postaction={decorate}] (1) -- (7);
    \end{scope}
  \end{tikzpicture}
  \caption{The loop momentum orientation of the heptagon.}
  \label{fig:ngon}
\end{figure}
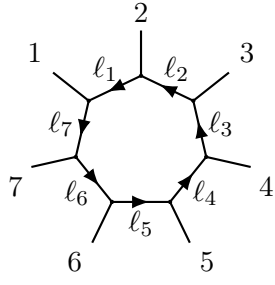

Moving on to seven points, we seed the heptagon, as shown in \cref{fig:ngon}, with
\cref{nptmaxcut} at $n=7$, just like in all the previous cases,
\begin{align}
\nmeven{1234567} &=
\big[ \pol_1\Cdot\ell_{1}   \pol_2\Cdot\ell_{2}  \pol_3\Cdot\ell_{3} \trVS(   f_4 f_5 f_6 f_7) + (1,2,3| 1,2,3,4,5,6,7)\big]
 \notag \\
 &\quad
- \big[ \pol_1\Cdot\ell_{1}   \pol_2\Cdot\ell_{2} \trVS(  f_3 f_4 f_5 f_6 f_7) + (1,2| 1,2,3,4,5,6,7)\big]
\notag \\
&\quad + \big[ \pol_1\Cdot\ell_{1} \trVS( f_2 f_3 f_4 f_5 f_6 f_7) +  \textrm{cyclic}(1,2,3,4,5,6,7)\big]
\notag \\
&\quad - \trVS(f_1 f_2 f_3 f_4 f_5 f_6 f_7) + \delta \nmeven{1234567} \, ,
\label{hepnum}
\end{align}
where $\delta \nmeven{1234567} $
collects all contact-term contributions
with a factor of $\ell_{j}^2$.
However, unlike at lower points, the cancellation of gauge variations
combined with color-kinematics relations no longer has a unique
solution for $ \delta \nmeven{1234567}$.
We organize the contact-term contributions according to the number of field strengths appearing in the $\trVS(\ldots)$
\begin{align}
\delta \nmeven{1234567} = \delta N^{\tr(f^5)}_{1234567}+\delta N^{\tr(f^4)}_{1234567}
\label{sit10d.1}
\end{align}
and find a 30-parameter freedom in the
terms $\delta N^{\tr(f^4)}_{1234567}$ with length-four traces.
The terms $\delta N^{\tr(f^5)}_{1234567}$ with length-five traces in turn are uniquely fixed by the color-kinematics duality and linearized gauge invariance, and are given by
\begin{align}
 \delta N_{1234567}^{\tr(f^5)}
  = & -\frac{1}{40}(\pol_1 \cdot \pol_2) (3 \ls{7} - 10 \ls{1} + 3 \ls{2}) \trVS(f_3f_4f_5f_6f_7) \notag \\
      &- \frac{1}{40} (\pol_1 \cdot \pol_3) (\ls{7} - 3 \ls{1} - 3 \ls{2} + \ls{3}) \trVS(f_2f_4f_5f_6f_7) \notag \\
    &+\frac{1}{80} (\pol_1 \cdot \pol_4)\Big[
      (-\ls{7} + 2 \ls{1} + 4 \ls{2} + 2 \ls{3} - \ls{4} + \ls{5} + \ls{6})
      \trVS(f_2f_3f_5f_6f_7) \notag \\
    & \quad + (\ls{4} - 2 \ls{5} + \ls{6})
      \trVS(f_2f_3f_6f_7f_5) \notag \\
      & \quad +(2 \ls{1} + 4 \ls{2} + 2 \ls{3} - \ls{4} + 2 \ls{5} - \ls{6})
      \trVS(f_2f_3f_7f_5f_6) \notag \\
  & \quad - (\ls{7} + 2 \ls{1} + 2 \ls{3} + \ls{4} + \ls{5} + \ls{6}) \trVS(f_2f_3f_7f_6f_5) \Big] \notag \\
  &+ \text{cyclic}(1,2,3,4,5,6,7)\,.
  \label{sit10d.2}
\end{align}
The free parameters in $\delta N^{\tr(f^4)}_{1234567}$ originate from
the following ansatz with 630
rational parameters $A_{i,j,k},B_{i,j,k}$
and $C_{i,j,k}$, 
\begin{align}
 \delta N_{1234567}^{\tr(f^4)}
  =&\; (\pol_1 \Cdot \pol_2)
  \sum_{k=3}^7\sum_{\substack{i,j=1 \\ i\neq k-1}}^{7}A_{i,j,k}\, (\pol_k \Cdot \ell_i) \, \ell_j^2 \, \trVS(f^4) \notag \\
    &+ (\pol_1 \Cdot \pol_3)
    \sum_{k\neq1,3}^7\sum_{\substack{i,j=1 \\ i\neq k-1}}^{7} B_{i,j,k}\,(\pol_k \Cdot \ell_i) \, \ell_j^2 \, \trVS(f^4) \notag \\
    &+(\pol_1 \Cdot \pol_4)
    \sum_{k\neq1,4}^7 \sum_{\substack{i,j=1 \\ i\neq k-1}}^{7}C_{i,j,k}\, (\pol_k \Cdot \ell_i)\, \ell_j^2 \, \trVS(f^4)\notag \\
    & +\text{cyclic}(1,2,3,4,5,6,7)\, ,
    \label{7pttrf4}
\end{align} 
where the $\trVS(f^4)$ are over the four field strengths not
appearing as separate polarization vectors, and the restricted sum over $i$ is due to $\pol_k \cdot \ell_{k-1} = \pol_k \cdot \ell_k$.
Imposing full crossing symmetry, color-kinematics duality 
and gauge invariance leads to 600 independent constraints on $A_{i,j,k},B_{i,j,k},C_{i,j,k}$ and still leaves 30 of the parameters undetermined. We have explicitly checked
that these remaining parameters cancel from all unitarity cuts and
color-ordered integrands of \msym, and they furthermore drop out from the gravity cuts obtained from the numerators in \cref{sec:4.4} and
the techniques reviewed in \cref{sec:grcut} below. 
Thus, these 30 parameters are attributed to the generalized gauge freedom which does not affect the resulting amplitudes.

Using the 30 free parameters, we can pick out a representation that
highlights point-to-point patterns.  One such choice is to force the
contact terms to match the six-point hexagon contact terms,
\cref{originhex}, as closely as possible and then finding a solution
that minimizes the size of the remaining terms
\begingroup
\allowdisplaybreaks
\begin{align*}
 \delta
N_{1234567}^{\tr(f^4)} 
   & = \frac{1}{40}(\pol_1 \cdot \pol_2)( 3 \ls{7} - 10 \ls{1} + 3 \ls{2})
     \Big[\pol_{3}\Cdot\ell_3\,\trVS(f_4f_5f_6f_7)+(3|3,4,5,6,7)\Big]\\ 
   & +\frac{1}{40}(\pol_1 \cdot \pol_3)(\ls{7} - 3 \ls{1} - 3 \ls{2} + \ls{3})\Big[\pol_{4}\Cdot\ell_4\,\trVS(f_2f_5f_6f_7)+(4|4,5,6,7)\Big]  \\
   & - \frac{1}{40}(\pol_1 \cdot \pol_4)(\ls{7} +\ls{1} + \ls{3} + \ls{4})
     \Big[\pol_{5}\Cdot\ell_5\,\trVS(f_2f_3f_6f_7)+(5|5,6,7)\Big] \\  
   &+ \frac{1}{40}(\pol_1 \cdot \pol_3) (\pol_2 \cdot \ell_2)(3 \ls{7} -5 \ls{1} -5 \ls{2} +3 \ls{3}) \trVS(f_4f_5f_6f_7)  \\
   & + \frac{1}{40}(\pol_1 \cdot \pol_4)(\ls{7} - 3 \ls{1} - 3 \ls{2} + \ls{3})\Big[\pol_{2}\Cdot\ell_2\,\trVS(f_3f_5f_6f_7)+(2\leftrightarrow 3)\Big] \\  
   & +\frac{1}{360} (\pol_1 \cdot \pol_2) \bigg[
   \trVS(f_{4}f_{5}f_{6}f_{7}) (9 (\pol_{3}\cdot\ell_{1}) (5 \ls{2}-3 \ls{7})+(\pol_{3}\cdot\ell_{7}) (60 \ls{7}-72 \ls{2})) \notag \\
   &\quad +\trVS(f_{3}f_{5}f_{6}f_{7}) ((\pol_{4}\cdot\ell_{1}) (27 \ls{2}-9\ls{7})+(\pol_{4}\cdot\ell_{7}) (30 \ls{7}-39 \ls{2})-3 \ls{7} (\pol_{4}\cdot\ell_{2})) \notag \\
   &\quad +\trVS(f_{3}f_{4}f_{6}f_{7}) (9 (\pol_{5}\cdot\ell_{1}) (\ls{2}+\ls{7})+2 (\pol_{5}\cdot\ell_{2}) (5\ls{2}-8 \ls{7})+(\pol_{5}\cdot\ell_{7}) (10 \ls{7}-16 \ls{2})) \notag \\
   &\quad +\trVS(f_{3}f_{4}f_{5}f_{7}) (-9 (\pol_{6}\cdot\ell_{1}) (\ls{2}-3 \ls{7})+(\pol_{6}\cdot\ell_{2}) (30 \ls{2}-39 \ls{7})-3 \ls{2}(\pol_{6}\cdot\ell_{7}))\notag \\
   &\quad+\trVS(f_{3}f_{4}f_{5}f_{6}) ((\pol_{7}\cdot\ell_{1}) (45 \ls{7}-27 \ls{2})+(\pol_{7}\cdot\ell_{2}) (60 \ls{2}-72 \ls{7}))
   \bigg]  \\
   & + \frac{1}{360} (\pol_1 \cdot \pol_3) \bigg[
  \trVS(f_{4}f_{5}f_{6}f_{7}) ((\pol_{2}\cdot\ell_{7}) (-27 \ls{1}+72 \ls{2}-42 \ls{3})\notag \\
  &\qquad +(\pol_{2}\cdot\ell_{3}) (72 \ls{1}-27 \ls{2}-42 \ls{7})) \notag \\
  &\quad +\trVS(f_{2}f_{5}f_{6}f_{7}) (3 (\pol_{4}\cdot\ell_{1})(20 \ls{1}-9 \ls{2}-6 \ls{3}-\ls{7})\notag \\
  &\qquad +(\pol_{4}\cdot\ell_{7}) (-60 \ls{1}+39 \ls{2}-18 \ls{3}+30 \ls{7}) \notag \\
  & \qquad +(\pol_{4}\cdot\ell_{2}) (9 (3 \ls{3}+\ls{7})-33 \ls{1})) \notag \\
  &\quad  +\trVS(f_{2}f_{4}f_{6}f_{7})   ((\pol_{5}\cdot\ell_{1}) (30 \ls{1}-19 \ls{2}-19 \ls{3}+17 \ls{7})\notag \\
  &\qquad +(\pol_{5}\cdot\ell_{2}) (-21 \ls{1}+10 \ls{2}+27 \ls{3}-10 \ls{7}) \notag \\
  &\qquad -2 (\pol_{5}\cdot\ell_{7}) (15 \ls{1}-8 \ls{2}+\ls{3}-5\ls{7})+(\pol_{5}\cdot\ell_{3}) (3 \ls{1}-10 \ls{2}+4 \ls{7})) \notag \\
  &\quad  +\trVS(f_{2}f_{4}f_{5}f_{7}) ((\pol_{6}\cdot\ell_{1}) (10 \ls{1}-21 \ls{2}-10 \ls{3}+27 \ls{7})\notag \\
  &\qquad +(\pol_{6}\cdot\ell_{2}) (-19\ls{1}+30 \ls{2}+17 \ls{3}-19 \ls{7})\notag \\
  &\qquad +2 (\pol_{6}\cdot\ell_{3}) (8 \ls{1}-15 \ls{2}+5 \ls{3}-\ls{7})+(\pol_{6}\cdot\ell_{7}) (-10 \ls{1}+3 \ls{2}+4 \ls{3})) \notag \\
  &\quad +\trVS(f_{2}f_{4}f_{5}f_{6}) (-3(\pol_{7}\cdot\ell_{2}) (9 \ls{1}-20 \ls{2}+\ls{3}+6 \ls{7})\notag \\
  &\qquad +(\pol_{7}\cdot\ell_{3}) (39 \ls{1}-60 \ls{2}+30 \ls{3}-18 \ls{7})\notag \\
  &\qquad +(\pol_{7}\cdot\ell_{1}) (9 (\ls{3}+3 \ls{7})-33 \ls{2}))
   \bigg] \\
   &+ \frac{1}{720}(\pol_1 \cdot \pol_4) \bigg[
   \trVS(f_{3}f_{5}f_{6}f_{7}) (-3 (\pol_{2}\cdot\ell_{7}) (6 \ls{1}-6 \ls{2}-28 \ls{3}+17 \ls{4}+9 \ls{6})\notag \\
   &\qquad +(\pol_{2}\cdot\ell_{4}) (36 \ls{1}-18 \ls{2}+20 \ls{4}-27\ls{5}+\ls{7})+(\pol_{2}\cdot\ell_{3}) (36 \ls{1}-20 \ls{4}-22 \ls{7})\notag \\
   &\qquad -27 (\pol_{2}\cdot\ell_{5}) (\ls{4}-2 \ls{5}+\ls{6})-27 (\pol_{2}\cdot\ell_{6}) (\ls{5}-2\ls{6}+\ls{7})) \notag \\
   &\quad +\trVS(f_{2}f_{5}f_{6}f_{7}) (3 (\pol_{3}\cdot\ell_{4}) (28 \ls{1}+6 \ls{2}-6 \ls{3}-9 \ls{5}-17 \ls{7})\notag \\
   &\qquad +(\pol_{3}\cdot\ell_{7}) (-18 \ls{2}+36 \ls{3}+\ls{4}-27 \ls{6}+20\ls{7})+(\pol_{3}\cdot\ell_{1}) (36 \ls{3}-22 \ls{4}-20 \ls{7}) \notag \\
   & \qquad -27 (\pol_{3}\cdot\ell_{5}) (\ls{4}-2 \ls{5}+\ls{6})-27 (\pol_{3}\cdot\ell_{6}) (\ls{5}-2 \ls{6}+\ls{7})) \notag \\
   &\quad +\trVS(f_{2}f_{3}f_{5}f_{7}) ((\pol_{6}\cdot\ell_{1}) (20 \ls{1}-54 \ls{2}+20 \ls{3}-\ls{4}+9 \ls{5}-9 \ls{6}+27 \ls{7})\notag \\
   & \qquad +(\pol_{6}\cdot\ell_{3}) (20 \ls{1}-54 \ls{2}+20 \ls{3}+27 \ls{4}-9 \ls{5}+9 \ls{6}-\ls{7})-54 (\pol_{6}\cdot\ell_{2}) (\ls{1}-2 \ls{2}+\ls{3})\notag \\
   &\qquad +(\pol_{6}\cdot\ell_{7}) (-20 \ls{1}-8 \ls{3}+22 \ls{4})+(\pol_{6}\cdot\ell_{4}) (-8 \ls{1}-20 \ls{3}+22\ls{7}))\notag \\
   &\quad +\trVS(f_{2}f_{3}f_{6}f_{7}) ((\pol_{5}\cdot\ell_{1}) (60 \ls{1}-54 \ls{2}+38 \ls{3}-24 \ls{4}-9 \ls{5}+7 \ls{7})\notag \\
   &\qquad -54 (\pol_{5}\cdot\ell_{2}) (\ls{1}-2 \ls{2}+\ls{3})+(\pol_{5}\cdot\ell_{3}) (-18 \ls{1}-54 \ls{2}+18 \ls{4}+9 \ls{5}+51 \ls{7})\notag \\
   &\qquad +4 (\pol_{5}\cdot\ell_{7}) (-15 \ls{1}+\ls{3}+6 \ls{4}+5 \ls{7}))\notag \\
   &\quad +\trVS(f_{2}f_{3}f_{5}f_{6})   ((\pol_{7}\cdot\ell_{3}) (38 \ls{1}-54 \ls{2}+60 \ls{3}+7 \ls{4}-9 \ls{6}-24 \ls{7})\notag \\
   &\qquad -54 (\pol_{7}\cdot\ell_{2}) (\ls{1}-2 \ls{2}+\ls{3})+4 (\pol_{7}\cdot\ell_{4}) (\ls{1}-15 \ls{3}+5 \ls{4}+6\ls{7})\notag \\
   &\qquad +(\pol_{7}\cdot\ell_{1}) (-54 \ls{2}-18 \ls{3}+51 \ls{4}+9 \ls{6}+18 \ls{7}))
   \bigg]\\
   & + \text{cyclic}(1,2,3,4,5,6,7)\,. \tag{\stepcounter{equation}\theequation}
\end{align*}
\endgroup
We also include the full heptagon numerator, with all free
parameters left unfixed, as an ancillary file in the arXiv submission.

We leave it to future work to identify the analogous contact terms $\ell_j^2$ in $(n \geq 8)$-point numerators that go beyond the simple expression (\ref{nptmaxcut}) for the maximal cuts. Already at eight points, we expect a hierarchy of contact terms -- contributions with either one factor of $(\pol_a \Cdot \pol_b) \ell_j^2$ or bilinears $(\pol_a \Cdot \pol_b) (\pol_c \Cdot \pol_d) \ell_j^2 \ell_k^2$.

\section{UV divergences}
\label{sec:UV}

In this section, we evaluate the UV divergences for \msym and type II supergravity up to and including six points that follow from the BCJ numerators given in the previous section. The subsequent UV divergences for \msym and type IIB supergravity in $d=8$ reproduce matrix elements of the supersymmetric counterterms ${\rm Tr}(F^4)$ and $R^4$, respectively. The $d=10$ UV divergences in turn reproduce matrix elements of ${\rm Tr}(D^2 F^4+F^5)$ for \msym and vanish for both type IIA and IIB supergravity. In this way, our numerators and their double copy are severely crosschecked.

We shall give a more detailed discussion of the maximally supersymmetric UV divergences in the critical dimension
$d=8$. Since the \msym numerators for $m$-gons are polynomials of degree
$\ell^{m-4}$, the $d=8$ UV divergences only receive contributions
from scalar boxes,
\begin{align}
\int \frac{ \dd^{D} \ell}{i\pi^{D/2}} \frac{1}{\ell^2 (\ell{+}K_1)^2 (\ell{+}K_{12})^2
(\ell{+}K_{123})^2} \, \Big|^{\rm UV}_{D=8-2\epsilon} = \frac{1}{3! \times \epsilon}\,,
\label{UV.01}
\end{align}
where the squares of the external momenta $K_1,K_2,K_3$ may be nonzero,
and we employ shorthands $K_{12\ldots j} = K_1{+}K_2{+}\ldots{+}K_j$. 
In supergravity, the double copy of \msym numerators leads to a maximum power
of $\ell^{2m-8}$ in $m$-gon numerators, such that all diagrams with $4 \leq m \leq n$ contribute to the
UV divergence at $n$ points. More specifically, pentagons exclusively contribute through their rank-two tensor part
$\sim T_{2}^{\mu \nu}$,
\begin{align}
\int \frac{ \dd^{D} \ell}{i\pi^{D/2}} \frac{T_0+\ell_\mu T^\mu_1+\ell_\mu \ell_\nu T_2^{\mu \nu}
}{\ell^2 (\ell{+}K_1)^2 (\ell{+}K_{12})^2
(\ell{+}K_{123})^2(\ell{+}K_{1234})^2} \, \Big|^{\rm UV}_{D=8-2\epsilon} = \frac{\eta_{\mu \nu} T_2^{\mu \nu}}{4! \times 2\epsilon}\,,
\label{UV.02}
\end{align}
and hexagons through the quartic order in $\ell$,
\begin{align}
&\int \frac{ \dd^{D} \ell}{i\pi^{D/2}} \frac{  \ell_\mu \ell_\nu  \ell_\lambda \ell_\rho
}{\ell^2 (\ell{+}K_1)^2 (\ell{+}K_{12})^2
(\ell{+}K_{123})^2(\ell{+}K_{1234})^2(\ell{+}K_{12345})^2} \, \Big|^{\rm UV}_{D=8-2\epsilon} \notag \\
&\ \  = \frac{1}{5! \times 4\epsilon}
(\eta_{\mu \nu}  \eta_{\lambda \rho} + \eta_{\mu \lambda}  \eta_{\nu \rho} + \eta_{\mu \rho}  \eta_{\nu \lambda} )\,.
\label{UV.03}
\end{align}
More generally, an $m$-gon contributes through the order of $\ell^{2m-8}$, 
\begin{align}
&\int \frac{ \dd^{D} \ell}{i\pi^{D/2}} \frac{  \ell_{\mu_1} \ell_{\mu_2} \ldots \ell_{\mu_{2m-8}}
}{\ell^2 (\ell{+}K_1)^2 (\ell{+}K_{12})^2
\ldots (\ell{+}K_{12\ldots (m-1)})^2} \, \Big|^{\rm UV}_{D=8-2\epsilon}\nonumber  \\
&\ \  = \frac{(2m{-}9)!!}{(m{-}1)! \times 2^{m-4}\epsilon} \, 
\eta_{(\mu_1 \mu_2}  \eta_{\mu_3 \mu_4}
\ldots \eta_{\mu_{2m-9} \mu_{2m-8})}
\,, \label{UV.04}
\end{align}
where the symmetrization of  $\eta_{(\mu_1 \mu_2}  
\ldots \eta_{\mu_{2m-9} \mu_{2m-8})}=( \eta_{\mu_1 \mu_2} 
\ldots \eta_{\mu_{2m-9} \mu_{2m-8}} + \ldots)/(2m{-}8)!$ is normalized such that $(2m{-}9)!! \eta_{(\mu_1 \mu_2}  
\ldots \eta_{\mu_{2m-9} \mu_{2m-8})}$ gathers all the $(2m{-}9)!!$ inequivalent permutations with unit coefficient.
Note that none of these leading UV contributions depends on the ordering
of the external momenta $K_1,K_2,\ldots,K_m$ of the $m$-gon since they
arise from the $M=0$ contribution of eq.~(6.16) in ref.~\cite{Edison:2021ebi}.

The evaluation of subleading UV divergences in dimensions $d=10,12,\ldots$ is combinatorially more involved but does not pose any conceptual challenges. These subleading divergences are sensitive to the
cyclic ordering of the $K_i$ as reviewed in section 6.3 of ref.~\cite{Edison:2021ebi}, and we will follow the conventions of the reference.

\subsection{UV divergences in maximally supersymmetric SYM}
\label{sec:4.2}

The $d=8$ UV divergences of \msym solely arise from box diagrams. Using \cref{UV.01} and the box numerators given in \cref{sec:maxSYM}, we can write the UV divergences as
\begingroup
\allowdisplaybreaks
\begin{align}
\epsilon A_{\textrm{\msym}}^{\textrm{1-loop}}(1,2,3,4) \, \Big|^{\rm UV}_{D=8-2\epsilon} 
&= \frac{1}{6} t_8(f_1,f_2,f_3,f_4) = \frac{1}{6}s_{12}s_{23}A^{\textrm{tree}}_{\textrm{YM}}(1,2,3,4)\,, \label{UV4p}\\
\epsilon A_{\text{\msym}}^{\text{1-loop}}(1,2,3,4,5) \, \Big|^{\rm UV}_{D=8-2\epsilon} 
&= - \frac{1}{6} \bigg[ \frac{ t_8(f_{12},f_3,f_4,f_5) }{s_{12}} + {\rm cyclic}(1,2,3,4,5) \bigg]\,, \label{UV5p} \\
\epsilon A_{\text{\msym}}^{\text{1-loop}}(1,2,3,4,5,6) \, \Big|^{\rm UV}_{D=8-2\epsilon} 
&=  \frac{1}{6} \bigg[ \frac{1}{s_{123}} \bigg( \frac{ \nmax{[[12]3]456} }{s_{12}}
+ \frac{ \nmax{[1[23]]456} }{s_{23}} \bigg) + \frac{ \nmax{[12][34]56} }{s_{12}s_{34}} \nonumber \\*
&\quad \quad \quad
+ \frac{ \nmax{[12]3[45]6} }{2s_{12}s_{45}}
+ \text{cyclic}(1,2,3,4,5,6)
\bigg]\,, \label{UV6p} 
\end{align}
\endgroup
together with a similar formula for seven points. At five points, we have used \cref{box5pt} for the box numerator. At six points, one can effectively substitute the box numerators with
\begin{align}
\nmax{[[12]3]456}
 &\rightarrow t_8(f_{123},f_4,f_5,f_6)\,, \notag \\
\nmax{[12][34]56}  &\rightarrow 
t_8(f_{12},f_{34},f_5,f_6)\,,
\label{UV.06} \\
\nmax{[12]3[45]6} &\rightarrow 
t_8(f_{12},f_3,f_{45},f_6)\,,
 \notag
\end{align}
within the $d=8$ UV divergence.
Namely, we only keep the $t_8$-tensors that depend on multiparticle field strengths in the parity even numerators given
in \cref{6ptboxnums} and discard the parity odd parts in \cref{nboxodd}. The contact terms in these numerators take the form of $\pol_{i}\Cdot \pol_j
t_8(f_a,f_b,f_c,f_d)$ as well as $\varepsilon_{10}(\pol_{i},\pol_{j},f_a,f_b,f_c,f_d)$, and they drop out in the assembly of the six-point UV divergence in \cref{UV6p}, with non-trivial cancellations between the parity even terms in the one-mass and two-mass boxes. 

The parity odd contributions also drop out from the ten-dimensional UV divergences of \msym at one loop. Physically, the dropouts of parity odd UV divergences in $d=8,10$ at all multiplicities are due to the absence of 
maximally supersymmetric counterterms
with parity odd gluon components at the mass dimensions of $\Tr(F^4)$ and $\Tr(D^2F^4{+}F^5)$.

To better organize the UV divergences, we introduce an $(n{-}3)!$ dimensional vector of tree-level and one-loop color-ordered \msym amplitudes,
\begin{align}\label{eq:Avec}
    & \pmb{A}^{\text{tree}}_{\text{YM}}(n) = A^{\textrm{tree}}_{\textrm{YM}}(1,\alpha,n{-}1,n)\,, & & \pmb{A}^{\text{1-loop}}_{\text{\msym}}(n) = A^{\textrm{1-loop}}_{\textrm{\msym}}(1,\alpha,n{-}1,n)\,,
\end{align}
with permutations $\alpha\in S_{n-3}$ of $2,3,\ldots,n{-}2$.\footnote{As a vector indexed by the permutation $\alpha(2),\alpha(3),\ldots,\alpha(n{-}2)$, the entries of $\pmb{A}^{\textrm{tree/1-loop}}$ are lexicographically ordered.} 
We have checked that, up to and including $n=7$ points, the residues of the $\frac{1}{\epsilon}$ poles in
\cref{UV4p,UV5p,UV6p} follow the generic pattern 
\begin{align}\label{eq:UV8sym}
    \epsilon \pmb{A}^{\text{1-loop}}_{\text{\msym}}(n)\,\Big|^{\rm UV}_{D=8-2\epsilon}=-\frac{1}{6} \, \pmb{P}_2(n)\cdot\pmb{A}^{\text{tree}}_{\text{YM}}(n)\,.
\end{align}
Furthermore, the $d=10$ UV divergences are given by 
\begin{align}\label{eq:UV10sym}
    \epsilon \pmb{A}^{\text{1-loop}}_{\text{\msym}}(n)\,\Big|^{\rm UV}_{D=10-2\epsilon}=\frac{1}{120} \, \pmb{M}_3(n)\cdot\pmb{A}^{\textrm{tree}}_{\text{YM}}(n)\,,
\end{align}
which is also checked up to seven points.\footnote{At seven points, we computed \cref{eq:UV8sym,eq:UV10sym} using only the parity even numerators given in \cref{sec:7pt}. Up to six points, we have also checked that the parity odd terms indeed drop out of the UV divergence in $d=8$ and $10$.} Here, $\pmb{P}_2(n)$ and $\pmb{M}_3(n)$ are $(n{-}3)!\times(n{-}3)!$ matrices whose entries are respectively degree $2$ and $3$ polynomials in Mandelstam variables with rational coefficients. The explicit form of $\pmb{P}_2$ and $\pmb{M}_3$ up to $n=7$ can be found in~\cite{MZVWebsite}. Namely, we have checked that the UV divergences of our BCJ representations of one-loop \msym amplitudes in $d=8$ and $d=10$ reproduce the $\alpha'^2 \zeta_2$- and $\alpha'^3\zeta_3$-contributions to open-superstring
tree amplitudes~\cite{Mafra:2011nv, Schlotterer:2012ny}. These subleading orders in the low-energy expansion are  well known  to feature the matrix elements of the supersymmetric ${\rm Tr}(F^4)$ and ${\rm Tr}(D^2F^4{+}F^5)$ interactions.

\subsection{Type IIA/IIB supergravity from double copy and UV divergences}
\label{sec:4.4}

Knowing the BCJ numerators for one-loop gauge amplitudes, we can directly construct the supergravity amplitudes through the double copy. By choosing either the same or opposite chirality for the fermions in the two copies of \msym in \cref{sgamplitude}, we obtain the numerators $\mathfrak{N}$ of type IIA or IIB supergravity,
\begin{align}
    \mathfrak{N}^{\text{IIA}} &= (\nmeven{} + \nmodd{})(\ntmeven{} - \ntmodd{})\,, \notag\\
    \mathfrak{N}^{\text{IIB}} &= (\nmeven{} + \nmodd{})(\ntmeven{} + \ntmodd{})\,. \label{eq:IIBdc}
\end{align}
The gravitational UV divergences coincide with the low-energy limits of the closed-string genus-one amplitudes \cite{Metsaev:1987ju, Green:2010sp, Pioline:2018pso}. 
The four-point string and supergravity amplitudes are known from \cite{Green:1982sw}, and superspace representations of higher-point string amplitudes encoding any combination of external bosons and fermions can be found in \cite{Green:2013bza} at five points and \cite{Mafra:2016nwr} at six points. As another non-trivial check to our main results, we verify the matching of UV divergences and low-energy limits for up to six external IIB gravitons, following the prescription~\eqref{eq:IIBdc} with $N=\widetilde{N}$. The UV divergences of these amplitudes in the critical dimension $d=8$ are checked to reproduce the $(n \leq 6)$-point matrix elements of the maximally supersymmetric $R^4$ operator.

We will write the UV divergences of supergravity in terms of
the coefficients of the highest tensor power of $\ell$ in pentagon and hexagon BCJ numerators, 
\begin{align}
 N^\mu &= \nmax{12345} \big|_{\ell_\mu} = \frac{\partial}{\partial\ell_{\mu}}\nmax{12345} \, , \notag\\
 N^{\mu \nu} &= \nmax{123456} \big|_{\ell_\mu \ell_\nu}=\frac{1}{2}\frac{\partial^2}{\partial\ell_{\mu}\partial\ell_{\nu}}\nmax{123456} \, , \label{UV.07} \\
N_{[12]}^\mu &= \nmax{[12]3456} \big|_{\ell_\mu}=\frac{\partial}{\partial\ell_{\mu}}\nmax{[12]3456}\,,\nonumber
\end{align}
where the $\ell$-derivatives act on all of $\ell_1,\ell_2,\ldots,\ell_n$ in the $n$-gon numerators. In the notation on the left-hand sides, we are exploiting the fact that the coefficients of the highest
powers of $\ell$ in our BCJ numerators do not depend on the ordering of external legs,
for example,\ $\nmax{123456} \big|_{\ell_\mu \ell_\nu} = \nmax{136425} \big|_{\ell_\mu \ell_\nu}$ 
and $\nmax{[12]3456} \big|_{\ell_\mu} = \nmax{5[12]364} \big|_{\ell_\mu}$. One would otherwise find that after using Jacobi identities to obtain the numerators of lower-gon topologies, the terms with highest power of $\ell$ do not cancel, which would conflict with the power counting of our BCJ numerators. 

At four and five points, we have
\begin{align}
\epsilon M_4^{\text{1-loop IIB}} \, \Big|^{\rm UV}_{D=8-2\epsilon} &=   \big| t_8(f_1,f_2,f_3,f_4)\big|^2\,,  \label{UV.08}
\\
\epsilon M_5^{\text{1-loop IIB}} \, \Big|^{\rm UV}_{D=8-2\epsilon} &= \bigg[ \frac{ \big|t_8(f_{12},f_3,f_4,f_5)\big|^2 }{s_{12}}
+(1,2|1,2,3,4,5) \bigg] + \frac{1}{2} N^\mu \eta_{\mu \nu} \widetilde{N}^\nu\,, \nonumber
\end{align}
where the factors of $\frac{1}{3!}$ and $\frac{1}{4!}$ in the box and pentagon
UV divergences \eqref{UV.01} and \eqref{UV.02} are compensated by the fact that all
the $3!$ and $4!$ cyclically inequivalent orderings of the box and pentagon corners 
yield the same UV contributions. Similarly, factors of $5!$ cancel in the hexagon contribution to
\begin{align}
\epsilon M_6^{\text{1-loop IIB}}\, \Big|^{\rm UV}_{D=8-2\epsilon} &= \bigg[  
\frac{ | \nmax{[[12]3]456} |^2 }{s_{12} s_{123}}
{+}\frac{  | \nmax{[[13]2]456} |^2 }{s_{13} s_{123}}
{+}\frac{  | \nmax{[[23]1]456} |^2 }{s_{23} s_{123}} 
+ (1,2,3|1,2,\ldots,6) \bigg]  \nonumber \\
&\quad + \bigg[
\frac{ |\nmax{[12][34]56}|^2 }{s_{12}s_{34}}
{+}\frac{ |\nmax{[13][24]56}|^2 }{s_{13}s_{24}}
{+}\frac{ |\nmax{[14][23]56}|^2 }{s_{14}s_{23}}
+(5,6|1,2,\ldots,6) \bigg]  \nonumber \\
&\quad + \frac{1}{2} \bigg[  \frac{ N^\mu_{[12]} \eta_{\mu \nu} \widetilde{N}^\nu_{[12]} 
}{s_{12}} + (1,2|1,2,\ldots,6) \bigg] \nonumber\\
&\quad
+ \frac{1}{4} N^{\mu \nu} (\eta_{\mu \nu}  \eta_{\lambda \rho} + \eta_{\mu \lambda}  \eta_{\nu \rho} + \eta_{\mu \rho}  \eta_{\nu \lambda} ) \widetilde{N}^{\lambda \rho} \,. \label{UV.09}
\end{align}
In \cref{UV.08,UV.09} and below, the notation $|t_8|^2$ and $|N|^2$ instructs to multiply the enclosed polarization-dependent expression by a second copy with $\pol_j \rightarrow \widetilde{\pol}_j$ and is unrelated to complex conjugation. 
The type IIA counterparts of \cref{UV.08,UV.09} can be obtained through a sign flip in the parity odd contributions to the $ \widetilde{N}$, cf.\ \cref{eq:IIBdc}. 

The eight-dimensional UV divergences \cref{UV.08,UV.09} apply to arbitrary combinations of external gravitons, $B$-fields and dilatons. We shall now specialize to external gravitons where the parity odd components and the UV divergences are particularly constrained.

The parity odd parts of the type IIA and IIB numerators arise from the crossterms in \cref{eq:IIBdc}, i.e.
\begin{align}
    \mathfrak{N}^{\text{IIA/B}} \, \big|_{\rm odd} &= 
    \nmodd{} \ntmeven{} \mp 
\nmeven{} \ntmodd{}\,. \label{eq:IIodd}
\end{align}
In the type IIB case, the parity odd terms $\nmeven{}\ntmodd{}+ \ntmeven{}\nmodd{}$ do not contribute to UV divergences of external gravitons (as opposed to $B$-fields or dilatons) in $d=8$.\footnote{The five-point UV divergences \cref{UV.08} in eight dimensions for instance comprise parity odd terms in the type IIA case with four external gravitons and one external $B$-field as well as in the type IIB case with three external gravitons and two external $B$-fields. In both case, we treat the one-loop amplitude as an analytic function in $D$ and defined the UV divergence at $d=8$ as the residue at $D=8-2\epsilon$. We refer to parity and $B$-field components in ten spacetime dimensions and do not perform a dimensional reduction on the external states.
%In both cases, we refer to parity and $B$-field components in ten spacetime dimensions and do not perform a dimensional reduction of the external states in determining the UV divergences from eight-dimensional internal states.
} At five points, this can be easily checked using the explicit form~\eqref{eq:pen5p} for the pentagon numerators. For six gravitons, we have checked numerically that parity odd terms indeed drop out from \cref{UV.09}. We expect that this feature holds to all multiplicities. In other words, it is sufficient to use
\begin{align}
    \mathfrak{N}^{\text{IIB}}
   \, \big|_{\rm even}= 
    |\nmeven{}|^2+|\nmodd{}|^2
\end{align} 
to compute the $d=8$ UV divergence for external gravitons. The double copy of the parity odd contributions will produce products of two Levi-Civita tensors, which can be turned into Gram determinants,
\begin{align}
    \varepsilon_{d}(v_1,\ldots,v_d)\varepsilon_{d}(w_1,\ldots,w_d)&=-\det\big[v_i\cdot w_j\big]_{d\times d}\,,\nonumber\\
    \varepsilon_{d}^{\mu}(v_1,\ldots,v_{d-1})\eta_{\mu\nu}\varepsilon_{d}^{\nu}(w_1,\ldots,w_{d-1})&=-\det\big[v_i\cdot w_j\big]_{(d-1)\times(d-1)}\,.
\end{align}
Note that the minus sign on the right-hand side is due to the Minkowskian signature.

We have checked up to six points that the purely gravitational UV divergence of type IIB supergravity in $d=8$ can be written compactly as
\begin{align}\label{eq:UV2b}
    \epsilon M_{n \ \text{gravitons}}^{\text{1-loop IIB}}\,\Big|^{\rm UV}_{D=8-2\epsilon} &= \big(\widetilde{\pmb{A}}{}_{\text{YM}}^{\text{tree}}(n)\big)^{T}\cdot\pmb{S}_0(n)\cdot\pmb{M}_3(n)\cdot\pmb{A}_{\text{YM}}^{\text{tree}}(n)\,,
\end{align}
for external gravitons. Here, $\widetilde{\pmb{A}}{}_{\text{YM}}^{\textrm{tree}}(n)$ is another $(n{-}3)!$ dimensional vector of color ordered \msym amplitudes, $\widetilde{\pmb{A}}{}_{\text{YM}}^{\textrm{tree}}(n) = A^{\textrm{tree}}_{\textrm{YM}}(1,\alpha,n,n{-}1)$ with permutations $\alpha\in S_{n-3}$ of $\{2,3,\ldots,n{-}2\}$.
Comparing with $\pmb{A}_{\text{YM}}^{\textrm{tree}}(n)$ given in \cref{eq:Avec}, we switch the last two fixed external legs $n{-}1,n$ such that the entries of $\pmb{S}_0(n)$, the field-theory KLT or momentum kernel~\cite{Kawai:1985xq, Bern:1998sv, Bjerrum-Bohr:2010pnr}, are given by polynomials of Mandelstam variables (with
all-multiplicity formulae in the references).

The kinematic factor in \cref{eq:UV2b} agrees with the $\alpha'^3\zeta_3$ coefficient of type IIB closed-superstring tree-level amplitudes~\cite{Schlotterer:2012ny}, which corresponds to a single insertion of a supersymmetric $R^4$ operator. At tree level, the expression \eqref{eq:UV2b} applies to the $\alpha'^3\zeta_3$ order for arbitrary external states of the type IIB supergravity multiplet including those configurations that violate the $U(1)$ R-symmetry of the $\alpha' \rightarrow 0$ limit. 
However, \cref{eq:UV2b} as a one-loop expression (for UV divergences in supergravity or low-energy limits of string amplitudes) only applies to external states of vanishing overall $U(1)$ charge such as $n$ gravitons. 
For $U(1)$-violating components such as UV divergences of one dilaton and $(n{-}1)$ gravitons, there are additional rational prefactors relative to the tree-level contribution $\sim \zeta_3$. Such relative factors play an essential role for S-duality properties of the low-energy effective action of type IIB superstrings, see for instance \cite{Green:2013bza, Green:2019rhz} and \cite{Gomez:2015uha, DHoker:2020tcq} for discussions at genus one and genus two, respectively.

We have verified up to six points that our integrands lead to vanishing ten-dimensional UV divergences for both type IIA and IIB supergravity, reflecting the absence of maximally supersymmetric counterterms $D^2R^4{+}R^5$. 
This is known to be a straightforward consequence of maximal supersymmetry, and it has been discussed from several perspectives in {\it e.g.}\ refs.~\cite{Drummond:2003ex, Kallosh:2009jb, Elvang:2010jv}.

As the final check to our gravity numerators, we have verified that our gravity integrands are gauge invariant on all the spanning cuts. At seven points,
we have furthermore verified that the cuts in type~IIA/B supergravity calculated using the double-copy numerators are independent of the free parameters discussed in \cref{sec:7pt}.
We compute the gravity cuts via the standard procedure of summing relabeled numerators along with uncut propagators for all diagrams that share a pole structure with the cut \cite{Bern:2010ue,Bern:2017yxu,Bern:2018jmv}.
It is explained with explicit examples in \cref{sec:grcut}.

\section{Results for half-maximally supersymmetric
theories}
\label{sec:half-max}

In this section, we present one-loop BCJ numerators in gauge theories with half-maximal supersymmetry (i.e.\ 8 supercharges)~\cite{Brink:1976bc}, later called \hsym for short. The resulting amplitudes capture supermultiplets with half-maximal supersymmetry running in the loop.
In six dimensions, half-maximal supersymmetry can be realized in a hypermultiplet that consists of one chiral fermion and two scalars, or in a vector multiplet that consists of one gluon and two chiral fermions. We write their contribution to the kinematic numerators as
\begin{align}
\label{discaround}
    & \text{hyp}_{(1,0)}: & &\nh{}=\nheven{}+\nhodd{}\,, & \text{vec}_{(1,0)}: & & \nv{}=\nveven{}+\nvodd{}\,,\nonumber\\
    & \overline{\text{hyp}}_{(0,1)}: & & \nhb{}=\nheven{}-\nhodd{}\,,& \text{vec}'_{(0,1)}: & & \nvb{}=\nveven{}-\nvodd{}\,,
\end{align}
where we have further separated the parity even and odd contributions. The multiplet hyp and $\overline{\text{hyp}}$ (vec and $\text{vec}'$) have opposite chiralities, where the subscript $(1,0)$ and $(0,1)$ denote the R-symmetry representations.\footnote{The $\text{hyp}_{(1,0)}$ multiplet is sometimes called ``half-hyper'' to emphasise that it encodes only two fermionic and two bosonic on-shell states.}
Meanwhile, the dimensional reduction of the \msym gauge multiplet to six dimensions is non-chiral and carries $(1,1)$ R-symmetry. It can be decomposed into two  hypermultiplets and one 
vector multiplet of the opposite chirality,
\begin{align}
    \nmax{}=\nv{}+2 \nhb{} \,,
\end{align}
where $\nmax{}= \nmeven{}$ is given in \cref{sec:maxSYM}. The contributions from vector and hypermultiplets are thus related by
\begin{align}\label{eq:decom}
    \nveven{}=\nmax{}-2\nheven{}\,,\qquad \nvodd{}=2\nhodd{}\,.
\end{align}
In the following, we will focus on the case of a hypermultiplet running in the loop since the contributions from vector multiplets can be reconstructed via \cref{eq:decom}.

\begin{figure}
    \centering
    \subfloat[][]{\begin{tikzpicture}[every node/.style={font=\footnotesize}]
        \draw [thick] (0,0) node[left=-1pt]{$1$} -- (1.5,0);
        \draw [thick] (0.5,0) -- ++ (0,0.5) node[above=-1pt]{$2$};
        \draw [thick] (1,0) -- ++ (0,0.5) node[above=-1pt]{$3$};
        \draw [thick] (1.5,0) to [out=70,in=110] (2.5,0);
        \draw [thick] (1.5,0) to [out=-70,in=-110] (2.5,0);
        \draw [thick] (2.5,0) -- ++ (0.5,0) node[right=-1pt]{$4$};
        \begin{scope}[yshift=1.2cm]
        \draw [thick] (0,0) node[left=-1pt]{$1$} -- (1.5,0);
        \draw [thick] (1,0) -- ++ (0,0.5) -- ++(135:0.5) node[above=-1pt]{$2$};
        \draw [thick] (1,0.5) -- ++ (45:0.5) node[above=-1pt]{$3$};
        \draw [thick] (1.5,0) to [out=70,in=110] (2.5,0);
        \draw [thick] (1.5,0) to [out=-70,in=-110] (2.5,0);
        \draw [thick] (2.5,0) -- ++ (0.5,0) node[right=-1pt]{$4$};
        \end{scope}
    \end{tikzpicture}\label{fig:exb}
    }
    \qquad
    \raisebox{1cm}{$\Longrightarrow$}
    \qquad
    \subfloat[ ][]{\begin{tikzpicture}[every node/.style={font=\footnotesize}]
        \draw [thick] (0,0) node[left=-1pt]{$1$} -- (1,0);
        \draw [thick] (0.5,0) -- ++ (0,0.5) node[above=-1pt]{$2$};
        \draw [thick] (1,0) -- ++ (0,0.5) node[above=-1pt]{$3$};
        \draw [thick] (1,0) to [out=70,in=110] (2,0);
        \draw [thick] (1,0) to [out=-70,in=-110] (2,0);
        \draw [thick] (2,0) -- ++ (0.5,0) node[right=-1pt]{$4$};
        \begin{scope}[yshift=1.2cm]
        \draw [thick] (0,0) node[left=-1pt]{$1$} -- (1,0);
        \draw [thick] (1,0) -- ++ (0,0.5) -- ++(135:0.5) node[above=-1pt]{$2$};
        \draw [thick] (1,0.5) -- ++ (45:0.5) node[above=-1pt]{$3$};
        \draw [thick] (1,0) to [out=70,in=110] (2,0);
        \draw [thick] (1,0) to [out=-70,in=-110] (2,0);
        \draw [thick] (2,0) -- ++ (0.5,0) node[right=-1pt]{$4$};
        \end{scope}\begin{scope}[xshift=3.5cm,yshift=0.6cm]
        \draw [thick] (0,0) node[left=-1pt]{$2$} -- (1,0);
        \draw [thick] (1,0) -- ++(150:1) node[left=-1pt]{$3$};
        \draw [thick] (1,0) -- ++ (-150:1) node[left=-1pt]{$1$};
        \draw [thick] (1,0) to [out=70,in=110] (2,0);
        \draw [thick] (1,0) to [out=-70,in=-110] (2,0);
        \draw [thick] (2,0) -- ++ (0.5,0) node[right=-1pt]{$4$};
        \end{scope}
    \end{tikzpicture}\label{fig:contact}
    }
    \caption{The BCJ numerators associated with the cubic graphs in the left panel (a) conspire to cancel the propagator $s_{123}^{-1}$ that diverges in the phase space of four on-shell momenta. After cancellation of their divergent propagator, the contributions of external bubbles may be visualized as ``snail graphs'' drawn in the right panel (b).}
    \label{fig:4p_exb}
\end{figure}
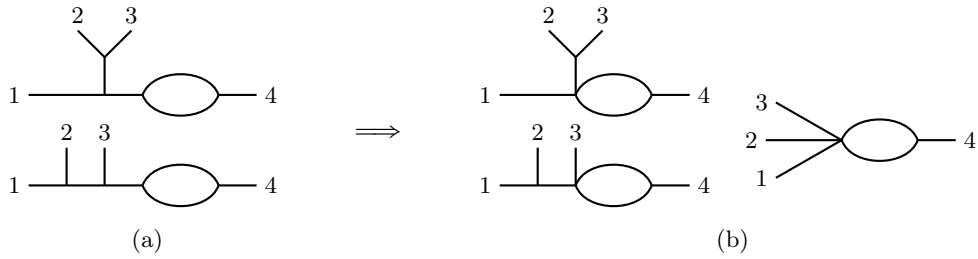

As we will see, the $n$-point numerators in \hsym take a form very similar to the $(n{+}2)$-point numerators in \msym. However, the presence of triangle and bubble topologies in \hsym causes additional subtleties. It is inevitable that Jacobi identities will relate triangles to bubbles in external legs with a formally divergent propagator, for example, $s^{-1}_{123}$ in the four-point diagrams of \cref{fig:exb}. They must be canceled by the same Mandelstam variable in the numerator and become the contact graphs shown in \cref{fig:contact}.
A consistent regularization prescription where these cancellations can be carried out was proposed by Minahan~\cite{Minahan:1987ha}:
we first relax the $n$-point momentum conservation by introducing an additional soft massless leg with momentum $p$,
\begin{align}\label{minahaning}
    k_1+k_2+\ldots+k_n=p\,,\qquad p^2=0\,.
\end{align}
As a result, we can only use the Mandelstam identity $s_{12\ldots n}=0$ of an $(n{+}1)$-point momentum phase space in the numerators until all the potentially divergent propagators like $s^{-1}_{12\ldots n-1}$ are canceled.\footnote{We can still use the original $n$-point momentum conservation in $\pol\Cdot k$, $\ell\Cdot k$ since they are irrelevant to cancelling the divergent propagators.} We can then set $p=0$ and arrive at a regular integrand that only contains contact graphs like in \cref{fig:contact}. 

\subsection{Kinematic building blocks and maximal cuts}\label{sec:trace_hm}

Just like in \msym, the kinematic numerators in \hsym can be conveniently expressed in terms of some particular combinations of vector and spinor traces. We define \cite{Edison:2020uzf}
\begin{align}\label{deftrhyp}
    \trhyp(f_1f_2\ldots f_m)&=-\frac{1}{2\times 4^m}f_1^{\mu_1\nu_1}f_2^{\mu_2\nu_2}\ldots f_m^{\mu_m\nu_m}\tr\Big(\gmsix{\mu_1 \nu_1}\gmsix{\mu_2 \nu_2}\ldots\gmsix{\mu_m \nu_m}\Big)\,,
\end{align}
with antisymmetrized combinations $\gmsix{\mu\nu}$ of $8\times 8$ Dirac gamma matrices $\gmsix{\mu}$ in six dimensions
subject to $\tr\big(\gmsix{\mu}\gmsix{\nu}\big)=8\eta_{\mu\nu}$. As the subscript suggests, $\trhyp$ will be used to express integrands with a hypermultiplet running in the loop. The examples up to length five are given below,
\begin{align}
    &\trhyp(f_1)=0\,,\quad \trhyp(f_1f_2)=-\frac{1}{2}\tr(f_1f_2)\,,\quad \trhyp(f_1f_2f_3)=-\frac{1}{2}\tr(f_1f_2f_3) \, ,\nonumber\\
    &\trhyp(f_1 f_2 f_3 f_4 )=-\frac{1}{2}
\tr(f_1f_2f_3f_4)+\frac{1}{4} t_8(f_1,f_2,f_3,f_4)\,,  \label{4ptexh} \\
&\trhyp(f_1 f_2 f_3 f_4 f_5)=-\frac{1}{2}
 {\rm tr}(f_1f_2f_3f_4f_5)
+\frac{1}{8}\Big[ t_8(f_1,[f_2,f_3],f_4,f_5)+(2,3|2,3,4,5)\Big]\,.
\nonumber
\end{align}
They will be used to express BCJ numerators up to five points.

The $n$-gon maximal cut contributed by a hypermultiplet can be obtained from the general formula~\eqref{eq:ngon-mc} by setting $(n_{\text{l}},n_{\text{r}})=(1,0)$ or $(0,1)$, $n_{\text{s}}=2$ and $n_{\text{v}}=0$. After converting the spinor traces in $d=6$ for the parity even contribution to vector traces via \cref{4ptexh}, we get dimension-agnostic expressions that also apply to four-dimensional SYM with eight supercharges\footnote{Similar to \cref{nptmaxcut} for \msym, we rescaled the numerator by a factor of $2$ compared with \cref{eq:ngon-mc}.}
\begin{align}\label{maxcuthalfs}
    \nheven{12\ldots n}&=\sum_{k=0}^{n-2}(-1)^{n-k}\bigg[\trhyp(f_{k+1}\ldots f_n)\prod_{j=1}^{k}\pol_{j}\cdot\ell_j+(1,2,\ldots,k|1,2,\ldots,n)\bigg] \nonumber\\
    &\quad +\mathcal{O}(\ell_i^2)\text{ terms that vanish on the maximal cut} \, .
\end{align}
Here, the parity even maximal cut has been identified as part of the $n$-gon numerator because it already manifests the full dihedral crossing symmetry. The nonzero length-two and length-three traces lead to the $\ell^{n-2}$ power counting for \hsym instead of $\ell^{n-4}$ for \msym. To determine the $\mathcal{O}(\ell_i^2)$ contact terms, one can make a generic ansatz and then solve it by matching all the lower-gon cuts constructed from tree amplitudes. Alternatively, just like in the \msym case, we can solve the ansatz by imposing linearized gauge invariance on the cuts. 
We note that the gauge variation of the maximal-cut part in \cref{maxcuthalfs} contains at least a length-two $\trhyp$. Since the $\mathcal{O}(\ell_i^2)$ contact terms are introduced to cancel the gauge variation on lower-gon cuts, it is natural to conjecture that the $\mathcal{O}(\ell_i^2)$ contact terms will also contain at least a length-two $\trhyp$.\footnote{
The regularization of external bubbles via \cref{minahaning} may require us to expand out certain traces such that not every term features a length-two trace, as we will see explicitly in \cref{bubble3pt,eq:contactbubble}. However, the ansatz proposed here is still sufficient since it turns out that bubble cuts will be automatically satisfied for all the cases we have considered. Thus we do not need new structures in the ansatz.
}

One can similarly consider a vector multiplet running in the loop. The SUSY decomposition~\eqref{eq:decom} implies that the $n$-gon maximal cut can be obtained by the replacement $\trhyp \rightarrow \trvec$ in \cref{maxcuthalfs}, where
\begin{align}
\trvec(f_1f_2\ldots f_m)  = \trVS(f_1f_2\ldots f_m) -  2\, \trhyp(f_1f_2\ldots f_m)\,.
\end{align}
At low multiplicities, we have,
\begin{gather}
    \trvec(f_1)=0\,,\quad \trvec(f_1f_2)=\tr(f_1f_2)\,,\quad \trvec(f_1f_2f_3)=\tr(f_1f_2f_3)\,,\nonumber\\ \label{eq:4pvec}
    \trvec(f_1f_2f_3f_4)=\frac{1}{2}t_8(f_1,f_2,f_3,f_4)+\tr(f_1f_2f_3f_4)\,.
\end{gather}
The approach that determines contact terms from gauge invariance on cuts is agnostic to whether we use $\trhyp$ or $\trvec$ in the maximal cut. It suggests that under our current assumption of $\mathcal{O}(\ell_i^2)$ contact terms, the prescription $\trhyp\rightarrow\trvec$ can be applied to the full numerator.
This holds for all the explicit examples shown later in this section. 
By construction, the union of one vector and two hypermultiplets in the loop will reproduce the parity even part of the \msym result~\eqref{nptmaxcut}.

\subsubsection{Three-point numerators and regularization}
\label{sec:3phalf}
Because of the power counting $\nh{12 \ldots n} \sim {\cal O}(\ell^{n{-}2})$, there are no $\mathcal{O}(\ell_i^2)$ contact terms at two and three points. Hence, the maximal cut already gives the full result for the bubble at two points and the triangle at three points known from \cite{Berg:2016fui}, 
\begin{align}\label{master2pt3pt}
   \nh{12} & = \trhyp{(f_1f_2)}\,,\\ 
   \nh{123} &  = \left[ \pol_1\Cdot\ell_{1} \trhyp (f_2f_3)+\text{cyclic}(1,2,3)\right]-\trhyp{(f_1f_2f_3)} -\frac{1}{4}
    \varepsilon_6(\ell,\pol_1,f_2,f_3)\,,
    \notag
\end{align}
where we already reinstated the parity odd contribution at three points from \cref{eq:ngon-mc}.
The numerator for the external bubble with $[12]$-dangling tree is \cite{Berg:2016fui}
\begin{align}
     \nh{[12]3}=\nh{123}-\nh{213}=-\trhyp(f_{12}f_3)\,,
     \label{3ptbub}
\end{align}
where no momentum-conservation relations have been used for the parity even part. This graph has a formally divergent propagator $s_{12}^{-1}$. To consistently cancel this spurious pole, we evaluate the trace of $f_{12}f_3$ in a momentum phase space with non-zero $s_{ij}$, corresponding to an additional soft leg in the regularization~\eqref{minahaning}.
In this way, the external bubble numerator~\eqref{3ptbub} takes the form \cite{Minahan:1987ha}
\begin{align}
\label{bubble3pt}
  \nh{[12]3}=\frac{1}{2}  s_{12} \, \epsilon _1\Cdot \epsilon _2 \, \epsilon _3 \Cdot k_1-\frac{1}{2} s_{123}\Big[ \epsilon _1\Cdot \epsilon _2 \, \epsilon _3 \Cdot k_1 + \text{cyclic}(1,2,3)\Big]\,.
\end{align}
Now the first term cancels the $s_{12}$ in the denominator from one of the bubble propagators, and the second term drops out as $s_{123}=0$ even in the presence of the additional soft leg. Finally, we collect all the graphs that are compatible with the ordering $(1,2,3)$ and arrive at the color ordered amplitude
\begin{align}
\label{loopintegrand3}
A^{\text{1-loop}}_{\text{hyp}}(1,2,3)=\int\frac{\dd^{D}\ell}{i\pi^{D/2}}
\left[\frac{\nh{123}}{\ls{1}\ls{2}\ls{3}}+ \frac{\epsilon _1\Cdot \epsilon _2\, \epsilon _3\Cdot k_1  }{2 \ls{2}\ls{3}}+ \frac{\epsilon _2\Cdot \epsilon _3 \, \epsilon _1\Cdot k_2  }{2\ls{1}  \ls{3}}+ \frac{\epsilon _3\Cdot \epsilon _1\, \epsilon _2\Cdot k_3  }{2\ls{1}  \ls{2}}\right]\,,
\end{align}
where $\ell_i=\ell{+}k_1{+}\ldots{+}k_i$. By virtue of the resolution of spurious poles in $s_{ij}$, this expression is regular and gauge invariant at the level of the loop integrand.

\subsection{The four- and five-point numerators}

Starting from four points, we need to work out the $\ls{i}$ contact terms in the $n$-gon numerators in \cref{maxcuthalfs}. We use the same strategy as in the \msym case: we make a general ansatz that respects the full dihedral crossing symmetry, scales as $\ell^{n-2}$ and contains at least one factor of $\ell_i^2$ in each term. All the other numerators are then obtained through Jacobi identities and relabelings, among which we require the tadpole numerators to vanish.

To solve for the remaining free parameters, we turn to the constraints imposed by lower-gon cuts. In particular, we require that all the lower-gon cuts are gauge invariant (for the parity odd terms, the cuts are taken strictly in $d=6$). The contact terms in the $n$-gon numerators are thus chosen to compensate the gauge variation when fewer loop momenta are on-shell. This consideration guides us to write down a natural ansatz. Our strategy is different from the usual unitarity-cut method, which requires the lower-gon cuts to match the product of tree amplitudes. Of course, we have done the calculation using both methods and found agreeing results.

Very interestingly, at both four and five points, the physically relevant part of our ansatz can be completely fixed after imposing the triangle cuts to be gauge invariant. In other words, bubble cuts will be satisfied automatically and thus do not give further constraints. At four points, there are no free parameters left; at five points, the remaining free parameters belong to the generalized gauge freedom and drop out in the color ordered integrands.\footnote{When constructing numerators for external-bubble graphs, we need to consistently apply the regularization~\eqref{minahaning} such that the divergent propagators in such graphs get canceled before matching the cuts and performing gauge variations. The simplest three-point example has been given in \cref{sec:3phalf}. The details for four and five points will be given in later sections.}

While the above procedure in principle works for both the parity even and odd parts, we will face the same technical difficulty for the parity odd parts encountered in \msym. We can only meaningfully match cuts (or check gauge invariance) in a minimal basis of kinematic variables. However, in the presence of Levi-Civita tensors and two (or more) Lorentz dot products, it is a challenge to attain a minimal basis by exhaustive application of Schouten identities \eqref{eq:schouten}. We will present the four-point parity odd result, and leave $n \geq 5$ points to the future.

\subsubsection{The four-point parity even numerators}
According to the all-multiplicity result \eqref{maxcuthalfs}, we write the parity even box numerator as 
\begin{align}
    \nheven{1234}&=\left[ \pol_1\Cdot\ell_{1}\pol_2\Cdot\ell_{2} \trhyp(f_3f_4)+(1,2|1,2,3,4)\right] \\
    &\quad -\left[ \pol_1\Cdot\ell_1 \trhyp(f_2f_3f_4) +{\rm cyclic}(1,2,3,4)\right] + \trhyp (f_1f_2f_3f_4) + \delta 
    \nheven{1234}\,,\nonumber
\end{align}
where $\delta \nheven{1234}$ contains all the $\mathcal{O}(\ls{i})$ contributions. Similar to the discussion for \msym given at the beginning of \cref{sec:maxSYM}, we start from an ansatz with $D_4$ crossing symmetry
\begin{align}
    \delta \nheven{1234} &= 
    \pol_{1} \Cdot \pol_{2}(a_1 \ell_1^2+a_2 \ell_2^2+a_2 \ell_4^2)  \trhyp (f_3f_4)
    \label{partiyevenhalf4pt} \\
    &\quad +  
    \pol_{1} \Cdot \pol_{3}a_3
    ( \ell_1^2+ \ell_2^2)  \trhyp (f_2f_4)+ {\rm cyclic}(1,2,3,4)\,,
    \notag
\end{align}
and only allow for inverse propagators $\ell_i^2$ adjacent to the polarization vectors in the dot product, cf.\ \cref{cont6pt}. The three rational parameters $a_1,a_2,a_3$ are uniquely fixed by imposing gauge invariance of the triangle cut,
\begin{align}
\text{Cut}\!\left[\!\!\!\vcenter{\hbox{
    \begin{tikzpicture}[scale=0.7,every node/.style={font=\scriptsize}]
    \pgfmathsetmacro{\r}{1.2}
        \pgfmathsetmacro{\l}{0.6}
        \pgfmathsetmacro{\y}{-2.2}
        \coordinate (1) at (180:\r);
        \coordinate (3) at (60:\r);
        \coordinate (4) at (-60:\r);
        \path (3) -- (4) coordinate[midway] (l3);
        \path (1) -- (3) coordinate[midway] (l2);
        \path (4) -- (1) coordinate[midway] (l4);
        \draw[cuts] (l3) ++ (-0.2,0) -- ++ (0.4,0) node [right=1pt]{$\ell_3$};
        \draw[cuts] (l2) ++ (-60:0.2) -- ++ (120:0.4) node [above=1pt]{$\ell_2$};
        \draw[cuts] (l4) ++ (60:0.2) -- ++ (-120:0.4) node [below=1pt]{$\ell_4$};
        \draw[thick] (1) -- ++ (120:\l) node[left=-1pt] {$2$};
        \draw[thick] (1) -- ++ (-120:\l) node[left=-1pt] {$1$};
        \draw[thick] (3) -- ++ (60:\l) node[right=-1pt]{$3$};
        \draw[thick] (4) -- ++ (-60:\l) node[right=-1pt]{$4$};
        \fill (1) circle (1pt);
        \draw[thick] (3) -- (1);
        \draw[thick] (4) -- (3);
        \draw[thick] (1) -- (4);
        \node at (0,0) {\scalebox{1.75}{$\circlearrowleft$}};
    \end{tikzpicture}
    }}\!\!\!\right]_{\pol_i\rightarrow k_i}
  =\bigg[\frac{\nheven{1234}}{(\ell_4+k_1)^2}+ \frac{\nheven{[12]34}}{(k_1+k_2)^2} \bigg]_{\substack{\ell^2=\ls{2}=\ls{3}=0\\ \pol_i\to k_i } } =0 \,,
\end{align}
leading to the following result for the parity even box numerator,
\begin{align}\label{eq:hmbox}
    \nheven{1234}&=\left[ \pol_1\Cdot\ell_{1}\pol_2\Cdot\ell_{2} \trhyp(f_3f_4)+(1,2|1,2,3,4)\right]\\
    \nonumber
    &\quad-\left[ \pol_1\Cdot\ell_1 \trhyp(f_2f_3f_4) +{\rm cyclic}(1,2,3,4)\right] + \trhyp (f_1f_2f_3f_4)
    \\
    &\quad-\frac{1}{24}\left[\pol_1\Cdot\pol_2(6\ls{1}-\ls{2}-\ls{4}) \trhyp(f_3f_4)+\text{cyclic}(1,2,3,4)\right]\nonumber\\
    &\quad- \frac{1}{24}\left[\pol_1\Cdot\pol_3(\ls{1}+\ls{2})\trhyp(f_2f_4)+\text{cyclic}(1,2,3,4)\right]\,.
        \nonumber
\end{align}
The numerators of lower-gon topologies can be obtained using kinematic Jacobi identities and conveniently expressed in
terms of Lorentz traces and multiparticle fields. 
The tadpole numerators obtained from repeated antisymmetrization of $\nheven{1234}$ vanish as expected. 
The triangle numerator with leg $1$ and $2$ in a dangling tree reads
\begin{align}
    \nheven{[12]34}&=-\ell_{\mu} \big[ \pol_{12}^{\mu} \trhyp(f_3 f_4)+\pol_{3}^{\mu}\trhyp(f_{12} f_4)+\pol_{4}^{\mu}\trhyp(f_{12} f_3) \big] \nonumber\\
   \nonumber
    &\quad
    +\frac{1}{6} \Big[ \ell\Cdot(k_1{+}k_2{-}k_3) \big(\pol_{1}\Cdot\pol_{3}\trhyp(f_2 f_4)-\pol_{2}\Cdot\pol_{3}\trhyp(f_1 f_4)\big) + (3\leftrightarrow 4)\Big]
       \\
   \nonumber
    &\quad
    +\frac{1}{12}s_{12}\Big[  \pol_{1}\Cdot\pol_{3}\trhyp(f_2 f_4)-\pol_{2}\Cdot\pol_{3}\trhyp(f_1 f_4) - (3\leftrightarrow 4)\Big]
     \\ &\quad
    +\frac{1}{2}\Big[\trhyp(f_{123} f_4)+\trhyp(f_{124} f_3)+\trhyp(f_{12} f_{34})  \Big]\,,
\end{align}
and the two-mass bubble with $[12]$ and $[34]$ dangling trees is given by
\begin{align}
 \nheven{[12][34]}=
 \trhyp({f}_{12}{f}_{34}) +\frac{s_{12}}{6}\Big[  \pol_{1}\Cdot\pol_{3}\,\trhyp(f_2f_4)-\pol_{2}\Cdot\pol_{3}\,\trhyp(f_1f_4) - (3\leftrightarrow 4)\Big]\,.
\end{align}
The triangles and two-mass bubbles do not contain any divergent propagators, and thus we can freely apply four-point momentum conservation to their numerators. 

Because of the divergent propagators in external-bubble graphs, we should follow the regularization~\eqref{minahaning} while computing their numerators from Jacobi relations. In particular, we cannot use four-point momentum conservation in the Mandelstam variables. For example, we consider the two external-bubble graphs that contain the divergent propagator $s_{123}^{-1}$,
\begin{subequations}\label{eq:extbubble4}
\begin{align}\label{eq:extbubblea}
\nheven{[[12]3]4}&=\nheven{1234}-\nheven{2134}-\nheven{3124}+\nheven{3214}\\
&=
\trhyp(f_{123}f_4)-\frac{1}{12}s_{123}\Big[   2\pol_1\Cdot\pol_3 \, \trhyp(f_2f_4)
+\pol_1\Cdot\pol_4 \, \trhyp(f_2f_3)
-(1\leftrightarrow 2)\Big]\,,\nonumber\\
\nheven{[1[23]]4}&=\nheven{1234}-\nheven{1324}-\nheven{2314}+\nheven{3214}=\nheven{[[12]3]4}\,\Big|_{1\leftrightarrow 3}\,.
\end{align}
\end{subequations}
The box numerators in the first line of \cref{eq:extbubblea} are obtained by relabeling \cref{eq:hmbox}, and we have used the regularization~\eqref{minahaning} in passing to the second line (which also carries over to the other bubble numerator $\nheven{[1[23]]4}$ related by a relabelling). By combining the two graphs in \cref{eq:extbubble4}, the divergent propagator $s_{123}^{-1}$ is cancelled, as depicted in \cref{fig:4p_exb},
\begin{align}\label{eq:contactbubble}
   &\frac{\nheven{[[12]3]4}}{s_{12}s_{123}} +  \frac{\nheven{[1[23]]4}}{s_{23}s_{123}} =\frac{1}{8}\Big(
   \epsilon _1\Cdot \epsilon _4 \epsilon _2\Cdot \epsilon _3+\epsilon _1\Cdot \epsilon _2 \epsilon _3\Cdot \epsilon _4   -2 \epsilon _1\Cdot \epsilon _3 \epsilon _2\Cdot \epsilon _4
\Big) \\
    &+ \frac{1}{24 s_{12}}\Big[6 \epsilon _4 \Cdot k_3 (\epsilon _1\Cdot \epsilon _2 \, \epsilon _3\Cdot k_1 -2 \epsilon _1\Cdot \epsilon _3 \, \epsilon _2\Cdot k_1) +  2\pol_1\Cdot\pol_3 \tr(f_2f_4)+\pol_1\Cdot\pol_4 \tr(f_2f_3)-(1\leftrightarrow 2) \Big]\nonumber\\
    &+ \frac{1}{24 s_{23}}\Big[6 \epsilon _4 \Cdot k_1 (\epsilon _2\Cdot \epsilon _3 \, \epsilon _1\Cdot k_3 -2 \epsilon _1\Cdot \epsilon _3 \, \epsilon _2\Cdot k_3) + 2\pol_1\Cdot\pol_3 \tr(f_2f_4)+\pol_3\Cdot\pol_4 \tr(f_1f_2)-(2\leftrightarrow 3) \Big]\,.
\nonumber
\end{align}
After removing the spurious divergence $s_{123}^{-1}$ we can freely apply four-point momentum conservation to contact bubble graphs like \cref{eq:contactbubble}.\footnote{We note that using four-point relations like $k_4=-k_1-k_2-k_3$ and $s_{13}=-s_{12}-s_{23}$ prematurely in earlier stages may cause a failure in removing the divergence of $s_{123}^{-1}$.} Note that the cancellation of $s_{123}^{-1}$ from the combination of the two diagrams in \cref{eq:contactbubble} has been studied in \cite{Berg:2016fui, Berg:2016wux} at the level of Berends-Giele currents. 

One can now verify that our integrand satisfies all the bubble cuts. Since we derive these numerators by only using information on the box and triangle cuts, the bubble cuts provide important checks to our results. Furthermore, we have also checked that the color-ordered loop integrand is gauge invariant, where the loop momentum is put at the same position across all the graphs involved. Both the bubble cuts and gauge invariance check can only be performed after removing divergent propagators in all the external-bubble graphs. 

\subsubsection{The four-point parity odd numerators}

For parity odd contributions to the four-point box numerator, the maximal cut is given by the last line of \cref{eq:ngon-mc}. To obtain the crossing symmetric numerators, one can make a general ansatz like the one in \cref{parityodd4an} while converting the ten-dimensional six-point expressions $\varepsilon_{10}(k^{a+2},\pol^{b+2})$ to six-dimensional four-point terms $\varepsilon_{6}(k^a,\pol^b)$. Alternatively, one can further exploit the similarity between $(n{+}2)$-point \msym and $n$-point \hsym and convert the ten-dimensional six-point numerators \cref{eq:N6ptoddsym} to an improved ansatz
\begin{align}
\nhodd{1234}&=a_1\Big[\pol_{1}\cdot\ell_1 \, \varepsilon_{6}(\ell_2,\pol_{2},f_3,f_4)+\pol_{2}\cdot\ell_2\, \varepsilon_{6}(\ell_1,\pol_{1},f_3,f_4)+(1,2|1,2,3,4)\Big]\nonumber\\
    &\quad +a_2\Big[\pol_{1}\cdot\ell_1 \,\varepsilon_{6}(f_2,f_3,f_4)+\textrm{cyclic}(1,2,3,4)\Big]\nonumber\\
    &\quad +a_3\Big[\Big(\varepsilon_{6}(\ell_1,\pol_{1},[f_2,f_3],f_4)+\textrm{cyclic}(1,2,3)\Big)+(1,2,3|1,2,3,4)\Big]\nonumber\\
    &\quad +a_4\Big[\ls{1} \, \varepsilon_{6}(\pol_{1},\pol_{2},f_3,f_4)+\textrm{cyclic}(1,2,3,4)\Big]\,.
\end{align}
We then impose the full dihedral crossing symmetry and solve the ansatz by matching the box and triangle cuts, which gives the unique solution,
\begin{align}
 \label{oddn4sym}
\nhodd{ 1234}=&-\frac{1}{16}\Big[\pol_{1}\cdot\ell_1 \,\varepsilon_{6}(\ell_2,\pol_{2},f_3,f_4)+\pol_{2}\cdot\ell_2 \, \varepsilon_{6}(\ell_1,\pol_{1},f_3,f_4)+(1,2|1,2,3,4)\Big]\nonumber\\
    &-\frac{1}{64}\Big[\pol_{1}\cdot\ell_1 \,\varepsilon_{6}(f_2,f_3,f_4)+\textrm{cyclic}(1,2,3,4)\Big]\nonumber\\
    &+\frac{1}{32}\Big[\Big(\varepsilon_{6}(\ell_1,\pol_{1},[f_2,f_3],f_4)+\textrm{cyclic}(1,2,3)\Big)+(1,2,3|1,2,3,4)\Big]\nonumber\\
    &-\frac{1}{16}\Big[\ls{1}\,\varepsilon_{6}(\pol_{1},\pol_{2},f_3,f_4)+\textrm{cyclic}(1,2,3,4)\Big]\,.
\end{align}
The triangle numerators are again obtained from Jacobi identities,
\begin{align}
    \nhodd{[12]34}&= \frac{1}{8} \, \varepsilon_{6}(\ell,\pol_{12},f_3,f_4)+\frac{1}{16}\Big[\varepsilon_{6}(\ell,\pol_{3},f_{12},f_4)+\varepsilon_{6}(\ell,\pol_{4},f_{12},f_3)\Big] \nonumber\\
    %&\quad-\frac{1}{32}(\ls{2}-\ls{4}-s_{12})\varepsilon_{6}(\pol_{1},\pol_{2},f_3,f_4)\,.
    &\quad-\frac{1}{16}\ell\Cdot(k_1{+}k_2) \,\varepsilon_{6}(\pol_1,\pol_2,f_3,f_4)\,,
\end{align}
and lead to vanishing two-mass bubble numerators $\nhodd{[12][34]}=0$. By carefully using the regularization~\eqref{minahaning} while applying the Jacobi identities, we get the external-bubble numerator,
\begin{align}
    \nhodd{ [[12]3]4}&=\nhodd{1234}-\nhodd{2134}-\nhodd{3124}+\nhodd{3214}=\frac{1}{32}s_{123}\,\varepsilon_{6}(\pol_{1},\pol_{2},f_3,f_4)\,,
    \label{oddbubb}
\end{align}
where the overall vanishing factor $s_{123}$ cancels the formally divergent propagator $s_{123}^{-1}$ of the external bubble and thus gives a regular integrand. We have checked that our regularized integrand satisfies all the bubble cuts, as well as the vanishing of tadpole numerators. 

As an important check to our result, we take the gauge variation $\pol_1\rightarrow k_1$ and indeed find the expected gauge anomaly in $D=6{-}2\epsilon$,
\begin{align}
     A^{\text{1-loop}}_{\text{hyp}}(1,2,3,4)\Big|_{\pol_{1}\rightarrow k_1}& =-\frac{1}{32}\varepsilon_{6}(f_2,f_3,f_4)I_4^{6-2\epsilon}\,, \label{toharm} \\ 
I_4^{6-2\epsilon}&=\int \frac{\dd^{6-2\epsilon}\ell}{i\pi^{(6-2\epsilon)/2}}\frac{(-\mu^2)
}{\ls{1}\ls{2}\ls{3}\ls{4}}=\frac{1}{3!} \,,
\nonumber
\end{align}
where $\mu^2=\ell^2{-}\ell_{(6)}^2$ is the extra-dimensional component of $\ell^2$. Because our numerators manifest crossing symmetry, we observe the same gauge anomaly
for all external legs.

\subsubsection{The five-point parity even numerator}
\label{sec:3.3}
At five points, the maximal-cut part of the parity even numerator can be directly read off from \cref{maxcuthalfs},
\begin{align}
     \nheven{12345}&=\left[ \pol_1\Cdot\ell_{1}\pol_2\Cdot\ell_{2} \pol_3\Cdot\ell_{3} \trhyp(f_4f_5)+(1,2,3|1,2,3,4,5)\right]\\
    &\quad-\left[ \pol_1\Cdot\ell_1 \pol_2\Cdot\ell_2 \trhyp (f_3f_4f_5) +(1,2|1,2,3,4,5)\right]\nonumber\\
    &\quad +\left[ \pol_1\Cdot\ell_{1}  \trhyp (f_2f_3f_4f_5)  
     +\text{cyclic}(1,2,3,4,5)\right]
     \nonumber
    \\
    &\quad -\trhyp(f_1f_2f_3f_4f_5) +\delta \nheven{12345}
    \,,
\nonumber
\end{align}
where $\delta \nheven{12345}$ gathers all contact terms with a contribution of $\ell_j^2$. Guided by the gauge-invariance argument, we consider the ansatz that features two or three linearized field strengths in a trace, 
\begin{align}
\delta \nheven{12345}=\delta N^{\tr(f^3)}_{12345}+\delta N^{\tr(f^2)}_{12345} \, .
\label{sit6.1}
\end{align} 
This resembles the organization of the contact terms in the maximally supersymmetric seven-point numerator in \cref{sit10d.1} with four or five factors of $f_i$ in $\trVS(\ldots)$. In the same way as the color-kinematics duality and gauge invariance were found to select a unique
$\tr(f^5)$ contribution \eqref{sit10d.2},
we find a unique answer for $\tr(f^3)$ contact terms in \cref{sit6.1},
\begin{align}
\delta N^{\tr(f^3)}_{12345} &= \frac{1}{24}\left[\pol_1\Cdot\pol_2(6\ls{1}-\ls{2}-\ls{5}) \trhyp(f_3f_4f_5)+\text{cyclic}(1,2,3,4,5)\right]
        \label{5pthalf}\\
    &\quad +\frac{1}{24}\left[\pol_1\Cdot\pol_3(\ls{1}+\ls{2}+2\ls{4})\trhyp(f_2f_4f_5)+\text{cyclic}(1,2,3,4,5)\right]
         \,.  \nonumber
\end{align}
The leftover terms can be organized in a 120-parameter ansatz analogous to \cref{7pttrf4},
\begin{align}
 \delta N_{12345}^{\tr(f^2)}
  =&\; (\pol_1 \Cdot \pol_2)
  \sum_{k=3}^5\sum_{\substack{i,j=1 \\ i\neq k-1}}^{5} \alpha_{i,j,k}\,(\pol_k \Cdot \ell_i)\, \ell_j^2\, \trhyp(f^2) \label{axpara} \\
    &+ (\pol_1 \Cdot \pol_3)
    \sum_{k\neq1,3}^5\sum_{\substack{i,j=1 \\ i\neq k-1}}^5 \beta_{i,j,k}\,(\pol_k \Cdot \ell_i)\, \ell_j^2\, \trhyp(f^2)  +\text{cyclic}(1,2,3,4,5)\, ,
    \notag
\end{align} 
where $\trhyp(f^2)$ in both lines refers to the field strengths that are not represented by the accompanying $(\pol_a \Cdot \pol_b)(\pol_k \Cdot \ell_i)$, and $i=k{-}1$ is again excluded by $\pol_k \Cdot \ell_{k-1}= \pol_k \Cdot \ell_{k}$. 
We then constrain the rational parameters $\alpha_{i,j,k},\beta_{i,j,k}$ in \cref{axpara} by imposing full dihedral symmetry, vanishing tadpoles, as well as gauge invariance on all the box and triangle cuts. This still leaves 13 of the $\alpha_{i,j,k},\beta_{i,j,k}$ parameters undetermined which may be viewed as the \hsym-analogue of the 30-parameter family of seven-point BCJ numerators for \msym described in \cref{sec:7pt}.

In spite of the $13$ free parameters for $\tr(f^2)$ contact terms, we proceed to computing bubble numerators using Jacobi identities under the regularization~\eqref{minahaning}. We find that even in the presence of these free parameters, all the potentially divergent propagators are canceled, and the bubble cuts are satisfied automatically. Thus they do not give further constraints, and we have checked that these free parameters indeed drop out of the gauge invariant color-ordered amplitude, already at the integrand level. 
Therefore, these $13$ parameters correspond to generalized gauge freedom.
Here we present the result with the 13 free parameters set to zero,
\begin{align}
&\delta N^{\tr(f^2)}_{12345}
= \frac{1}{48}\Big[\pol_1\Cdot\pol_2 \Big(2 (\ls{2} - 6 \ls{1}) \epsilon _5\Cdot\ell_5 \trhyp(f_3f_4)  -2 (\ls{2}-3 \ls{5}) \epsilon _5\Cdot \ell_1 \trhyp(f_3f_4) 
 \nonumber
         \\& \quad\quad
          -
 \ls{2} \epsilon _4 \Cdot \ell_5 \trhyp(f_3f_5)  
 -\ls{5} \epsilon_4\Cdot\ell_2\trhyp(f_3f_5)  -2 (6 \ls{1}-\ls{2}-\ls{5}) \epsilon _4\Cdot\ell_4 \trhyp(f_3f_5)  
 \nonumber
         \\& \quad\quad
         +2 (\ls{2}+\ls{5}) \epsilon _4\Cdot \ell_1\trhyp(f_3f_5) 
 -2 (6 \ls{1}-\ls{5}) \epsilon _3\Cdot \ell_3 \trhyp(f_4f_5) 
 \nonumber
 \\& \quad\quad
 +2 (3 \ls{2}-\ls{5})\epsilon _3\Cdot\ell_1 \trhyp(f_4f_5)  
    \Big)+\text{cyclic}(1,2,3,4,5)\Big]
\nonumber
\\ 
&\quad +\frac{1}{48}\Big[\pol_1\Cdot\pol_3 \Big(
2 (\ls{4}-\ls{1}) \epsilon _5\Cdot\ell_2 \trhyp(f_2f_4) 
-2 (\ls{1}+\ls{2}+2 \ls{3})  \epsilon _5\Cdot\ell_5 \trhyp(f_2f_4) 
 \nonumber
         \\&\quad\quad
         + (2 \ls{1}-2 \ls{4}+\ls{5})\epsilon _4\Cdot\ell_2 \trhyp(f_2f_5)  
         +(2 \ls{2}+\ls{3}-2 \ls{4})\epsilon _5\Cdot\ell_1 \trhyp(f_2f_4)  
          \nonumber
         \\&\quad\quad
         +\ls{2} \epsilon _4\Cdot\ell_5 \trhyp(f_2f_5)  -2 (\ls{2}-\ls{4})  \epsilon _4\Cdot\ell_1 \trhyp(f_2f_5) 
         +\ls{1} \epsilon _5\Cdot\ell_3 \trhyp(f_2f_4) 
       \nonumber  \\ 
         &\quad\quad -2 (\ls{1}+\ls{2}+2 \ls{5}) \epsilon _4\Cdot\ell_4 \trhyp(f_2f_5)  -\ls{3}  \epsilon _2\Cdot\ell_5 \trhyp(f_4f_5) 
          -\ls{5} \epsilon _2\Cdot\ell_3 \trhyp(f_4f_5)    \nonumber
         \\& \quad\quad
       -2 (3 \ls{1}+3 \ls{2}-2 \ls{3}-2\ls{5}) \epsilon _2\Cdot\ell_2 \trhyp(f_4f_5)  
    \Big)+\text{cyclic}(1,2,3,4,5)\Big]\,.
    \label{final5pt}
\end{align}
The full result that includes all the $13$ free parameters is given in an ancillary file of the arXiv submission.

\subsubsection{UV divergences of half maximally SYM}
As a crosscheck for our results, we first show that the integrands reproduce the correct UV divergences of \hsym in $d=4$ and $d=6$.
Since our $m$-gon numerators are polynomials of degree
$\ell^{m-2}$, the UV divergence in $d=4$ only receives contributions from bubbles. By a naive dimensional analysis based on our bubble numerators and the measure $\dd^D \ell$, we expect it to be absorbed by the field renormalization of $\Tr(F^2)$. The UV divergence in $d=6$ receives contributions also from triangles. However, it is bound to vanish since the only counterterm compatible with dimensional analysis is $\Tr(F^3)$, which cannot be supersymmetrized. In the $n$-gon basis numerator, the terms that can source a bubble or triangle topology are proportional to $\ell^{n-2}$ and $\ell^{n-3}$, which only involve traces of length two and three, respectively. For these terms, $\trhyp$ and $\trvec$ only differ by a factor of $-2$ according to \cref{4ptexh,eq:4pvec}. We have verified for $n\leq 5$ that our integrands indeed give the expected UV behavior~\cite{Elvang:2011fx}, 
\begin{align}\label{eq:UV8symh}
    & \epsilon A^{\text{1-loop}}_{\text{\hsym}}(1,2,\ldots,n)\,\Big|^{\text{UV}}_{D=4-2\epsilon}=-\frac{n{-}2}{2}(n_{\text{hyp}}{-}2n_{\text{vec}})A^{\text{tree}}_{\text{YM}}(1,2,\ldots,n)\,,\nonumber\\
    & \epsilon A^{\text{1-loop}}_{\text{\hsym}}(1,2,\ldots,n)\,\Big|^{\text{UV}}_{D=6-2\epsilon}=0\,,
\end{align}
where $n_{\text{hyp}}$ and $n_{\text{vec}}$ are the numbers of hyper and vector multiplets in the loop.
We note that the parity odd contribution drops out from the UV divergences in $d=4,6$ due to the absence of counterterms as we have explicitly verified at $n=3,4$ using \cref{master2pt3pt} and \cref{oddbubb}. Thus we do not need to specify the chirality in \cref{eq:UV8symh}. 
%in this equation.

\subsection{Half-maximal supergravities and UV divergences}
\label{sec:uvhalf}

We can double copy BCJ numerators of \hsym to construct loop integrands for half-maximal supergravities (with 16 supercharges). By choosing different gauge-theory copies, we obtain supergravities with different particle contents in the loop~\cite{Johansson:2017bfl}:
\begin{align}\label{eq:hmdc}
    & \text{hyp}_{(1,0)}\otimes\text{hyp}_{(1,0)}=\text{tensor}_{(2,0)}\,, & & \text{vec}_{(1,0)}\otimes\text{vec}_{(1,0)}=\text{graviton}_{(2,0)}+\text{tensor}_{(2,0)}\,,\nonumber\\
    & \text{hyp}_{(1,0)}\otimes\overline{\text{hyp}}_{(0,1)}=\text{vec}_{(1,1)}\,, & & \text{vec}_{(1,0)}\otimes\text{vec}'_{(0,1)}=\text{graviton}_{(1,1)}\,.
\end{align}
The two inequivalent graviton multiplets $\text{graviton}_{(2,0)}$ and $\text{graviton}_{(1,1)}$ in the right column are chiral and non-chiral, respectively, and both columns feature a
self-dual tensor multiplet $\text{tensor}_{(2,0)}$. The two hypermultiplets in the double copy $\text{hyp}_{(1,0)}\otimes\text{hyp}_{(1,0)}$ both transform in a non-complex representation of the gauge group in order to avoid overcounting degrees of freedom~\cite{Chiodaroli:2015wal}.

The double copy between a half-maximal vector and adjoint hypermultiplet results in a gravitino multiplet. For adjoint fermions, supersymmetry is compatible with the color-kinematics duality only if they reside in a vector multiplet, and consequently the gravitino from double copy must be in a graviton multiplet~\cite{Chiodaroli:2013upa}. In other words, local supersymmetry is incompatible with stand-alone massless gravitinos, unless made massive through some symmetry breaking. Here, we merely view them as formal objects that interpolate the difference in particle contents between maximal and half-maximal supergravities.

We start with the double copy of two hypermultiplets both with the same chirality, $\text{hyp}\otimes\text{hyp}$, and the opposite chirality, $\text{hyp}\otimes\overline{\text{hyp}}$,
\begin{align}
    \mathfrak{N}^{1/2\text{-IIA}} &= (\nheven{} + \nhodd{})(\ntheven{} - \nthodd{})\,, \notag\\
    \mathfrak{N}^{1/2\text{-IIB}} &= (\nheven{} + \nhodd{})(\ntheven{} + \nthodd{})\,. \label{eq:IIBdch}
\end{align}
We label the resulting theories as $1/2\text{-IIA}$ and $1/2\text{-IIB}$ instead of their particle contents as in \cref{eq:hmdc} since the integrands can be obtained as the low-energy limit of the type IIA and IIB string amplitudes compactified on a K3 manifold~\cite{Gregori:1997hi, Bianchi:2006nf, Bianchi:2015vsa, Berg:2016fui, Berg:2016wux}. The gravity integrand after double copy still has graphs with bubbles on external legs, which contains divergent propagators. Similar to \hsym, the divergent propagators will cancel under the regularization~\eqref{minahaning} among the graphs that share the same divergence. For example, at four points, the three dangling trees $[[12]3]$, $[1[23]]$ and $[[31]2]$ will conspire to cancel the divergent propagator $s_{123}^{-1}$. The fact that gravity integrands are free of such divergences is already a highly nontrivial check to our results. We have also verified down to bubbles that the cuts of the supergravity integrand obtained through double copy of the five-point numerators in  \cref{sec:3.3} are gauge invariant and independent of the free parameters, %in the five-point numerators, 
see \cref{sec:grcut} for further details.

The double copy of \hsym leads to a maximal power of $\ell^{2m-4}$ in $m$-gon numerators, which contributes to the UV divergence in $d=4$ through the coefficient of $\ell^{\mu_1}\ldots\ell^{\mu_{2m-4}}$.
Even though the double copies \cref{eq:IIBdch} yield supergravity numerators for any combination of external gravitons, $B$-fields and dilatons, we shall focus on the $n$-graviton components in this section.
When calculating the UV divergence for pure gravitons, the parity odd part of $\mathfrak{N}^{1/2\text{-IIA/IIB}}$ drops out, so that we can effectively use
\begin{align}
    \mathfrak{N}_{n\text{ gravitons}}^{1/2\text{-IIA}} &=  |\nheven{}|^2 - |\nhodd{}|^2\,,\nonumber\\
    \mathfrak{N}_{n\text{ gravitons}}^{1/2\text{-IIB}} &=  |\nheven{}|^2 + |\nhodd{}|^2\,. \label{eq:IIBdch2}
\end{align}
Up to four points, the contributions from $|\nheven{}|^2$ and $|\nhodd{}|^2$ to the four-dimensional UV divergence for external gravitons differ only by a sign. Both of them are proportional to matrix elements of the supersymmetrizable $R^2$ counterterm, which can be realized as a double copy between (super-)Yang-Mills and a single insertion of the $\Tr(F^3)$ operator~\cite{Broedel:2012rc},\footnote{We normalize the operator such that $A^{\text{tree}}_{F^3}(1,2,3)=-2(\pol_{1}\Cdot k_2)(\pol_{2}\Cdot k_3)(\pol_{3}\Cdot k_1)$.} 
\begin{align}
M_{R^2}^{\text{tree}}(n) =
\big(\widetilde{\pmb{A}}{}_{\text{YM}}^{\text{tree}}(n)\big)^{T}\cdot\pmb{S}_0(n)\cdot\pmb{A}_{F^3}^{\text{tree}}(n) \, .
\label{r2mel}
\end{align}
The $(n{-}3)!$-component vector $\pmb{A}^{\text{tree}}_{F^3}(n) = A^{\textrm{tree}}_{F^3}(1,\alpha,n{-}1,n)$ 
with $\alpha \in S_{n-3}$ refers to a BCJ basis of color-ordered matrix elements of $\Tr(F^3)$, and the objects $\widetilde{\pmb{A}}{}_{\text{YM}}^{\text{tree}}$$(n)$
and $\pmb{S}_0(n)$ are specified below \cref{eq:UV2b}.

Similar to the gauge-theory case, only length-two and length-three traces over linearized field strengths contribute to the gravity UV divergences in $d=4$ and $6$, respectively. After including the vector-multiplet loop using \cref{4ptexh,eq:4pvec,eq:decom}, we have verified up to four points that
\begin{subequations}
\label{eq:5UV}
\begin{align}\label{eq:UVhsg}
    \epsilon M_{n\text{ gravitons}}^{1/2\text{-IIA}}\,\Big|^{\text{UV}}_{D=4-2\epsilon}&=(n_{\overline{\text{hyp}}}-2n_{\text{vec}})^2 M_{R^2}^{\text{tree}} \, \big|_{n\text{ gravitons}} \,,\\
    \epsilon M_{n\text{ gravitons}}^{1/2\text{-IIB}}\,\Big|^{\text{UV}}_{D=4-2\epsilon}&=0\,,\label{eq:UVhsg2}
\end{align}
\end{subequations}
where the hyper and vector multiplets are of opposite chiralities. 
In addition, we have checked that the parity even contribution $|\nheven{}|^2$ for five external gravitons yields the $R^2$ matrix element \eqref{r2mel} with coefficient
$\frac{1}{2}(n_{\overline{\text{hyp}}}-2n_{\text{vec}})^2$, consistent with the average of both lines in \cref{eq:5UV}. 

On the other hand, the UV divergence in $d=6$ from both $|\nheven{}|^2$ and $|\nhodd{}|^2$ is expected to vanish due to the absence of supersymmetrizable $R^3$ counterterms:
\begin{align}
\epsilon M_{n\text{ gravitons}}^{1/2\text{-IIA}}\,\Big|^{\rm UV}_{D=6-2\epsilon}=  \epsilon M_{n\text{ gravitons}}^{1/2\text{-IIB}}\,\Big|^{\rm UV}_{D=6-2\epsilon} &= 0 \,,   
\end{align}
which we have checked up to four points for $|\nhodd{}|^2$ and five points for $|\nheven{}|^2$.

\section{Conclusion and outlook}
\label{sec:6}
In this paper, we have fulfilled a long-standing quest in constructing one-loop higher-multiplicity BCJ numerators with desirable properties in various SYM and supergravity theories.
We present the first BCJ numerators for six- and seven-point SYM amplitudes with maximal supersymmetry, where the supergravity double copy can be correctly performed using conventional Feynman propagators quadratic in loop momenta.
The key to achieve this was to identify simple $D$-dimensional kinematic building blocks for generic gauge theories that manifest many of the properties that we require. Our main results can be summarized as follows:
\begin{itemize}
\item Using basic building blocks, we provided compact all-multiplicity formulae for the maximal-cut contribution to one-loop $n$-gon numerators by sewing three-point amplitudes; see \cref{eq:ngon-mc} for the expression in generic (S)YM theories 
as well as \cref{nptmaxcut,maxcuthalfs} for the maximally and half-maximally supersymmetric cases.

\item We constructed for the first time one-loop BCJ numerators at multiplicites $n=6,7$, for \msym valid in $D\leq 10$ dimensions, which have standard (quadratic) Feynman propagators,  respect all graph symmetries and double copy to consistent supergravity numerators; via kinematic Jacobi identities they give numerators for other cubic graphs (see {\it e.g.} \cref{originhex,eq:N6ptoddsym} for the $n=6$ case).
\item Similarly, we constructed one-loop BCJ numerators for \hsym in $D\leq 6$ dimensions at multiplicities $n=4,5$ (see {\it e.g.}\ \cref{eq:hmbox,oddn4sym}), and we discussed double copies to several supergravity
theories. 
\item As a check of the supergravity integrands obtained from the double copy of our SYM numerators, we explicitly computed their UV divergences: eight- and ten-dimensional ones up to $n=6$ in the \msym case (see for instance \cref{eq:UV2b}) as well as four- and six-dimensional ones up to $n=5$ in the \hsym case (see {\it e.g.}\ \cref{eq:5UV}).
\end{itemize}

Our results open up various new directions for further investigations. With compact one-loop integrands available, an important next step would be to analytically integrate them, beyond the UV divergences computed here. This would provide valuable new data for the analytic structure of one-loop (supersymmetric) gauge-theory and gravity amplitudes in general dimensions. 

It would also be very interesting to apply our method to higher-point numerators, {\it e.g.}\ $n\geq 8$ in \msym and $n\geq 6$ in \hsym. Starting from the all-$n$ formulae for the maximal-cut contributions, we expect that the remaining terms can found such that the numerators become compatible with kinematic Jacobi identities, linearized gauge invariance and all non-maximal cuts. However, the increased computational complexity in fixing higher-multiplicity numerators from an ansatz for the contact terms makes it desirable to refine the approach: can one optimally compose the information from unitarity, color-kinematics duality and gauge invariance?

On general grounds, we expect that explicit expressions for non-maximal cuts are not needed in the construction, instead imposing \emph{gauge invariance} for these cuts appears to be sufficient for fixing the amplitude. Indeed, we explicitly confirmed this property for the parity even contributions of the amplitudes that we computed for \msym and \hsym. Only imposing gauge invariance on all non-maximal cuts down to boxes and bubbles, respectively,  guarantees that the integrands produce correct cuts. It would be desirable to study how one could even further boost the power of gauge invariance and color-kinematics duality for constructing one-loop integrands in various (S)YM theories.

Another open question is to connect our results to other one-loop numerators such as those obtained from pure-spinor methods~\cite{Mafra:2014gja, Bridges:2021ebs} or ambitwistor strings/forward-limits of CHY formulae~\cite{Geyer:2015bja, He:2015yua, Geyer:2015jch, He:2017spx, Geyer:2017ela, Edison:2020uzf}:
\begin{itemize}
\item It would be interesting to embed the maximal-cut part of our $n$-gon numerators for $D=10$ SYM into pure-spinor superspace and to connect with the BRST-covariant building blocks of \cite{Mafra:2014gsa, Mafra:2018nla}. In this way, one could infer valuable extra information on the contributions beyond the maximal cut from spacetime supersymmetry and generalize the field-strength traces in our numerators to external fermions. 
\item Traditional ambitwistor-string formulae for one-loop integrands involve propagators linear in the loop momentum $\ell$, and their conversion to quadratic propagators has been discussed from a variety of perspectives \cite{Gomez:2016cqb, Gomez:2017lhy, Gomez:2017cpe, Ahmadiniaz:2018nvr, Agerskov:2019ryp, Farrow:2020voh, Edison:2021ebi, Feng:2022wee}. We hope that our numerators may help to clarify how the recombination of linearized propagators to quadratic ones can be made to preserve both the color-kinematics duality and locality.
\end{itemize}
Last but not least, the methods of this work call for applications to higher-point integrands at two loops in general spacetime dimensions. Two-loop five-point BCJ numerators are available in terms of four-dimensional spinor-helicity variables \cite{Carrasco:2011mn} and pure-spinor superspace expressions \cite{Mafra:2015mja}. The latter can be converted to dimension-agnostic gluon-polarizations that share the characteristic $t_8$-tensors of our one-loop numerators through the ``effective components'' of ref.\ \cite{DHoker:2020tcq}. At six points, however, the planar two-loop integrands in the literature are limited to four-dimensional expressions without known BCJ representations \cite{Bourjaily:2019iqr, Bourjaily:2019gqu, Bourjaily:2021iyq}.

As the first step towards constructing dimension-agnostic two-loop six-point integrands with manifest color-kinematics duality, we have computed double-pentagon and hexa-box numerators quadratic in loop momenta that obey maximal cuts by uplifting our sewing procedure for three-point amplitudes. The real challenge is then to go beyond these maximal cuts via ans\"{a}tze based on $t_8$-tensors, color-kinematics duality and linearized gauge invariance. The two-loop showcases of terms beyond the maximal cuts will be much richer than the one-loop examples in this work, so we expect the ongoing studies at two loops to provide crucial guidance for the general structure of multiloop BCJ numerators. Based on the reasoning of ref.~\cite{Geyer:2021oox}, it is even conceivable that the aspired two-loop numerators could fuel string-amplitude computations, going beyond the recent five-point genus-two results in \cite{DHoker:2020prr, DHoker:2020tcq, DHoker:2021kks}.

\acknowledgments

We are grateful to Filippo Balli, Marcus Berg, Freddy Cachazo, Eric D'Hoker, Jin Dong, Michael Haack, Aidan Herderschee, Carlos Mafra, Radu Roiban, Tianheng Wang and Yaoqi Zhang for inspiring discussions and collaboration on related topics. Moreover, we thank Marcus Berg, Zvi Bern, Michael Haack, Carlos Mafra and Radu Roiban for helpful comments on the draft.
This research was supported in part by the National Science Foundation under Grant No.\ NSF PHY-1748958, and we are grateful to the KITP Santa Barbara as well as the organizers of the workshop ``High-precision gravitational waves'' for providing stimulating atmosphere, support and hospitality.
Similarly, this work was supported in part by the Munich Institute for Astro-, Particle and BioPhysics (MIAPbP), which is funded by the Deutsche Forschungsgemeinschaft (DFG, German Research Foundation) under Germany's Excellence Strategy --EXC-2094--390783311.

AE is supported in part by the Knut and Alice
Wallenberg Foundation under KAW 2018.0116, by Northwestern University via the Amplitudes and Insight Group, Department of Physics and Astronomy, and Weinberg College of Arts and Sciences, and by the US Department of Energy under contract DE-SC0015910.
The research of SH is supported in
part by Key Research
Program of CAS, Grant No.\ XDPB15 and National Natural Science Foundation of China under Grant No.\ 11935013, 11947301, 12047502, 12047503 and 12225510. 
The research of HJ is supported by the Knut and
Alice Wallenberg Foundation under grants KAW 2018.0116 ({From Scattering Amplitudes to Gravitational Waves}) and KAW 2018.0162, and the Ragnar S\"{o}derberg Foundation (Swedish Foundations’ Starting Grant).  
The research of OS is supported by the European Research Council under ERC-STG-804286 UNISCAMP. 
The research of FT is supported by the US Department of Energy under Grant No.\ DE-SC00019066. 
The research of YZ was supported in part by a grant from the Gluskin Sheff/Onex Freeman Dyson Chair in Theoretical Physics and by Perimeter Institute. Research at Perimeter Institute is supported in part by the Government of Canada through the Department of Innovation, Science and Economic Development Canada and by the Province of Ontario through the Ministry of Colleges and Universities.

\appendix

\section{Constructing gravitational cuts}\label{sec:grcut}

In this appendix, we provide a more detailed description of the unitarity checks of the supergravity integrands in \cref{sec:4.4,sec:uvhalf}. Verifications of loop integrands via unitarity cuts follow well-established techniques, see for instance \cite{Bern:2010ue,Bern:2017yxu,Bern:2018jmv} or section 3 of the lecture notes \cite{Carrasco:2015iwa}, and we review the key ideas relevant to our one-loop examples for the sake of a self-contained presentation.

Given a set of gravity numerators through double copy, one efficient way to check their validity is to verify if all the cuts of the gravity amplitudes satisfy the expected physical properties. In fact, we only need to consider the spanning set of cuts, which contains all the information of the other cuts. Since our numerators have manifest power counting and no-triangle (no-tadpole) properties, the quadruple cuts or box cuts (two-particle cuts or bubble cuts) are the spanning sets for \msym (\hsym). Similar to the color-ordered cuts in gauge theories discussed in \cref{sec:cuts}, we can construct cuts by assembling all the numerators together with the uncut propagators of the graphs that contribute to the cut. The only difference in gravity is that the cut is now unordered, such that we need to sum over permutations of external legs with loop momentum fixed by the cut condition.

To illustrate the idea, we consider the following one-mass box cut at six points as an example. This is one of the three permutation-inequivalent topologies of spanning cuts for maximal supergravities at six points,
\begin{align}\label{eq:1mboxcut}
    \text{Cut}\!\left[\!\!\!\vcenter{\hbox{
    \begin{tikzpicture}[scale=0.7,every node/.style={font=\scriptsize}]
    \pgfmathsetmacro{\r}{1.1}
        \pgfmathsetmacro{\l}{0.6}
        \pgfmathsetmacro{\y}{-2.2}
        \coordinate (1) at (135:\r);
        \draw[cuts] (90:{\r-0.6}) -- (90:{\r});
        \node at ( 180: {\r + 0.35} ) {$\ell$};
        \draw[thick] (1) -- ++ (-\l,0) node[label={[label distance=-6pt]180:1}] {};
        \draw[thick] (1) -- ++ (0,\l) node[label={[label distance=-6pt]90:3}] {};
        \draw[thick] (1) -- ++ (135:\l) node[label={[label distance=-8pt]135:2}] {};
        \fill (1) circle (1pt);
        \foreach \x in {4,5,6} {
            \pgfmathsetmacro{\a}{135-90*(\x-3)}
            \pgfmathsetmacro{\b}{90-90*(\x-3)}
            \coordinate (\x) at ( \a : \r);
            \fill (\x) circle (1pt);
            \draw[thick] (\x) -- ++ ( \a : \l) node[label={[label distance=-8pt]\a:\x}] {};
            \draw[cuts] (\b:{\r-0.6}) -- (\b:{\r});
        }
        \draw[thick] (4) -- (1);
        \draw[thick] (5) -- (4);
        \draw[thick] (6) -- (5);
        \draw[thick] (1) -- (6);
        \node at ( 90: {\r + 0.35} ) {$\ell_3$};
        \node at ( 0: {\r + 0.35} ) {$\ell_{4}$};
        \node at ( -90: {\r + 0.35} ) {$\ell_{5}$};
        \node at (0,0) {\scalebox{1.75}{$\circlearrowleft$}};
    \end{tikzpicture}
 }}\!\!\!\right]&=\left[\frac{ | \nmax{123456} |^2 }{(\ell{+}k_1)^2(\ell{+}k_{12})^2}+\text{perm}(1,2,3)\right]_{\ell^2=\ell_3^2=\ell_4^2=\ell_5^2=0}\nonumber\\[-0.7cm]
    &\quad +\left[\frac{ | \nmax{[12]3456} |^2 }{s_{12}(\ell{+}k_{12})^2}+\frac{ | \nmax{1[23]456} |^2}{s_{23}(\ell{+}k_{1})^2}+\text{cyclic}(1,2,3)\right]_{\ell^2=\ell_3^2=\ell_4^2=\ell_5^2=0}\nonumber\\
    &\quad + \left[\frac{ | \nmax{[[12]3]456} |^2 }{s_{12}s_{123}}+\text{cyclic}(1,2,3)\right]_{\ell^2=\ell_3^2=\ell_4^2=\ell_5^2=0}\,,
\end{align}
where $\ell_i=\ell{+}k_{12\ldots i}$. There is a total of 15 graphs in the sum, corresponding to the cubic-diagram expansion of the five-point tree-level amplitude $M(\ell,k_1,k_2,k_3,-\ell_3)$ in the upper-left corner of \cref{eq:1mboxcut}. We follow the notation of \cref{UV.09} for the gravity numerators $| \nmax{\ldots} |^2$, and the gauge-theory numerators are given in \cref{sec:3.1,sec:3.2}. Note that the parity odd contributions from the $\tilde \pol_i$ enter with an additional minus sign for type IIA supergravity.

In the 15 graphs of \cref{eq:1mboxcut}, we have lined up the labelling of loop momenta such that the edge between legs $1$ and $6$ carries momentum $\ell$. All the other numerators are then obtained by permuting external legs but with the position of $\ell$ fixed. For instance, $\nmax{1[23]456}$ is obtained from $\nmax{[12]3456}$ via cyclic permutation $1\rightarrow 2,\ 2\rightarrow 3, \ldots,\  6\rightarrow 1$ with $\ell\rightarrow\ell{+}k_1$, following \cref{crossing_symmetry}.

One can similarly construct the other two topologies of six-point box cuts,
\begin{align}
\label{other2cuts}
    \text{Cut}\!\left[\!\!\!\vcenter{\hbox{
    \begin{tikzpicture}[scale=0.7,every node/.style={font=\scriptsize}]
    \pgfmathsetmacro{\r}{1.1}
        \pgfmathsetmacro{\l}{0.6}
        \pgfmathsetmacro{\y}{-2.2}
        \coordinate (1) at (135:\r);
        \coordinate (2) at (45:\r);
        \coordinate (3) at (-45:\r);
        \coordinate (4) at (-135:\r);
        \draw[thick] (4) -- (1) -- (2) -- (3) -- cycle;
        \node at ( 180: {\r + 0.35} ) {$\ell$};
        \draw[thick] (1) -- ++ (-\l,0) node[label={[label distance=-6pt]180:1}] {};
        \draw[thick] (1) -- ++ (0,\l) node[label={[label distance=-6pt]90:2}] {};
        \draw[thick] (2) -- ++ (0,\l) node[label={[label distance=-6pt]90:3}] {};
        \draw[thick] (2) -- ++ (\l,0) node[label={[label distance=-6pt]0:4}] {};
        \draw[thick] (3) -- ++ ( -45 : \l ) node[label={[label distance=-6pt]-45:5}] {};
        \draw[thick] (6) -- ++ ( -135 : \l ) node[label={[label distance=-6pt]-135:6}] {};
        \foreach \x in {1,2,3,4} {
            \pgfmathsetmacro{\a}{135-90*(\x-1)}
            \pgfmathsetmacro{\b}{90-90*(\x-1)}
            \fill (\x) circle (1pt);
            \draw[cuts] (\b:{\r-0.6}) -- (\b:{\r});
        }
        \node at ( 90: {\r + 0.35} ) {$\ell_2$};
        \node at ( 0: {\r + 0.35} ) {$\ell_{4}$};
        \node at ( -90: {\r + 0.35} ) {$\ell_{5}$};
        \node at (0,0) {\scalebox{1.75}{$\circlearrowleft$}};
    \end{tikzpicture}
    }}\!\!\!\right]\,,\qquad \text{Cut}\!\left[\!\!\!\vcenter{\hbox{
    \begin{tikzpicture}[scale=0.7,every node/.style={font=\scriptsize}]
    \pgfmathsetmacro{\r}{1.1}
        \pgfmathsetmacro{\l}{0.6}
        \pgfmathsetmacro{\y}{-2.2}
        \coordinate (1) at (135:\r);
        \coordinate (2) at (45:\r);
        \coordinate (3) at (-45:\r);
        \coordinate (4) at (-135:\r);
        \draw[thick] (4) -- (1) -- (2) -- (3) -- cycle;
        \node at ( 180: {\r + 0.35} ) {$\ell$};
        \draw[thick] (1) -- ++ (-\l,0) node[label={[label distance=-6pt]180:1}] {};
        \draw[thick] (1) -- ++ (0,\l) node[label={[label distance=-6pt]90:2}] {};
        \draw[thick] (2) -- ++ (45:\l) node[label={[label distance=-6pt]45:3}] {};
        \draw[thick] (3) -- ++ (0:\l) node[label={[label distance=-6pt]0:4}] {};
        \draw[thick] (3) -- ++ ( -90 : \l ) node[label={[label distance=-6pt]-90:5}] {};
        \draw[thick] (6) -- ++ ( -135 : \l ) node[label={[label distance=-6pt]-135:6}] {};
        \foreach \x in {1,2,3,4} {
            \pgfmathsetmacro{\a}{135-90*(\x-1)}
            \pgfmathsetmacro{\b}{90-90*(\x-1)}
            \fill (\x) circle (1pt);
            \draw[cuts] (\b:{\r-0.6}) -- (\b:{\r});
        }
        \node at ( 90: {\r + 0.35} ) {$\ell_2$};
        \node at ( 0: {\r + 0.35} ) {$\ell_{3}$};
        \node at ( -90: {\r + 0.35} ) {$\ell_{5}$};
        \node at (0,0) {\scalebox{1.75}{$\circlearrowleft$}};
    \end{tikzpicture}
    }}\!\!\!\right]\,,
\end{align}
both of which are contributed by 9 graphs, where the numerators are evaluated under the cut conditions $\ell^2=\ell_2^2=\ell_4^2 = \ell_5^2=0$, respectively $\ell^2=\ell_2^2=\ell_3^2 = \ell_5^2=0$ in the routing of loop momenta in \cref{other2cuts}. Since our numerators are crossing symmetric, the spanning set of cuts can be formed from relabelings of the above three box cuts. 

For half-maximal supergravities, the spanning cuts consist of bubbles. For example, at four points the following two topologies of bubble cuts form the spanning set,
\begin{align}
    \text{Cut}\!\left[\!\!\!\vcenter{\hbox{
    \begin{tikzpicture}[scale=0.7,every node/.style={font=\scriptsize}]
    \pgfmathsetmacro{\r}{1}
        \pgfmathsetmacro{\l}{0.7}
        \pgfmathsetmacro{\y}{-2.2}
        \coordinate (a) at ( 180: \r );
        \coordinate (b) at ( 0: \r );
        \draw[thick] (a) to[bend left=70] (b);
        \draw[thick] (a) to[bend right=70] (b);
        \draw[thick] (a) -- ++(-135: \l) node[label={[label distance=-10pt]-135:1}] {};
        \draw[thick] (a) -- ++(135: \l) node[label={[label distance=-10pt]135:2}] {};
        \draw[thick] (b) -- ++(-45: \l) node[label={[label distance=-10pt]-45:4}] {};
        \draw[thick] (b) -- ++(45: \l) node[label={[label distance=-10pt]45:3}] {};
        \filldraw (a) circle (1pt);
        \filldraw (b) circle (1pt);
        \draw[cuts] (0,-0.8) -- (0,-0.4) (0,0.4) -- (0,0.8);
        \node at ( -90: {\r + 0.1} ) {$\ell$};
        \node at ( 90: {\r + 0.1} ) {$\ell_2$};
        \node at (0,0) {\scalebox{1.75}{$\circlearrowleft$}};
    \end{tikzpicture}
    }}\!\!\!\right]\,,\qquad\text{Cut}\!\left[\!\!\!\vcenter{\hbox{
    \begin{tikzpicture}[scale=0.7,every node/.style={font=\scriptsize}]
    \pgfmathsetmacro{\r}{1}
        \pgfmathsetmacro{\l}{0.7}
        \pgfmathsetmacro{\y}{-2.2}
        \coordinate (a) at ( 180: \r );
        \coordinate (b) at ( 0: \r );
        \draw[thick] (a) to[bend left=70] (b);
        \draw[thick] (a) to[bend right=70] (b);
        \draw[thick] (a) -- ++(-135: \l) node[label={[label distance=-10pt]-135:1}] {};
        \draw[thick] (a) -- ++(180: \l) node[label={[label distance=-6pt]180:2}] {};
        \draw[thick] (a) -- ++(135: \l) node[label={[label distance=-10pt]135:3}] {};
        \draw[thick] (b) -- ++(0: \l) node[label={[label distance=-6pt]0:4}] {};
        \filldraw (a) circle (1pt);
        \filldraw (b) circle (1pt);
        \draw[cuts] (0,-0.8) -- (0,-0.4) (0,0.4) -- (0,0.8);
        \node at ( -90: {\r + 0.1} ) {$\ell$};
        \node at ( 90: {\r + 0.1} ) {$\ell_3$};
        \node at (0,0) {\scalebox{1.75}{$\circlearrowleft$}};
    \end{tikzpicture}
    }}\!\!\!\right]\,,
    \label{bubcuts}
\end{align}
under the requirement of manifest crossing symmetry. We note that the regularization in \cref{minahaning} is needed when computing the external-bubble cut to reach a finite result.

At tree level, double copy leads to a theory that is separately invariant under the gauge transformations $\pol_{\mu}\rightarrow \pol_{\mu}+k_{\mu}$ (at fixed $\tilde \pol_{\nu}$) and $\tilde \pol_{\nu}\rightarrow \tilde \pol_{\nu }+k_{\nu}$ (at fixed $\pol_{\mu}$). The symmetric (antisymmetric) combination of these transformations corresponds to linearized diffeomorphisms of the graviton polarization (linearized gauge transformations of the B-field polarization) within the tensor product $\pol_{\mu} \tilde \pol_{\nu}$.
At loop level, these invariances hold explicitly on cuts~\cite{Chiodaroli:2017ngp}.

We have checked that the double copy of the six- and seven-point numerators of \msym in \cref{sec:maxSYM} and the four- and five-point numerators of \hsym in \cref{sec:half-max} gives rise to gauge invariant gravity cuts. More specifically, this has been verified for every topology of quadruple cuts in the maximally supersymmetric case including \cref{eq:1mboxcut,other2cuts} and every topology of bubble cuts in case of half-maximal supersymmetry including \cref{bubcuts}. Given the non-trivial gauge variation of the contact terms proportional to $\pol_a \Cdot \pol_b \ell_j^2$ in our basis numerators, gauge invariance of a spanning set of cuts strongly validates our results.

\section{Fermion \texorpdfstring{$n$}{n}-gon cut}
\label{sec:fcut}

We consider the same $n$-gon cut as in \cref{eq:ngon-mc}, but now with a chiral fermion running in the loop. Using the three-point tree-level amplitude given in \cref{eq:a3f}, where both fermions are represented by the $\chi$ spinor, we get
\begin{align}
    \text{Cut}_{\text{$n$-gon}}^{\text{fermion}}&=-\sum_{\text{states}}\prod_{i=1}^{n}A(\ell_{i-1}^{\psi},i^{g},-\ell_i^{\psi})\nonumber\\
    &=-\left(-\frac{1}{2}\right)^n \sum_{\text{states}} \prod_{i=1}^{n}\bar{\chi}_{\ell_{i-1}}\slashed{\pol}_i\chi_{-\ell_i}=-\frac{1}{2^n}\tr \Big(P_{\pm}\prod_{i=1}^{n}\slashed{\pol}{}_i\slashed{\ell}{}_i\Big)\,,
\end{align}
where $\bar{\chi}_{\ell_{0}}=\bar{\chi}_{\ell_{n}}$, and the chiral projectors $P_{\pm}$ are defined in the discussion around \cref{eq:g5}.
Choosing two consecutive terms in this chain of gamma matrices, we can show that
\begin{align}
    \slashed{\pol}{}_{i-1}\slashed{\ell}{}_{i-1}\slashed{\pol}{}_i\slashed{\ell}{}_i&=\slashed{\pol}{}_{i-1}\slashed{\ell}{}_{i-1}\Big[(2\pol_i\Cdot\ell_i)\ids-\slashed{k}{}_i\slashed{\pol}{}_i\Big]-\ell_{i-1}^2\slashed{\pol}_{i-1}\slashed{\pol}_i\nonumber\\
    &=2\,\slashed{\pol}{}_{i-1}\slashed{\ell}{}_{i-1}\Big[(\pol_i\Cdot\ell_i)\ids-\fsl_i\Big]-\ell_{i-1}^2\slashed{\pol}_{i-1}\slashed{\pol}_i\,.
\end{align}
Since $\ell_{i-1}^2=0$ on the maximal cut, we effectively have the prescription to replace all but one $\slashed{\pol}_i\slashed{\ell}_i$ by $2\big[(\pol_i\Cdot\ell_i)\ids-\fsl_i\big]$. This leads to 
\begin{align}
    \text{Cut}_{\text{$n$-gon}}^{\text{fermion}}&=-\frac{1}{2}\tr\Big[P_{\pm}\slashed{\pol}{}_1\slashed{\ell}{}_1\prod_{i=2}^{n}\Big((\pol_i\Cdot\ell_i)\ids - \fsl_i\Big)\Big] \label{altderiv}\\
    &=-\frac{1}{4}\tr\Big[\slashed{\pol}{}_1\slashed{\ell}{}_1\prod_{i=2}^{n}\Big((\pol_i\Cdot\ell_i)\ids - \fsl_i\Big)\Big] \mp \frac{1}{4}\tr\Big[\Gamma\slashed{\pol}{}_1\slashed{\ell}{}_1\prod_{i=2}^{n}\Big((\pol_i\Cdot\ell_i)\ids - \fsl_i\Big)\Big]\,,\nonumber
\end{align}
which gives the parity odd contribution as shown in \cref{eq:ngon-mc}. There are no further obvious simplifications that can be used. The parity even contribution, however, can be written in a more symmetric form.  We need to use the following identity that holds on the maximal cut,
\begin{align}
\tr\Big[\slashed{\pol}{}_1\slashed{\ell}{}_1\prod_{i=2}^{n}\Big((\pol_i\Cdot\ell_i)\ids - \fsl_i\Big)\Big]=\tr\Big[\prod_{i=1}^{n}\Big((\pol_i\Cdot\ell_i)\ids - \fsl_i\Big)\Big] + \mathcal{O}(\ell_i^2)\,.
\end{align}

In fact, we can obtain the parity even contribution in a more direct manner by using both types of spinors $\chi$ and $\xi$ in the three-point amplitudes in \cref{eq:a3f}. For this choice, the state sum is given by
\begin{align}
    \sum_{\text{states}}\xi{}_{-\ell}\bar{\chi}{}_{\ell}=-\frac{1}{2 q\Cdot\ell}\slashed{q}\sum_{\text{states}}\chi{}_{-\ell}\bar{\chi}_{\ell}=\frac{\slashed{q}\slashed{\ell}}{2 q\Cdot\ell}
\end{align}
with lightlike reference momentum $q$. The $n$-gon cut is given by
\begin{align}
    \text{Cut}_{\text{$n$-gon}}^{\text{fermion}}&=-\sum_{\text{states}}\prod_{i=1}^{n}\Big[(\pol_i\Cdot\ell_i)\bar{\chi}_{\ell_{i-1}}\xi_{-\ell_i}-\bar{\chi}_{\ell_{i-1}}\fsl_i\xi_{-\ell_i}\Big]\nonumber\\
    &=-\tr\bigg[ P_{\pm} \prod_{i=1}^{n}\Big((\pol_i\Cdot\ell_i)\ids-\fsl_i\Big)\frac{\slashed{q}\slashed{\ell}_i}{2q\Cdot\ell_i}\bigg]\,.
\end{align}
We now pick two consecutive terms from the product in the above equation and commute the first $\slashed{q}$ to become adjacent to the second one such that $\slashed{q} \slashed{q}=0$,
\begin{align}
    &\Big((\pol_{i-1}\Cdot\ell_{i-1})\ids-\fsl_{i-1}\Big)\frac{\slashed{q}\slashed{\ell}_{i-1}}{2q\Cdot\ell_{i-1}}\Big((\pol_{i}\Cdot\ell_{i})\ids-\fsl_{i}\Big)\frac{\slashed{q}\slashed{\ell}_{i}}{2q\Cdot\ell_{i}}\nonumber\\
    &=\Big((\pol_{i-1}\Cdot\ell_{i-1})\ids-\fsl_{i-1}\Big)\Big((\pol_{i}\Cdot\ell_{i})\ids-\fsl_{i}\Big)\frac{\slashed{q}\slashed{\ell}_{i}}{2q\Cdot\ell_{i}} + \mathcal{O}(q\Cdot k_i)+\mathcal{O}(q\Cdot\pol_i)\,,
\end{align}
where $\mathcal{O}(q\Cdot k_i)$ and $\mathcal{O}(q\Cdot\pol_i)$ collect terms proportional to $q\Cdot k_i$ and $q\Cdot \pol_i$, which are generated by the commutator $[\slashed{q},\fsl_{i+1}]=q\Cdot k_{i+1}\slashed{\pol}_{i+1}-q\Cdot\pol_{i+1}\slashed{k}_{i+1}$.
First setting aside these terms, we effectively have $\sum_{\text{states}}\bar{\chi}\xi\rightarrow \ids$ for all internal fermions but the last one. Consequently, the parity even contribution can be written as
\begin{align}
    \text{Cut}_{\text{$n$-gon}}^{\text{fermion}}\,\Big|_{\text{even}}
    &=-\frac{1}{2}\tr\bigg[\frac{\slashed{q}\slashed{\ell}_n}{2q\Cdot\ell_n} \prod_{i=1}^{n}\Big((\pol_i\Cdot\ell_i)\ids-\fsl_i\Big)\bigg] + \mathcal{O}(q\Cdot k_i)+\mathcal{O}(q\Cdot\pol_i)\nonumber\\
    &=-\frac{1}{4}\tr\bigg[\prod_{i=1}^{n}\Big((\pol_i\Cdot\ell_i)\ids-\fsl_i\Big)\bigg]\,.
\end{align}
To obtain the final result, we commute $\slashed{q}$ to the end and cycle it back to the original position. 
 This process will generate a manifestly  $q$-independent contribution $ \{\slashed{q},\slashed{\ell}_n \}/(2q\Cdot\ell_n)=1$ and potentially also
% the independent terms $\mathcal{O}(q\Cdot k_i)+\mathcal{O}(q\Cdot\pol_i)$ 
separate terms of the form $\mathcal{O}(q\Cdot k_i)$ and $\mathcal{O}(q\Cdot\pol_i)$ 
 from the commutators of $\slashed{q}$ with $\fsl_i$. However, all the $q$ dependence from $q\Cdot k_i$ and $q\Cdot\pol_i$ must cancel out identically, since cuts are gauge invariant quantities and cannot depend on the reference vector.
Thus for the last state sum, we effectively have $\sum_{\text{states}}\bar{\chi}\xi\rightarrow \frac{1}{2}\ids$ and reproduce the parity even part of \cref{altderiv}.
This is exactly the state-sum rule used by ref.~\cite{Edison:2020uzf} to obtain the forward limit.

\section{The explicit form of the seven trace \texorpdfstring{$\trVS$}{trmax}}
\label{app:7trace}

In order to unpack the maximal-cut contribution \cref{hepnum} to the heptagon numerator of \msym, one can decompose the seven-point instance of the $\trVS$ defined in \cref{deftrvs} into
\begin{align}
    &\trVS(f_1f_2f_3f_4f_5f_6f_7)=\frac{1}{2}\Big[t_{12}(f_1,f_{[2,3]},f_4,f_5,f_6,f_7)+(2,3|2,3,4,5,6,7)\Big]\nonumber\\
    &\qquad+\frac{1}{12}\Big[t_8(f_1,f_{[[[2,3],4],5]},f_6,f_7)+t_8(f_1,f_{[[2,[3,4]],5]},f_6,f_7)+t_8(f_1,f_{[2,[[3,4],5]]},f_6,f_7)\nonumber\\
    &\qquad\qquad\quad+t_8(f_1,f_{[2,[3,[4,5]]]},f_6,f_7)+(2,3,4,5|2,3,4,5,6,7)\Big]\nonumber\\
    &\qquad+\frac{1}{12}\Big[\Big(t_8(f_1,f_{[[2,3],4]},f_{[5,6]},f_7)+t_8(f_1,f_{[[2,3],4]},f_{[5,7]},f_6)\nonumber\\
    &\qquad\qquad\quad+t_8(f_1,f_{[[2,3],4]},f_{[6,7]},f_5)+(2\leftrightarrow 4)\Big)+(2,3,4|2,3,4,5,6,7)\Big]\nonumber\\
    &\qquad+\frac{1}{8}\Big[t_8(f_1,f_{[2,3]},f_{[4,5]},f_{[6,7]})+t_8(f_1,f_{[2,3]},f_{[4,6]},f_{[5,7]})\nonumber\\
    &\qquad\qquad+t_8(f_1,f_{[2,3]},f_{[4,7]},f_{[5,6]})+(3|3,4,5,6,7)\Big]\,,
\end{align}
where the $t_{12}$-tensor is given by \cref{t12ten}.
Here we have used the shorthand notation $f_{[i,j]}=[f_i,f_j]$, $f_{[[i,j],k]}=[[f_i,f_j],f_k]$, etc., to compactify the expression. See \cref{5ptex,6ptex} for its five- and six-point counterparts.

\section{Analysis of the \texorpdfstring{$\mu^2$}{mu2} term freedom of the hexagon numerator} 
\label{sec:mu_term}

In this appendix, we shall demonstrate that $\mu^2$-term deformations of the parity odd hexagon numerator \cref{eq:N6ptoddsym} of \msym are incompatible with manifest crossing symmetry. 
There are five independent Levi-Civita identities that contain $\ell^2$-terms that may affect the anomaly considered in \cref{sec:3.2}. We have the 11-term Schouten identity \eqref{eq:schouten}
\begin{equation}
0 =\ell^\mu \varepsilon_{10}(\pol_1,\pol_2,\pol_3,\pol_4,\pol_5,\epsilon_6,k_3,k_4,k_5,k_6) \,+\, {\rm cyclic}(\ell,\pol_1,\pol_2,\pol_3,\pol_4,\pol_5,\epsilon_6,k_3,k_4,k_5,k_6)\,,
\end{equation}
plus four more similar ones obtained by swapping the four momenta $k_3,k_4,k_5,k_6$ for any of the five independent momenta. The loop momentum involved in this relation is strictly in ten dimensions. Thus if we contract it with another $\ell_{\mu}$ in the context of dimensional regularization, we get
\begin{align}
0 &= \ell^2_{(10)} \varepsilon_{10}(\pol_1,\pol_2,f_3,f_4,f_5,f_6)- \pol_1\Cdot\ell \, \varepsilon_{10}(\ell, \pol_2,f_3,f_4,f_5,f_6)+ \pol_2\Cdot\ell \, \varepsilon_{10}(\ell, \pol_1,f_3,f_4,f_5,f_6) \nn \\    
& \quad + 2\Big[ \pol_3\Cdot\ell\,\varepsilon_{10}(\ell, \pol_1,\pol_2,k_3, f_4,f_5,f_6) - k_3\Cdot \ell\, \varepsilon_{10}(\ell, \pol_1,\pol_2,\pol_3, f_4,f_5,f_6) + \text{cyclic}(3,4,5,6) \Big]\,,
\end{align}
where $\ell^2_{(10)}=\ell^2{-}\mu^2$ such that the identity holds after integration.
However, when considering the strict $d=10$ integrand, we can freely add the Levi-Civita identities without the $\mu^2$ term, since the strict $d=10$ cuts are not sensitive to the extra-dimensional momentum. Or, equivalently, we can freely add the $\mu^2$ terms. Thus it is legitimate to add the following five terms with free parameters $\alpha_i$ to the hexagon numerator,
\begin{align}
& \delta  \nmodd{123456} = \mu^2 \Big[  \alpha_2 \varepsilon_{10}(\pol_1,\pol_2,f_3,f_4,f_5,f_6) + \alpha_3 \varepsilon_{10}(\pol_1,\pol_3,f_2,f_4,f_5,f_6)  \\ 
    & \quad + \alpha_4 \varepsilon_{10}(\pol_1,\pol_4,f_2,f_3,f_5,f_6)+ \alpha_5 \varepsilon_{10}(\pol_1,\pol_5,f_2,f_4,f_3,f_6)+ \alpha_6 \varepsilon_{10}(\pol_1,\pol_6,f_2,f_4,f_3,f_5)\Big]\,.
     \nn
\end{align}
If we require that $\delta  \nmodd{123456}$ respects the crossing symmetry of the hexagon numerator, then it sets all the $\alpha_i=0$. Thus, if the crossing symmetry is manifest in the hexagon numerator (without relying on Levi-Civita identities), there is no freedom to modify the $\mu^2$ terms and the anomaly will be unique.

\bibliography{ref}
\bibliographystyle{JHEP}

\end{document}